\documentclass[final,5p,times,twocolumn]{elsarticle}
\usepackage{graphicx}
\usepackage{amssymb}
\usepackage{amsmath}
\usepackage{amsthm}
\usepackage{lineno,hyperref}
\usepackage{color}
\hypersetup{colorlinks=true,linkcolor=red,citecolor=cyan,}
\modulolinenumbers[200]

\journal{Journal Name}

\begin{document}

\begin{frontmatter}

\title{Quasiperiodic oscillations, quasinormal modes and shadows of \\ Bardeen-Kiselev Black Holes}

\author[address2,address3,address4,address5]
{Javlon Rayimbaev}
\ead{javlon@astrin.uz}

\author[address7]
{Bushra~Majeed}
\ead{bushra.majeed@ceme.nust.edu.pk}
\author[address8,address9,address10]
{Mubasher~Jamil}
\ead{bushra.majeed@ceme.nust.edu.pk}
\author[address11]
{Kimet~Jusufi}
\ead{kimet.jusufi@unite.edu.mk}

\author[address12]
{Anzhong~Wang}
\ead{anzhong_wang@baylor.edu}

\address[address2]{Ulugh Beg Astronomical Institute, Astronomy Str. 33, Tashkent 100052, Uzbekistan}
\address[address3]{Department of Engineering, Akfa University, Milliy Bog Street 264, Tashkent 111221, Uzbekistan}
\address[address4]{Institute of Fundamental and Applied Research of National Research University TIIAME, Kori Niyaziy Street
39, Tashkent 100000, Uzbekistan}

\address[address5]{National University of Uzbekistan, Tashkent 100174, Uzbekistan}

\address[address7]{College of Electrical and Mechanical Engineering (CEME), National University of Sciences and Technology, Islamabad, 44000, Pakistan.}

\address[address8]{Institute for Theoretical Physics and Cosmology, Zhejiang University of Technology, Hangzhou 310023 China}

\address[address9]{School of Natural Sciences, National University of Sciences and Technology, Islamabad, 44000, Pakistan}

\address[address10]{Canadian Quantum Research Center, 204-3002, 32 Ave, Vernon, BC, V1T 2L7, Canada}

\address[address11]{Department of Physics, State University of Tetovo  Ilinden nn, 1200 Tetova, Macedonia}

\address[address12]{GCAP-CASPER, Physis Department, Baylor University, Waco, Texas 76798-7316, USA}

\date{Received: date / Accepted: date}

\begin{abstract}
In this article, we study the particle dynamics around a static and spherically symmetric Bardeen-Kiselev black hole (BK BH) which is a solution of the Einstein-non-linear Maxwell field equations along with a quintessential field. We discuss its essential geometrical properties such as scalar invariants and size of innermost stable circular orbits. Dynamics of test particles around the BH is also studied. Moreover, we also computed the fundamental frequencies of a test particle orbiting the BH in a slightly perturbed orbit. Further, the degeneracy relations between the spin of rotating Kerr BH and magnetic charge of the BK BH at the values of the quintessential parameter $\omega_q=-1/3$ and  in terms of the same values of innermost stable circular orbits (ISCOs) radius, energy efficiency, twin-peaks quasiperiodic oscillations (QPOs) frequencies and impact parameter for photon are also discussed in detail. Finally, we have investigated the relationship between the shadow radius and the real part of the quasinormal mode (QNM) frequency.

\end{abstract}


\end{frontmatter}

\linenumbers

\section{Introduction}

In the last two decades the astrophysical research (based on the detection of the gravitational waves by the merger of BHs \cite{AbbottPRL2016} and the BH shadow of the M87 \cite{EHTColl2019ApJL,EHTColl2019ApJL1,EHTColl2019ApJL3,EHTColl2019ApJL4,EHTColl2019ApJL5,EHTColl2019ApJL6}) has provided the strong evidences for the existence of BHs in the center of many galaxies, such as the spiral Milky Way and the elliptical M87 galaxy \cite{2017ApJ}. These astronomical observations are very helpful to understand the Universe on large scale structure. The Shadow of a BH is one of the most  interesting topics of research in recent years, and a number of interesting results have been found \cite{2019JCAP, 2019PhRvD.100b4014A, 2020CQGra, 2020EPJC}.
 In particular, Synge and Luminet \cite{1966MNRAS,1979A&A}  studied the shadow of a Schwarzschild BH and later Bardeen  analysed the shadow of a Kerr BH. The  non-rotating BHs have shadows which are in circular form, while the spin parameter of rotating BHs and the specific inclination angle of view cause a deformation  in the boundary of the shadow \cite{1974IAUS, 1998mtbh.book,Jusufi2019PhRvD}.
Other interesting topics concerning the physical nature of a BH's photon ring and the shadow have been recently discussed in the literature \cite{2012PhRvD, Bambi09, Bambi2019PhRvD}.
 Kiselev derived a new solution of Einstein’s field equations for a BH surrounded by energy-matter called quintessential field \cite{Kiselev03a,Kiselev03}. The components of stress-energy tensor for the Kiselev BH solution are \cite{Kiselev03}
\begin{eqnarray}\label{tmunu}
    T^{t}_{t}&=&T^{r}_{r}=\rho_q,\nonumber\\
    T^{\theta}_{\theta}&=&T^{\phi}_{\phi}=-\frac{1}{2}\rho_q (1+3 \omega_q).
\end{eqnarray}

There are binary systems, known as microquasars, consisting of a BH and a companion star. An accretion disk and relativistic jets form, as matter drifts from the companion star onto the BH. Friction in the matter of the accretion disk produces electromagnetic radiations and X-rays in the surrounding of the BH horizon. The QPOs or profiled spectral lines observed in the X-ray spectra are counted as one of the most effective tests in the strong gravity regime. Spectroscopy (frequency distributions of photons) and timing (time dependence of the photon number) for particular microquasars are the methods commonly used to obtain  useful information about the range of the system parameters  \cite{remillard2006x}. 
 Based on the observed frequencies of QPOs, ranging from few mHz up to 0.5 kHz, different categories of QPOs were recognized. Most prominent ones are the high frequency (HF) and low frequency (LF) QPOs with frequencies up to 500 Hz and 30 Hz, respectively. The HF QPOs are generally identified with the twin-peaks which have approximate  frequency ratio  3:2 as observed in the galactic BH microquasars GRS 1915+105, XTE 1550-564 and GRO J1655-40. Since the detection of QPOs many models have been developed, in the framework of general  theory of relativity or alternative theories of gravity. These models are proposed to fit the data with the observed QPOs, some of which are the disko-seismic models, warped disk model, hot-spot models and many versions of resonance models, such as forced or Keplerian. Geodesic oscillatory models are the extended ones where the observed frequencies   are associated to the frequencies of the geodesic orbital and epicyclic motion. In these models, the characteristic frequencies of HF QPOs are  approximately the same as the values of the frequencies of test particles and geodesic epicyclic oscillations near the ISCO. Based on this observation, this  model is developed, which involves the frequencies of oscillations of the orbital motion around Kerr BHs \cite{stuchlik2013multi}. However, the physical mechanism behind the generation of HF QPOs is not exactly  known yet, as none of these models can match exactly with the observational data from different sources \cite{bursa2005high}.

In this work we will discuss the dynamics of both test neutral and  electrically charged test particles around a
regular Bardeen-Kiselev BH (BKBH) immersed in an asymptotically uniform external magnetic field, and the analysis will be carried out by studying shadows, quasinormal modes (QNMs) and QPOs using the BKBH as a source. An outline of this article is as follows: In section II, the spacetime of a BKBH  and the corresponding scalar invariants are discussed. In section III,  dynamics of test particles around a BKBH is studied. The analysis for inner stable circular orbits (ISCO) and energy efficiency is also discussed. In section IV, the fundamental frequencies and Harmonic oscillations of the test particles are investigated. In section V, the motion of massless particles is discussed. In section VI astrophysical applications of the BH in comparison to the Kerr BH are discussed, we specifically focus on the differences of ISCO, bolometric luminosity and QPO frequencies between the BK  and Kerr BHs. In section VII, we present our main results. The metric signature throughout the work is $(-,+,+,+)$ and ${\rm c}=1,G=1$.

\section{Regular Bardeen-Kiselev BH}

The Bardeen BH emerges as a regular BH which satisfies the weak energy condition, and is obtained by coupling  the Einstein’s gravity minimally with the non-linear electrodynamics. 
However, we assume the generalized action for this theory as: 
\begin{equation}
    \mathcal{S}=\frac{1}{16\pi}\int{d^4x \sqrt{-{\rm g}} \left(R- \mathcal{L}(F)+\mathcal{L}_m\right)},
\end{equation}
where $R$ is the scalar curvature, and $\mathcal{L}_m$ is the anisotropic matter field whose non-vanishing stress-energy tensor components are given by Eq. (\ref{tmunu}) and \begin{equation}
    \mathcal{L}(F)=\frac{3}{2sg^2}\Bigg(\frac{\sqrt{2g^2F}}{1+\sqrt{2g^2F}}\Bigg)^{\frac{5}{2}},
\end{equation}   with
\begin{equation}
    F= \frac{1}{4}F_{\mu \nu }F^{\mu \nu}, \qquad F_{\mu \nu} =2 \triangledown_{[\mu} A_{\nu]},
\end{equation} 
where $F_{\mu \nu }$ denotes  the Maxwell electromagnetic field tensor
and $s=\frac{\rvert g\rvert }{2m}$, with $m$ and $g$ being free parameters, denoting the mass and magnetic charge of a nonlinear self gravitating monopole respectively. A BK spacetime 
satisfying the weak energy condition can be obtained by coupling Einstein’s
gravity to a nonlinear electrodynamics field \cite{Ayon-Beato99a,Bronnikov00,Bronnikov01}. This model is described by
the metric given by \cite{Ghaderi2016ApSS,ghaderi2018thermodynamics}
\begin{eqnarray}\label{1}
ds^{2}&=&-f(r)dt^2+ \frac{1}{f(r)}dr^2 +r^2(d\theta^2 +\sin^2\theta
d\phi^2),
\end{eqnarray}
where
\begin{equation}
\label{metric}f(r)=1-\frac{2M r^2}{(r^2+g^2)^{\frac{3}{2}}}-\frac{c}{r^{3\omega_q +1}}.
\end{equation}
Here $M$ is total mass of the BH, 
 $\omega_q$ is the state parameter with $-1< \omega _q<-1/3$ and $\rho_q$ is the density   which is always positive and  $\rho_q=-3\omega_q(c/2)/r^{3(1+\omega_q)}$ and $c>0$ is a normalization factor. One can easily see that the solution (\ref{metric}) reflects the Schwarzschild metric when $g=c=0$ and as $c=0$ (without quintessential field) it becomes the pure regular Bardeen BH solution. When the parameter $\omega_q=-1$ the spacetime represents anti de-Sitter character with $\Lambda=3c$. 

\subsection{Horizon's Structure}

\begin{figure*}[ht!]
   \centering
  \includegraphics[width=0.298\linewidth]{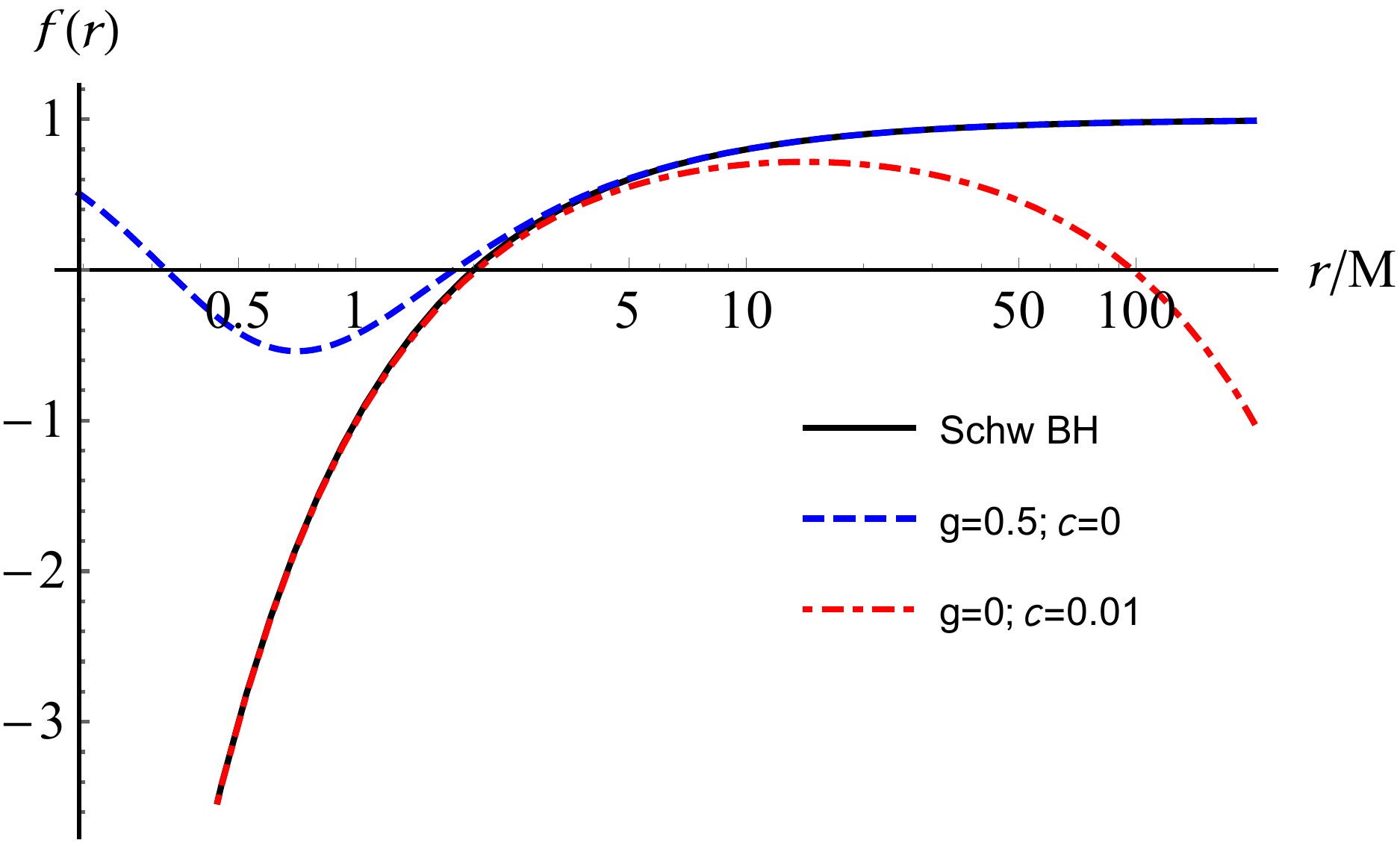}
    \includegraphics[width=0.298\linewidth]{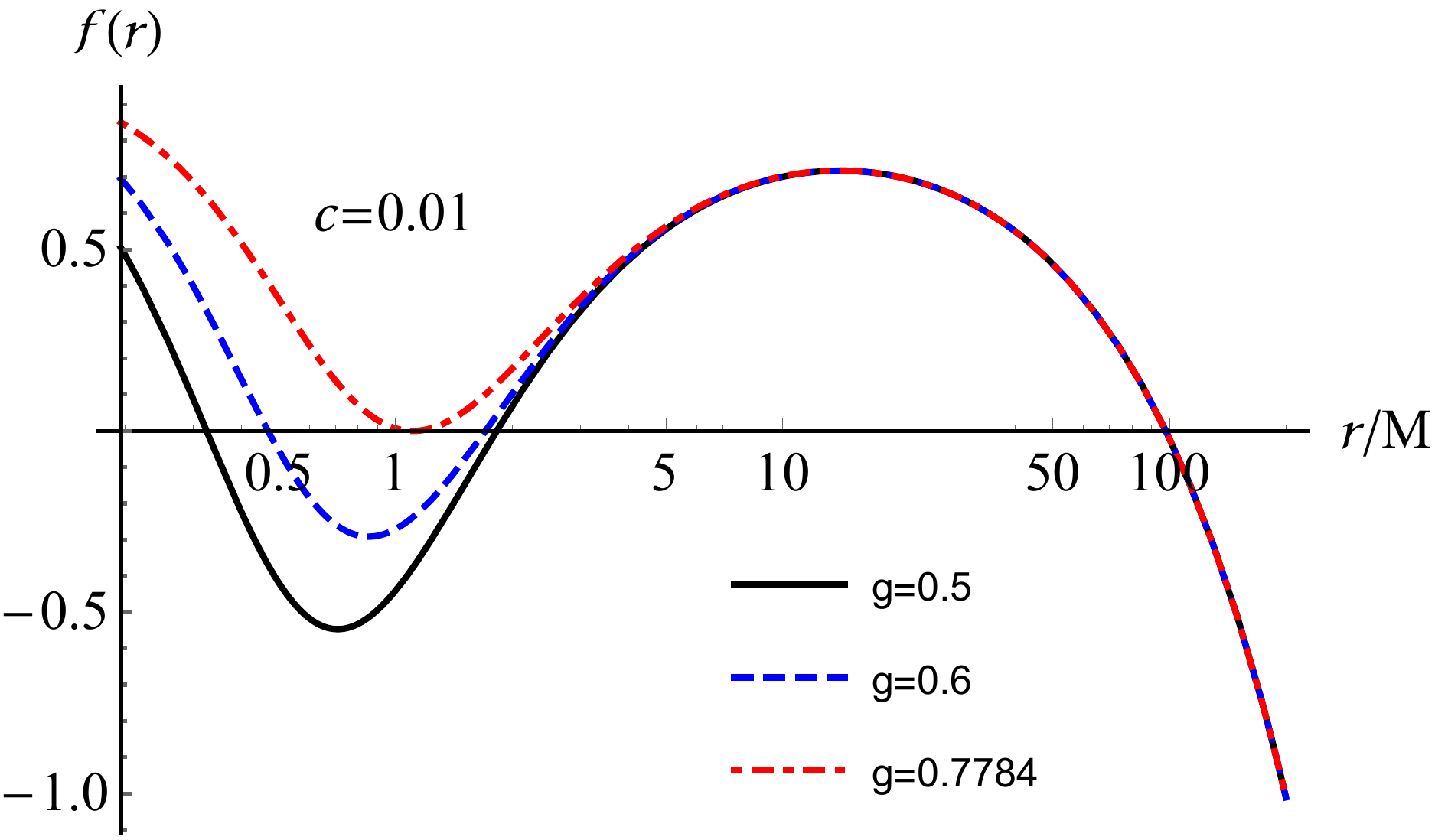}
    \includegraphics[width=0.298\linewidth]{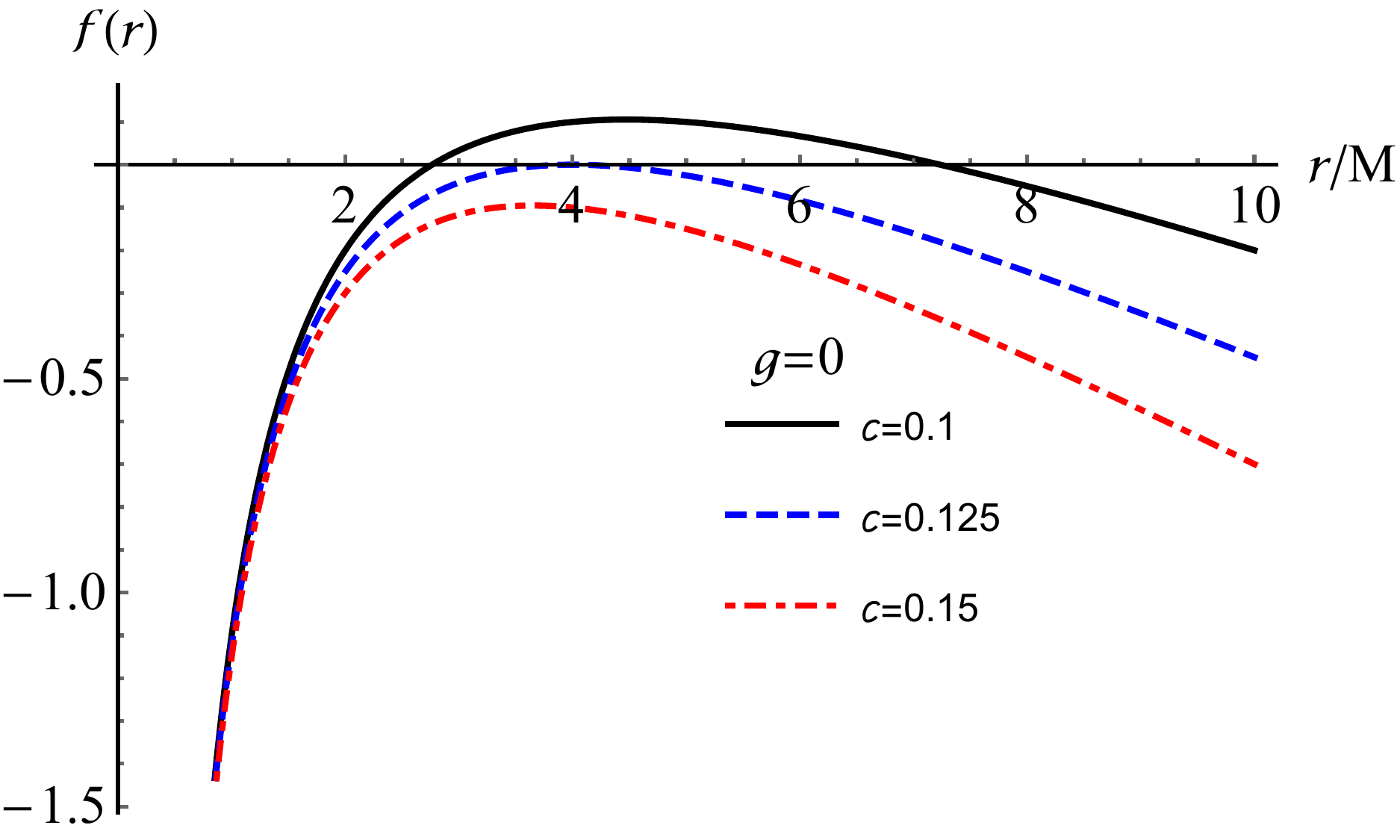}
	\caption{Lapse function}. \label{lapsef}
\end{figure*}

The horizon of the BH (denoted by $r_h$) is defined as a null-hypersurface where the lapse function vanishes.
In the case of $\omega_q=-2/3$, the lapse function contains the term proportional to the radial coordinate $\sim c r$ (where $c$ is a parameter with unit $1/M$).  When $g=0$, the spacetime has two horizons: the BH horizon at $r_h$ and quintessential cosmological horizon $r_q$ (which is inverse proportional to the parameter $c$). There are upper limits in the parameters of magnetic charge and the quintessential field at which has horizon in the spacetime. When the magnetic charge $g\neq 0$ ($g<g_{extr}$), there are three horizons: two BH horizons (inner and outer) and the quintessential horizon. In case of Schwarzschild-Kiselev BH ($g=0,\ c\neq0$) only two horizons related to the BH and quintessential field. In Fig.\ref{lapsef}, we have graphically shown that there are extreme values of parameters $g$ and $c$. That means the lapse function have three zeros when $g<g_{extr}$ and $c<c_{extr}$,  and in the case when both magnetic charge and the quintessential parameters equals to their extreme values, the BH will have only two horizons (out and inner horizons of the BH coincide creates only one BH horizon) (see Fig.\ref{lapsef}).  

\begin{figure}[ht!]
   \centering
  \includegraphics[width=0.9\linewidth]{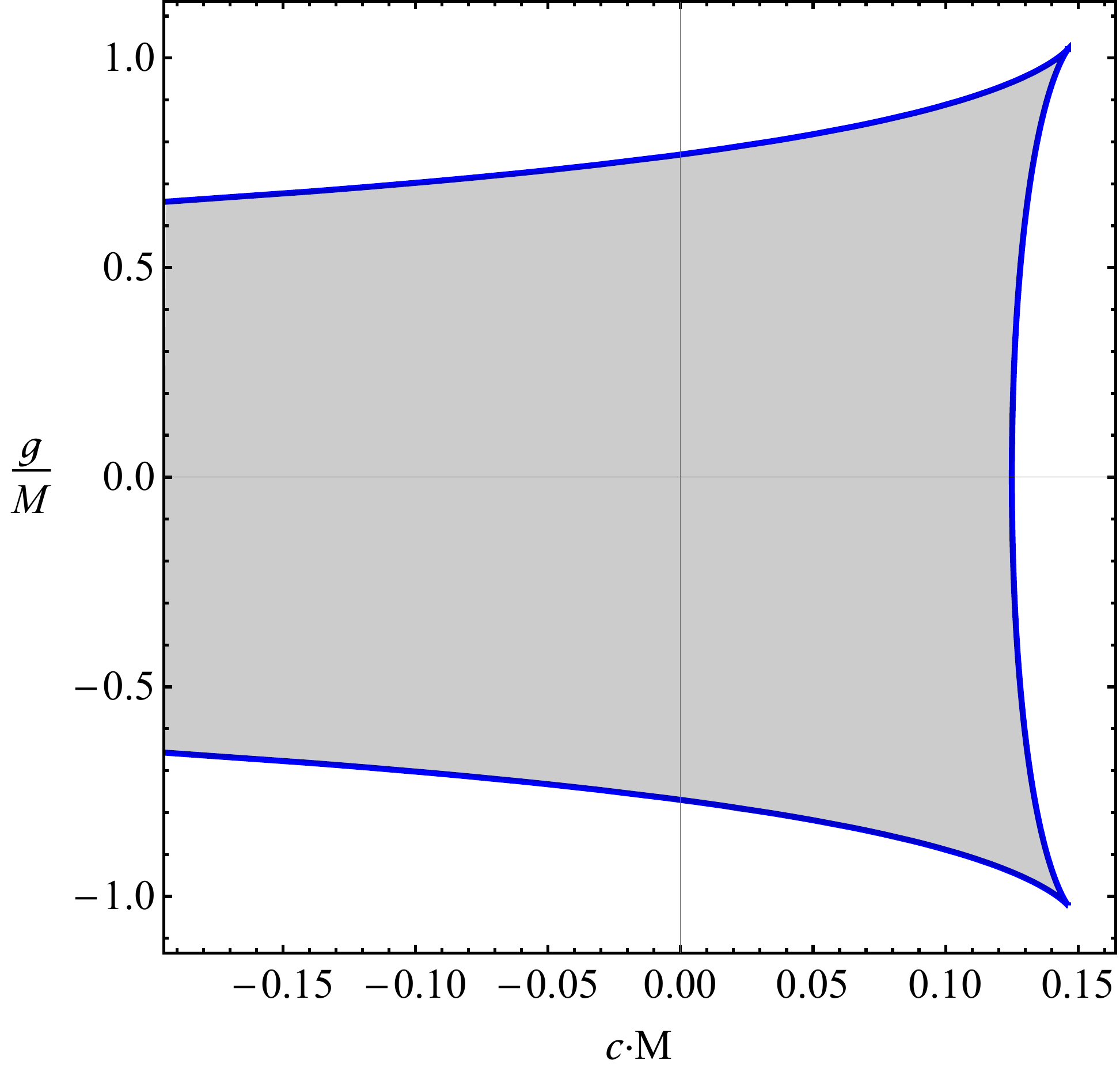}
    	\caption{Relation between magnetic charge and the quintessential field parameters for the existence of BH horizon. The grey-coloured area indicates values of the magnetic charge and quintessential parameters for the existence of BH horizon.} Blue lines correspond to extreme values of the parameters $g$ and $c$ \label{gvscfig}
\end{figure} 

In Fig.\ref{gvscfig}, we have presented values of magnetic charge of the BH and the parameter  $\omega_q=-2/3$. Here, the shaded area corresponds to values of magnetic charge and the parameter $c$ corresponding to $\omega_q=-2/3$ which allow the existence of the BH horizon. So, outside the shaded area, there is no (outer or quintessential or cosmological) horizon. The extreme value of $c$ at $g=0$ has $c_{extr}=1/8$, and this value slightly increases with the increase of the magnetic charge and reaches up to $c=0.15$. Similarly, the extreme of magnetic charge of the Bardeen BH is $g=(4 M)/(3 \sqrt{3})\simeq 0.7698 M$ in the absence of quintessential field, and when parameter $c$ increases up to its extreme value, the extreme magnetic charge reaches $g=M$ .

\begin{figure}[h!]
   \centering
  \includegraphics[width=0.98\linewidth]{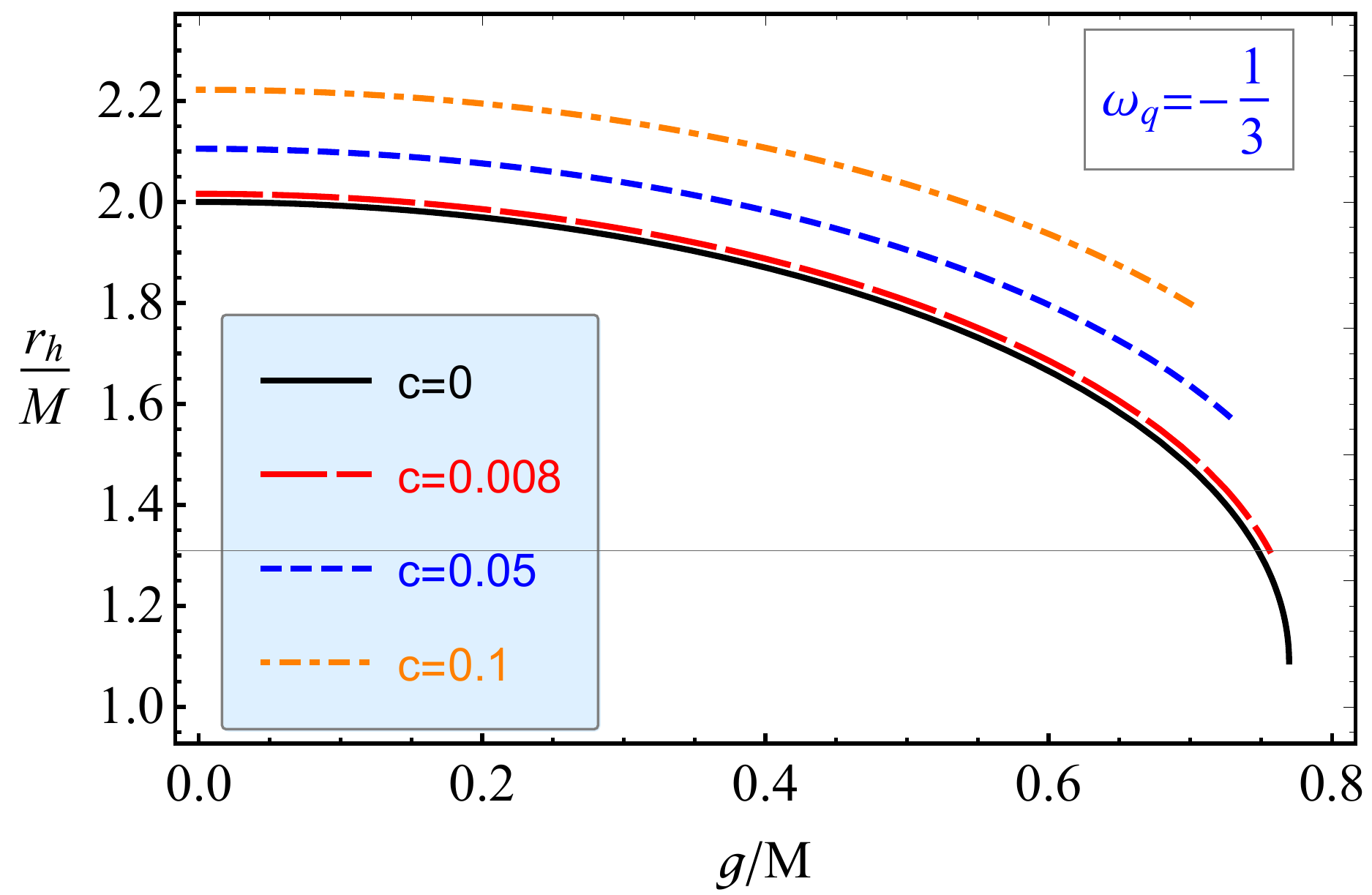}
    \includegraphics[width=0.98\linewidth]{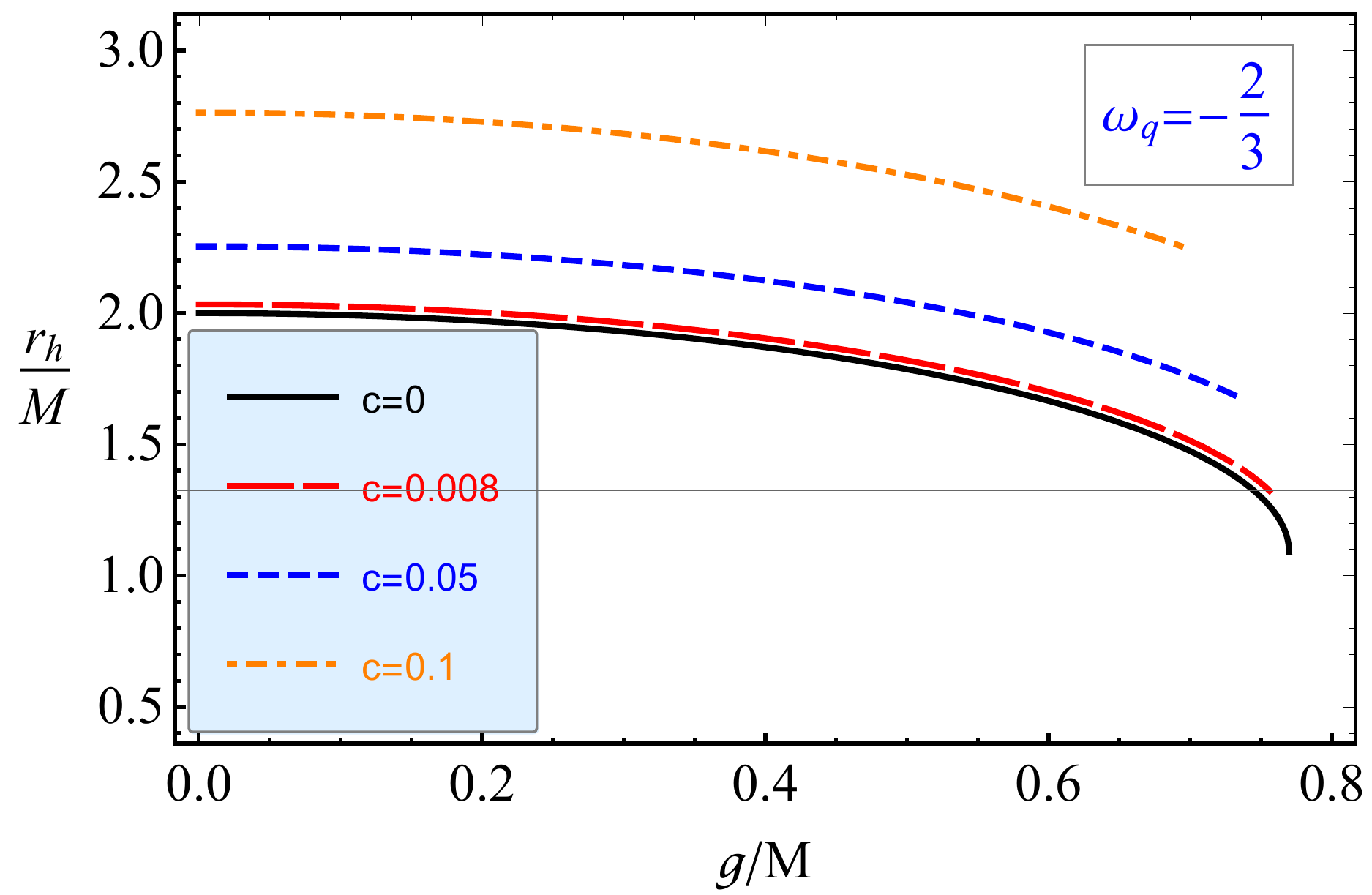}
	\caption{Event outer horizon radius versus the magnetic charge of the BKBH, for different values of the normalization parameter $c$.}. \label{horizon}
\end{figure}

For a BKBH, we numerically solve  $f(r)=0$ for $r_h$ by choosing different values of the parameters $c$ and $g$, and the behaviour of the horizon is shown graphically in Fig.(\ref{horizon}). It can be seen that as the value of $c$ increases, the horizon is stretched farther than compared to the case when $c$ is small. Also, the horizon of the BH becomes smaller as $g$ is increased. The equation of state parameter $\omega_q$ also has an important role to determine the location of $r_h$. It can be observed from the figure that when the value of $\omega_q$ is large (small) the horizon is at a smaller (larger) radius in the top  and bottom panel respectively in Fig.~\ref{horizon}.

\begin{table}[ht]
\caption{ Minimal radius of the event horizon and the extreme value of magnetic charge for different values of the parameters $c$ and $\omega_q$. \label{Tab}}
\centering
\begin{tabular}{|c |c | c | c | c | c | c | c |}
\hline
$\omega_q \to $ &\multicolumn{2}{c|}{$-\frac{1}{3}$}  & \multicolumn{2}{c|}{$-\frac{2}{3}$} \\ 
 \hline $c$ $\downarrow$& $\frac{(r_h)_{min}}{M}$ & $\frac{g_{\rm extr}}{M}$ &  $\frac{(r_h)_{min}}{M}$ & $\frac{g_{\rm extr}}{M}$ \\ [1.0ex]
\hline
0& $1.08866$ & $0.7698$ & $1.08866$ & $0.7698$ \\[1.0ex]
\hline
0.001 & $1.09074$ & $0.77064$& $1.08975$ & $0.77057$ \\[1.0ex]
\hline
0.005 & $1.09922$ & $0.77405$ & $1.09413$ & $0.77367$ \\[1.0ex]
\hline
0.010 & $1.11017$ & $0.7784$ & $1.09966$ & $0.77758$ \\[1.0ex]
\hline
0.050 & $1.21585$ & $0.81831$ & $1.14596$ & $0.81032$ \\[1.0ex]
\hline
0.100 & $1.43367$ & $0.88861$ & $1.20962$ & $0.85533$\\[1.0ex]
\hline
0.120 & $1.59027$ & $0.929699$ & $1.23712$ & $0.87477$ \\[1.0ex]
\hline
0.130 & $3.68405$ & $0.961325$ & $1.25134$ & $0.88483$ \\[1.0ex]
\hline
\end{tabular}
\end{table}

\subsection{Scalar Invariants}

The curvature invariants are the quantities which help to understand  the properties of spacetime of a geometrical object like a BH. Well known scalar invariants include the Ricci scalar, the square of the Ricci tensor and the
Kretschmann scalar (the square of the Riemann curvature tensor). In this section, we calculate and analyse these quantities by observing their behaviors graphically.

\subsubsection{Ricci Scalar}

It can be shown that now the Ricci scalar for the BH metric (\ref{metric}) is given by,
\begin{eqnarray} \label{b1}
R\equiv g^{\mu \nu}R_{\mu \nu}=6 M g^2  \frac{4 g^2-r^2}{\left(g^2+r^2\right)^{7/2}}+3 \omega_q c\frac{3 \omega_q -1}{r^{3 (\omega_q +1)}}.
\end{eqnarray}

One can immediately see that in the case when $g=c=0$, the Ricci scalar also becomes zero. In pure Bardeen limit, it takes the form: $24 M/g^3$. Below, we analyse the effect of parameters $c, \omega_q$ graphically.
 
\begin{figure}[ht!]
   \centering
  \includegraphics[width=0.98\linewidth]{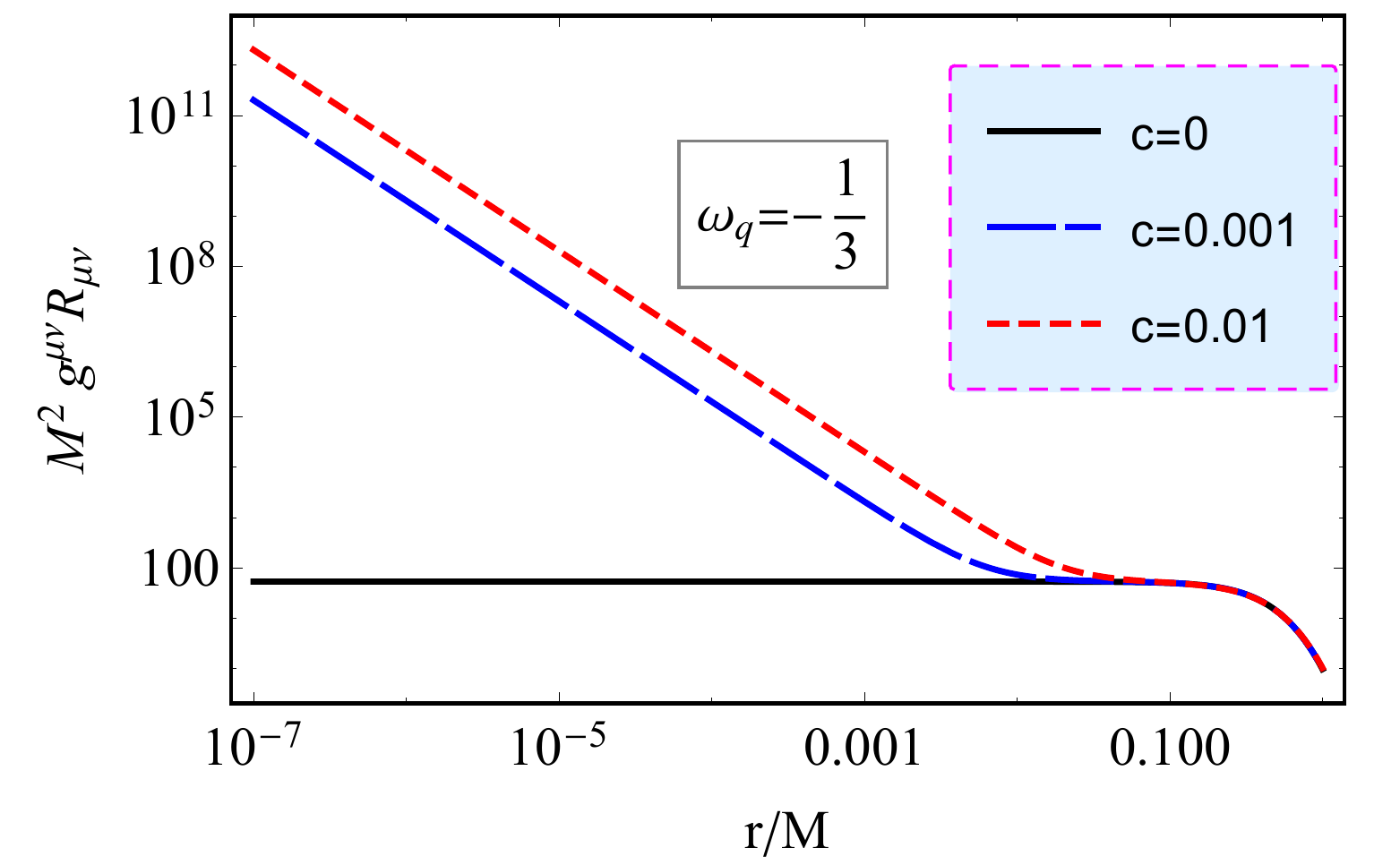}
  \includegraphics[width=0.98\linewidth]{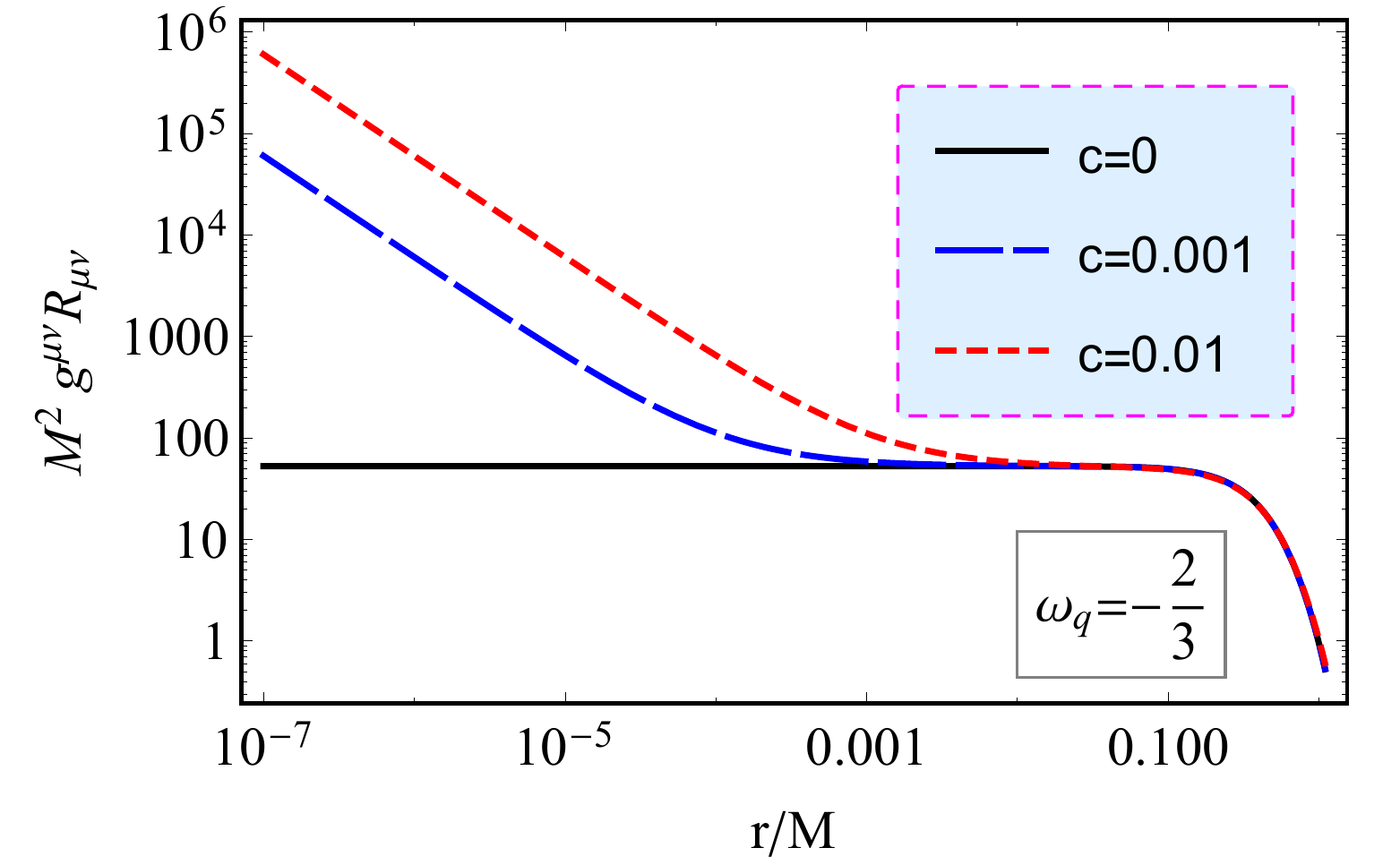}
   	\caption{Radial profiles of the Ricci scalar of the spacetime given in Eq.(\ref{metric}) for different values of the normalization parameter $c$ and at the extreme value of the magnetic charge of a regular BH.  \label{ricci}}
\end{figure} 

In Fig.\ref{ricci}, we plot Eq.(\ref{b1}), from which  we find that $R$ is well-defined near the central point $r=0$, hence the BH solution is regular at the center ($r\rightarrow 0$),  at the parameter $c=0$ (in the case of pure regular Bardeen BH) \cite{ayon2000bardeen}. However, for $\omega_q>-1$ (in the presence of the quintessential field), the Ricci scalar goes to infinity at $r\to 0$. It implies that the spacetime around Bardeen-Kiselev BH is not Ricci flat one. Fig.\ref{ricci} depicts the Ricci scalar of the spacetime. One may see that the Ricci scalar decreases with the increase of  $\omega_q$, while $c$ causes it to increase. 

\subsubsection{Square Ricci Tensor}

The next quantity to be studied is the square of Ricci tensor ${\cal R}=R^{\mu \nu}R_{\mu \nu}$. For the spacetime given in Eq.(\ref{Omega}) it is  given by, 
\begin{eqnarray}\label{SQR}
\nonumber
\frac{2}{9} {\cal R}&=&\frac{c^2 \omega_q^2}{r^{6 \left(\omega_q+1\right)}} \left(9 \omega_q^2+6 \omega_q+5\right) \\\nonumber&+&\frac{4 c g^2 M \omega_q }{r^{3 \left(\omega_q+1\right)}}\frac{2g^2 \left(3 \omega_q-1\right)-r^2 \left(9 \omega_q+7\right)}{\left(g^2+r^2\right)^{7/2}}\\&+&\frac{4 M^2 }{\left(g^2+r^2\right)^7\left(8 g^8-4 g^6 r^2+13 g^4 r^4\right)}.
\end{eqnarray}

For $c=0$ case, the square of the Ricci tensor takes the form $144M^2/g^6$. Due to the complicated form of the expression, we provide the analyses by plotting Eq.(\ref{SQR}). 

\begin{figure}[ht!]
   \centering
     \includegraphics[width=0.98\linewidth]{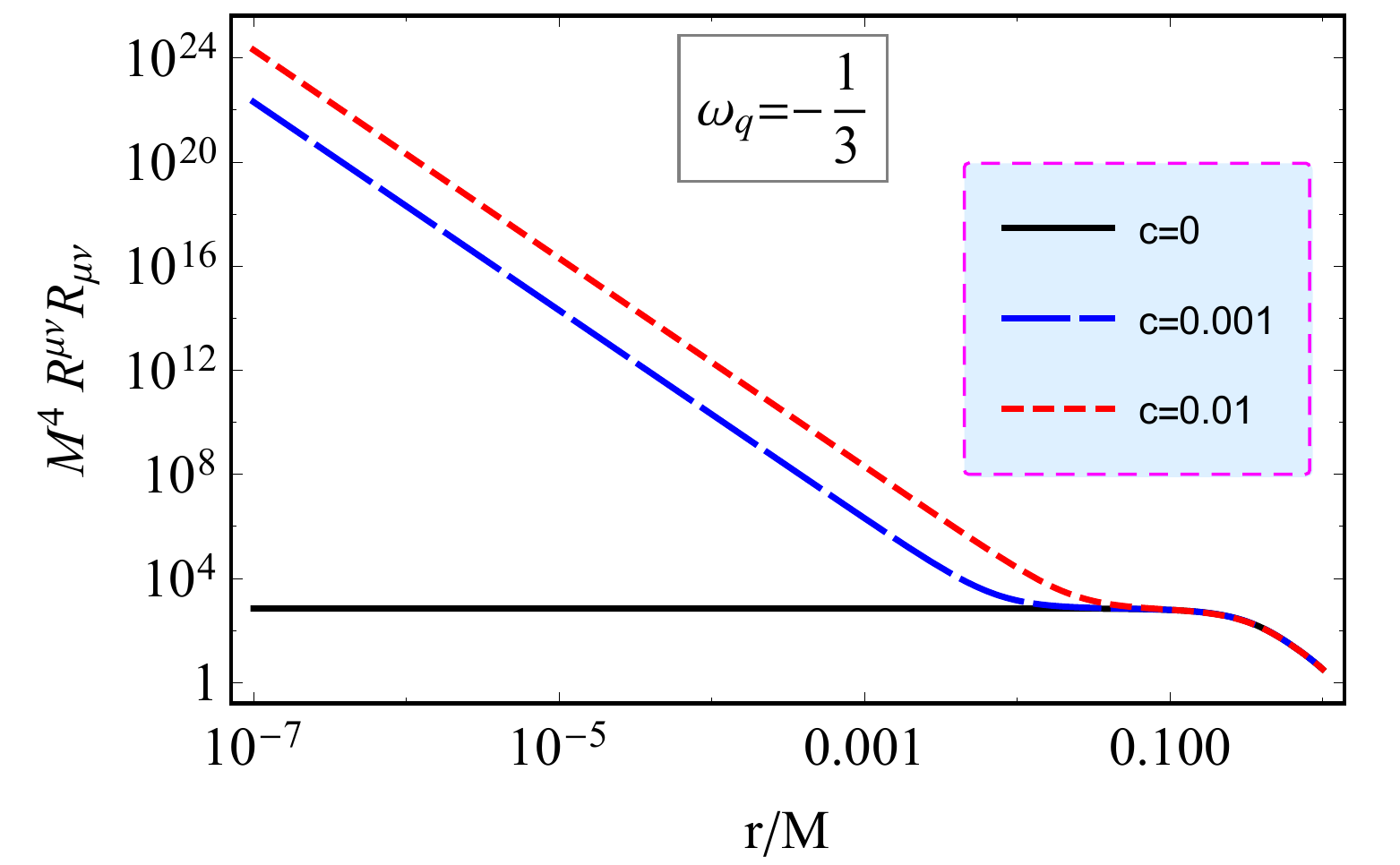}
   \includegraphics[width=0.98\linewidth]{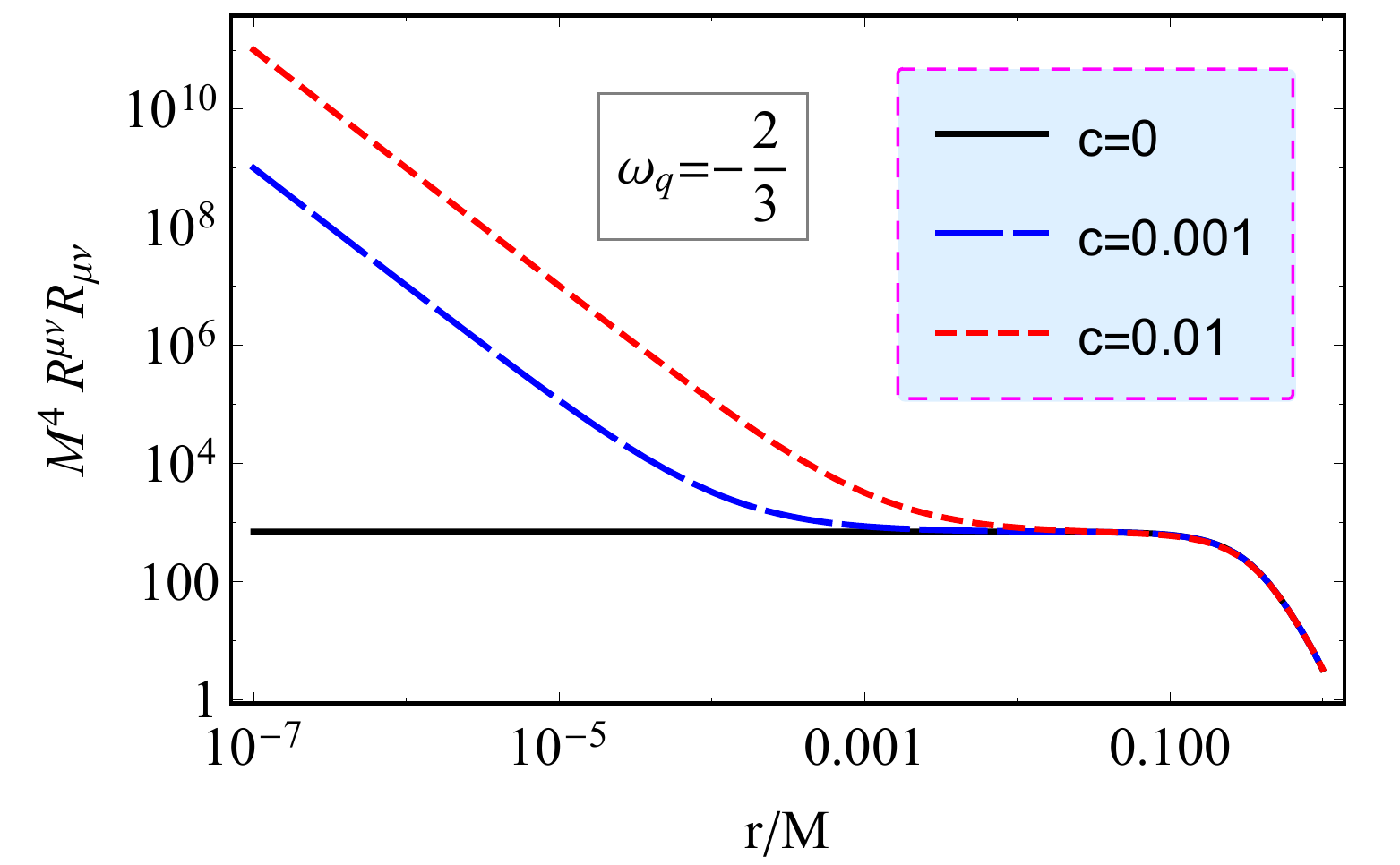}
	\caption{The square of Ricci tensor with the same choice of the parameters as those given in Fig.\ref{ricci}.}.  \label{RR}
\end{figure} 

This quantity is plotted in Fig.\ref{RR}, which shows that it is also well-defined at the centre $r=0$ at $c=0$, and all the features of the square of the Ricci tensor are just like the Ricci scalar. 

\subsubsection{Kretchmann scalar}

The Kretschmann scalar is another scalar invariant, which provides more information about the curvature of the spacetime, in particular,  it does not vanish  in a Ricci flat spacetime. The expression for the Kretschmann scalar with the spacetime of a BKBH (\ref{metric}) is
\begin{eqnarray}
&&\frac{1}{3}{\cal K}=\frac{4 c M }{r^{3 \left(\omega _q+1\right)}\left(g^2+r^2\right)^{7/2}}\Big\{2 g^4 \omega _q \left(3 \omega _q-1\right)\\\nonumber&&-g^2 r^2 \left[\omega _q \left(33 \omega _q+37\right)+6\right]+2 r^4 \left(\omega _q+1\right) \left(3 \omega _q+2\right)\Big\}\\\nonumber&&+\frac{c^2 }{r^{6 \left(\omega _q+1\right)}}\Big\{\omega _q \left[3 \omega _q \left(9 \omega _q \left(\omega _q+2\right)+17\right)+20\right]+4\Big\} \\\nonumber&&+\frac{4 M^2}{\left(g^2+r^2\right)^7} \left(8 g^8-4 g^6 r^2+47 g^4 r^4-12 g^2 r^6+4 r^8\right)
\end{eqnarray}

It is seen that the Kretchmann scalar is $96M^2/g^6$ when $c=0$. 

\begin{figure}[ht!]
   \centering
     \includegraphics[width=0.98\linewidth]{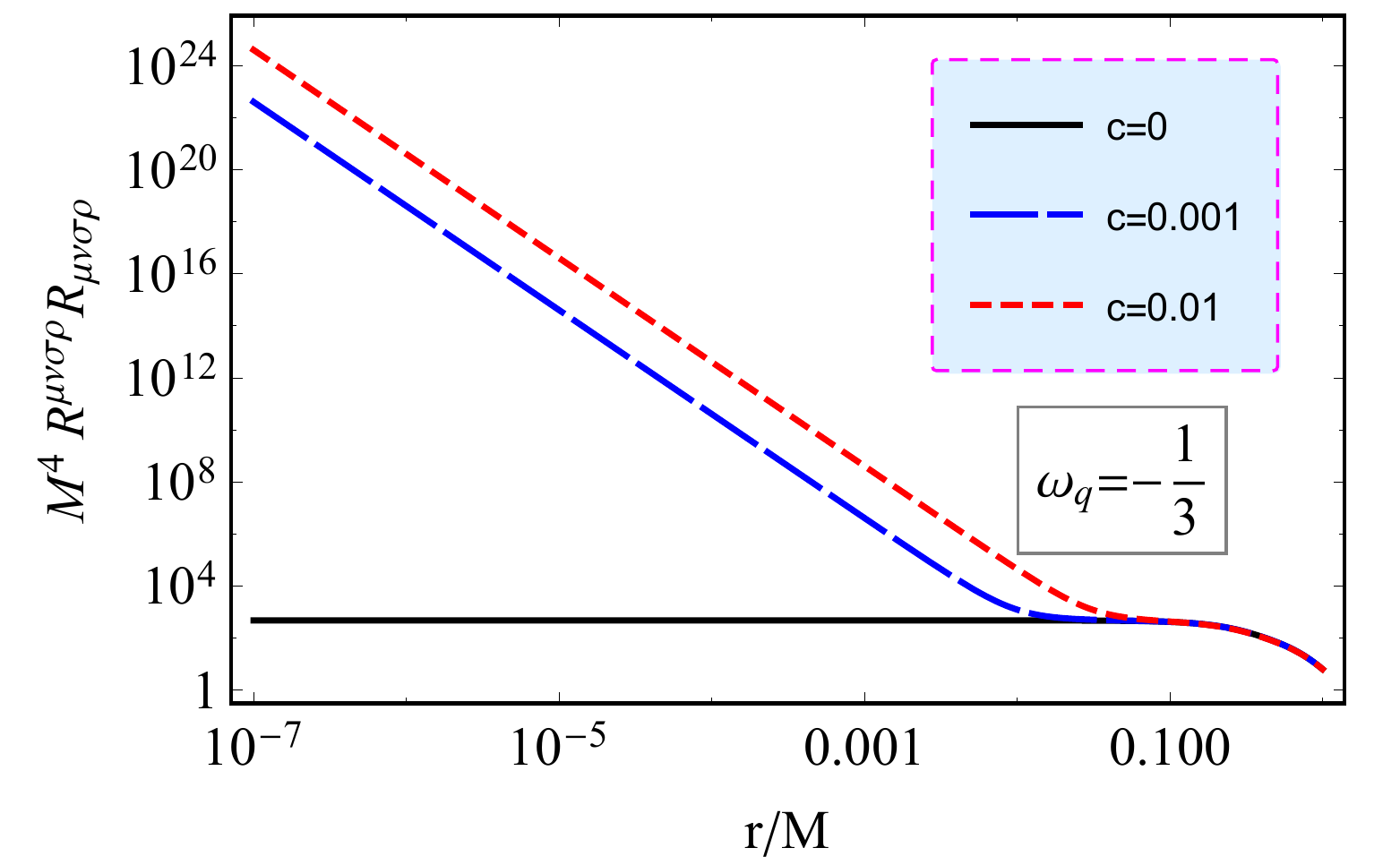}
   \includegraphics[width=0.98\linewidth]{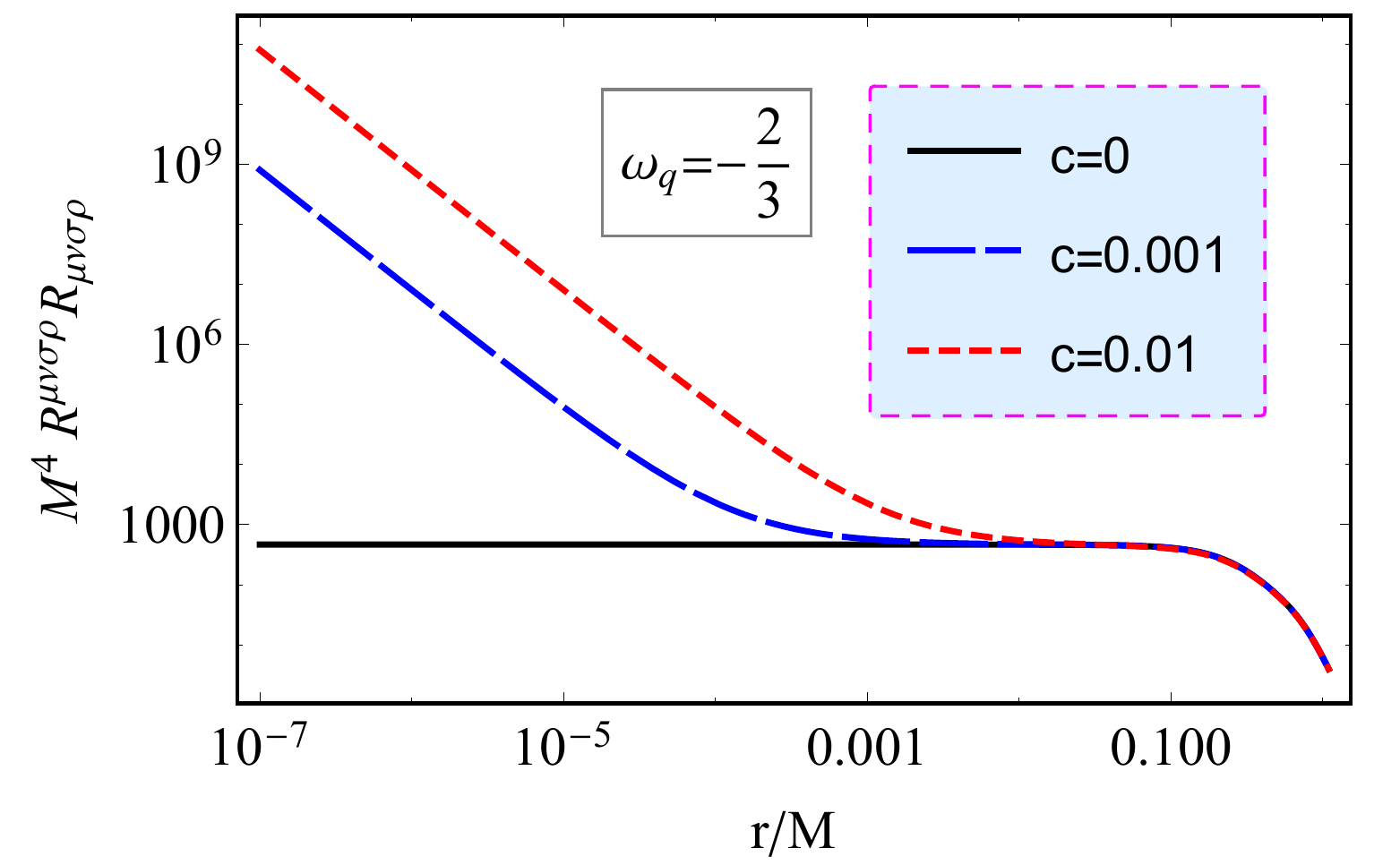}
	\caption{The Kretchmann scalar, with the same choice of the parameters as those given in Fig.\ref{ricci}.}.  \label{Kretchmanfig}
\end{figure} 

The same can be said about the properties of the Kretchmann scalar, that it is similar to the properties of other scalar invariants as discussed above.

\section{Test particle dynamics around a BKBH }
\subsection{Equation of Motion}
Useful symmetries of the static and spherically symmetric BH metric are the time translation and spatial rotation around the symmetry axis. The corresponding constants of motion can be calculated using the Killing vectors, given respectively by
\begin{equation}
\xi_{(t)}^{\mu}\partial_{\mu}=\partial_{t} , \qquad
\xi_{(\phi)}^{\mu}\partial_{\mu}=\partial_{\phi},
\end{equation}
here
$\xi_{(t)}^{\mu}=(1,~0,~0,~0)$ and $\xi_{(\phi)}^{\mu}=(0,~0,~0,~1)$.
The corresponding conserved quantities are the total energy
$\mathcal{E}$  and the azimuthal angular momentum $L_{z}={\cal L}$ of the moving
particle
\begin{equation}\label{2}
\dot{t}=\frac{\mathcal{E}}{f(r)}, \qquad \dot{\phi}=\frac{{\cal L}}{r^2}.
\end{equation}
Here, an over-dot denotes the differentiation with respect to the proper-time coordinate
$\tau$. From the normalization condition $u^{\mu}u_{\mu}=-1$, we have
\begin{eqnarray}\label{4}
\dot{r}^{2}=\mathcal{E}^2- V_\text{eff};\qquad
V_\text{eff}(r,\theta)= f(r)\Big(1 +\frac{{\cal L}^2}{r^2\sin^2\theta}\Big).
\end{eqnarray}
From now onward, the motion of the test particles at the equatorial plane will be considered by constant plane, choosing $\theta=\pi/2$. 

\begin{figure}[ht!]
   \centering
     \includegraphics[width=0.98\linewidth]{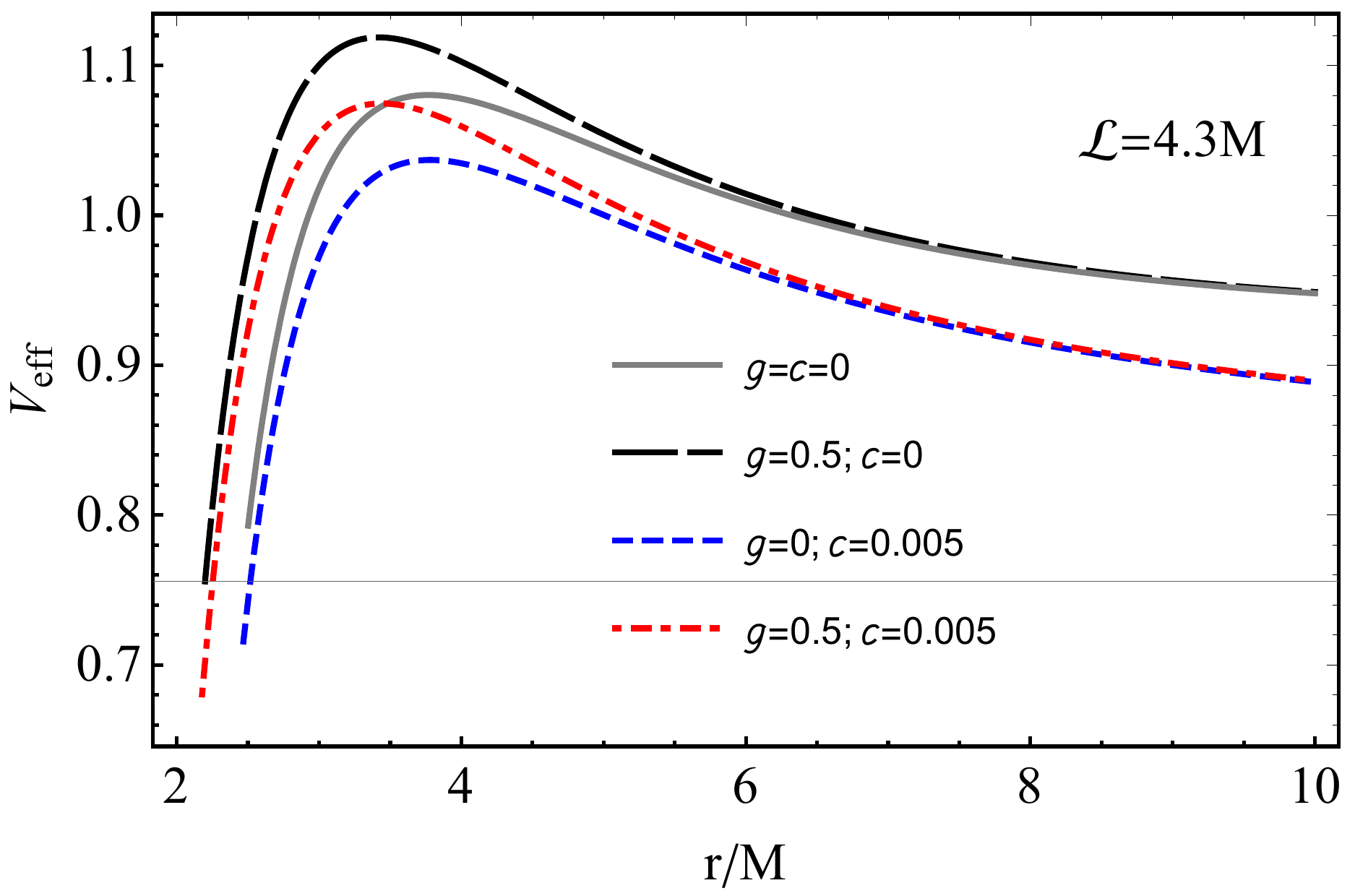}
   	\caption{Profiles of effective potential for radial motion of test particles around the BK BH for different values of parameter $c$ and the magnetic charge at $\omega_q=-2/3$.}.  \label{eff:fig}
\end{figure} 

In Fig.\ref{eff:fig}, we analyse effective potential of radial motion of test particles around the BK BH for different values of the BH charge and the parameter $c$. It is noticed that the presence of the BH charge (parameter $c$) increases (decreases) the maximum values of the effective potential.

The motion of test particles in circular orbits can be studied by solving the equation $\partial_rV_{\text{eff}}=0$, which gives the critical values of the angular momentum of the particles in the circular motion ${\cal L}_{cr}={\cal L}$.

Now one may consider the conditions for the circular motion for which the radial velocity vanishes i.e., $\dot{r}=0$ and furthermore, no external forces act in the radial direction, $\ddot{r}=0$, which together provide the radial profiles of the specific angular momentum and the specific energy of particles in circular orbits at the equatorial plane ($\theta=\pi/2$) in the following forms,
\begin{eqnarray}\label{LandE} 
{\cal L}^2=\frac{r^3 \left(\frac{4 M r}{\left(g^2+r^2\right)^{3/2}}-\frac{6 M r^3}{\left(g^2+r^2\right)^{5/2}}-\frac{c \left(3 \omega_q+1\right)}{r^{3 \omega _q+2}}\right)}{\frac{3 c (\omega_q +1)}{r^{3 \omega_q -1}}+\frac{6 M r^4}{\left(g^2+r^2\right)^{5/2}}-2},
\end{eqnarray}
\begin{eqnarray}\label{Energy}
\nonumber
{\cal E}^2&=&\left(1-\frac{2 M r^2}{\left(g^2+r^2\right)^{3/2}}-\frac{c}{r^{3 \omega _q+1}}\right) \\ & \times & \left(1+\frac{\frac{4 M r}{\left(g^2+r^2\right)^{3/2}}-\frac{c \left(3 \omega _q+1\right)}{r^{3 \omega _q+2}}-\frac{6 M r^3}{\left(g^2+r^2\right)^{5/2}}}{\frac{3 c \left(\omega _q+1\right)}{r^{3 \omega _q-2}}+\frac{6 M r^3}{\left(g^2+r^2\right)^{5/2}}-\frac{2}{r}}\right).
\end{eqnarray}

\begin{figure}[h!]
   \centering
\includegraphics[width=0.98\linewidth]{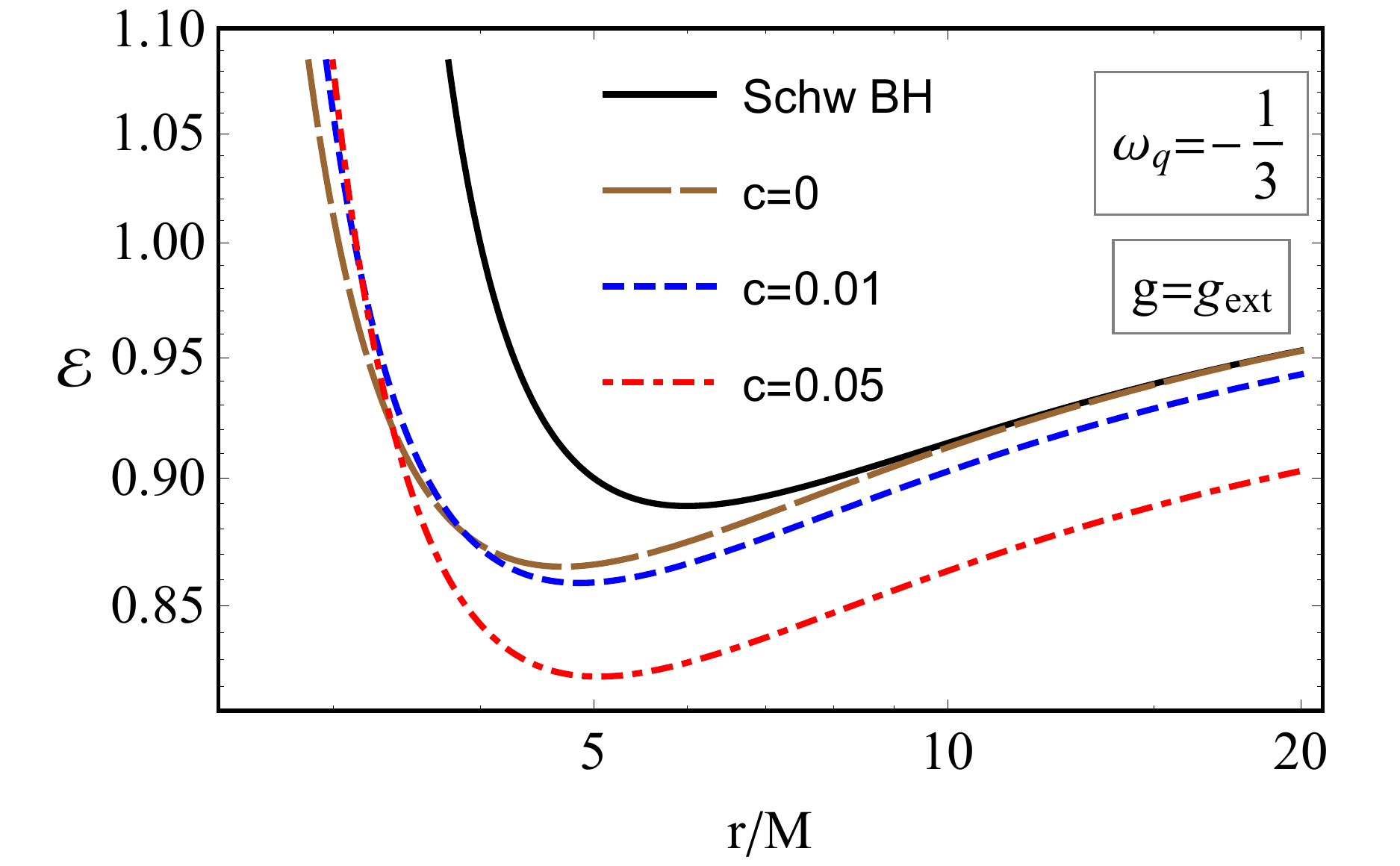} \includegraphics[width=0.98\linewidth]{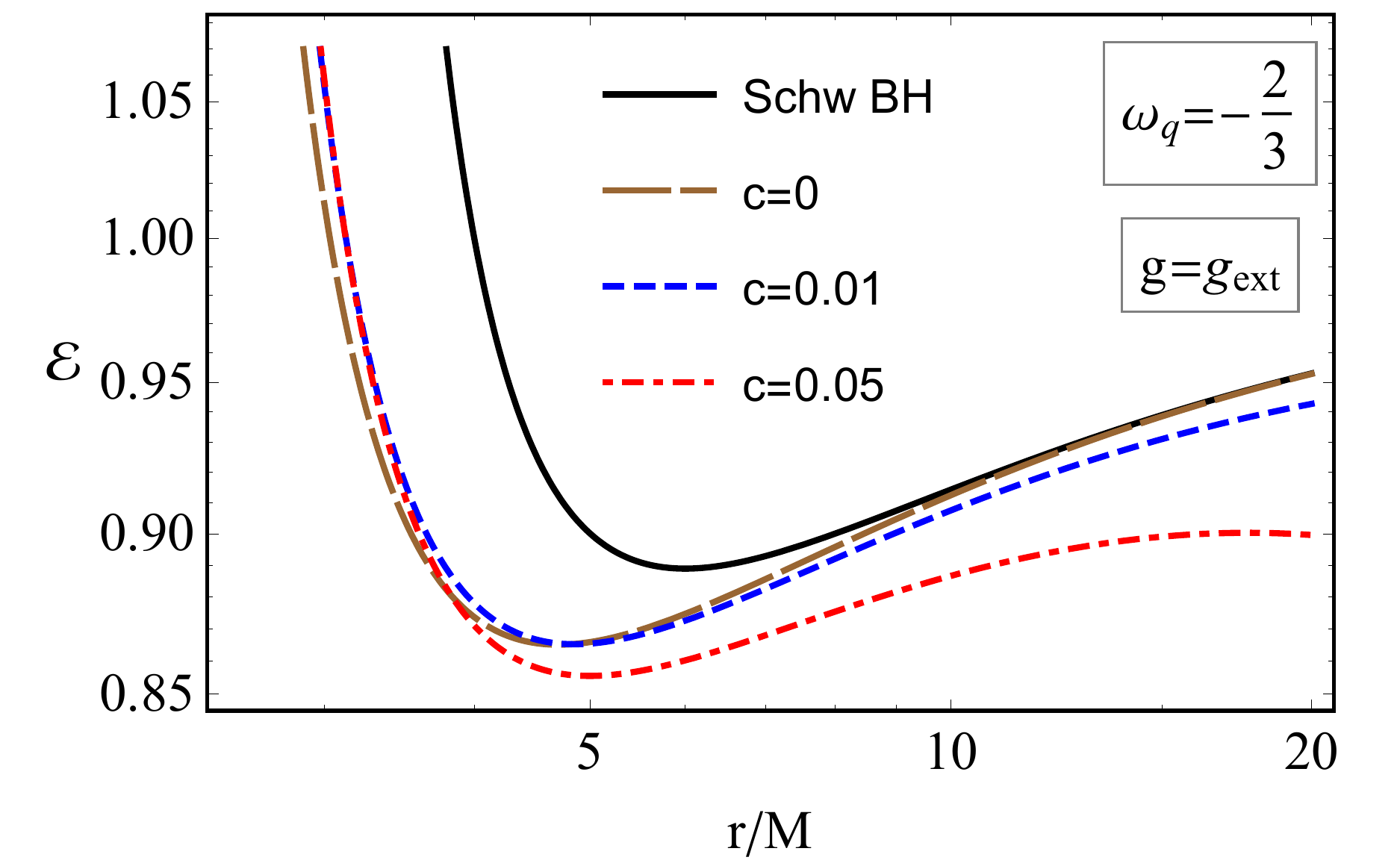}
	The radial dependence of specific energy of particle at circular orbits around the BK BH for different values of parameters $c$ and $g$ for the parameter $\omega_q=-1/3$ (top panel) and $\omega_q=-2/3$ (bottom panel). \label{landefign1}
\end{figure}

In Fig.\ref{landefign1}, the specific energy of an electrically neutral test particle around a BKBH is plotted, for the values $\omega_q=-1/3$ (top panel) and $\omega_q=-2/3$ (bottom panel) along with a comparison for the Schwarzschild BH case. The figure shows that the specific energy of the test particle is smaller when $\omega_q$ is also small, as compared to the case when the value of $\omega_q$ is larger. The behaviour of the specific angular momentum of a charged particle for $\omega_q=-1/3$ (top panel) and for $\omega_q=-2/3$ (bottom panel) is studied against the radial coordinate for some specific values of $c$ and $g=g_{\rm ext}$. Note that as the values of $c$ increases, the specific energy of the particle decreases.

\begin{figure}[h!]
   \centering
\includegraphics[width=0.98\linewidth]{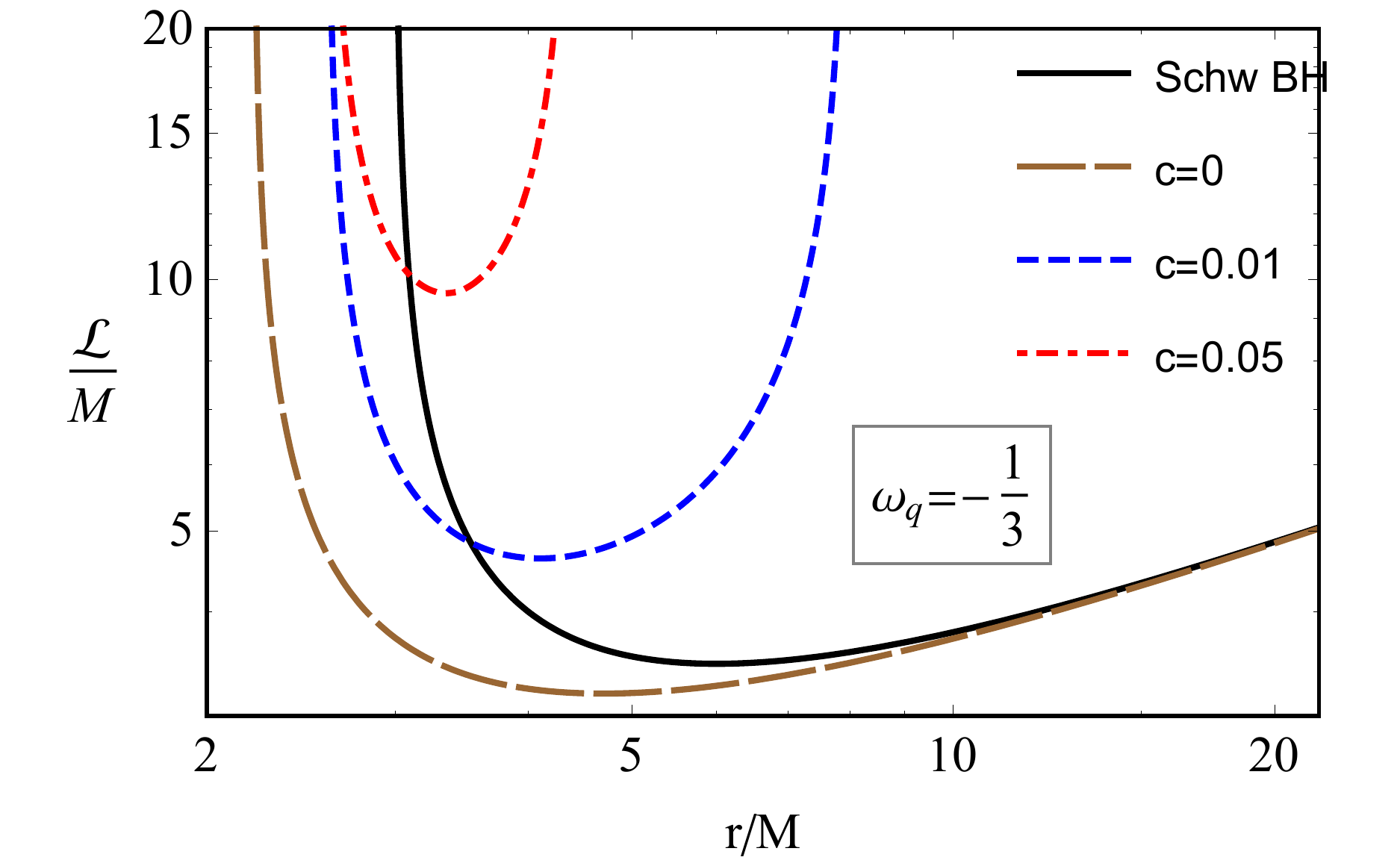} \includegraphics[width=0.98\linewidth]{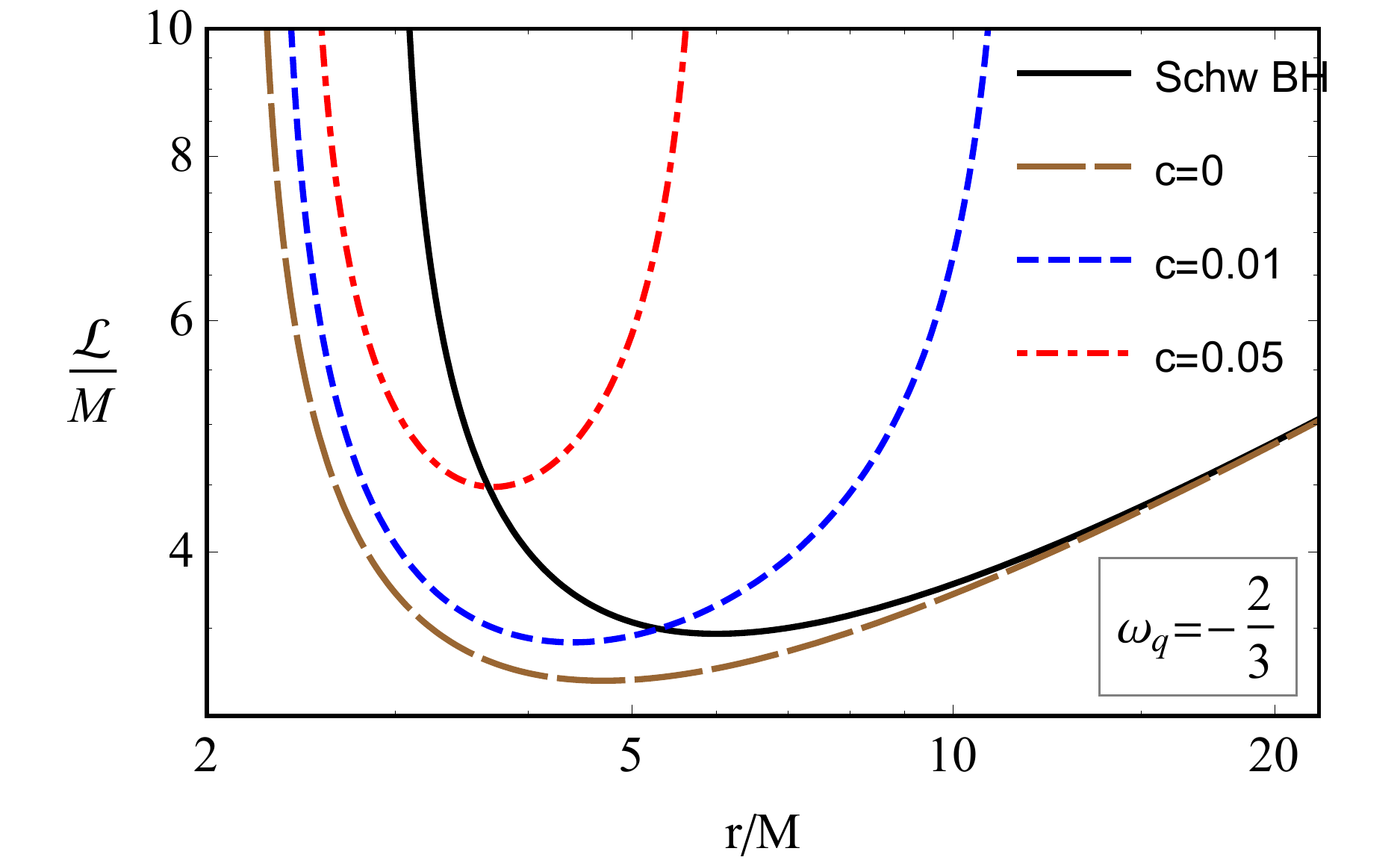}
	\caption{The same figure with Fig.\ref{landefign1} but for specific angular momentum.\label{landefign2}}
\end{figure}

In Fig. \ref{landefign2} we display the specific angular momentum of the particle corresponding to the circular motion as a function of radial coordinate for different values of $c$ but for a fixed value of $g$. Likewise, the plots are given for $\omega_q=-1/3$ (top panel) and $\omega_q=-2/3$ (bottom panel) in Fig. \ref{landefign2}. Note that as the value of $c$ increases, the relevant specific angular momentum of the particle also increases, and  the minimum radius also decreases for the circular orbits. It shows that as the values of  $c$ increases, the specific angular momentum of the particle becomes large, and it becomes very close to the BH as compared to the case when $c$ is smaller. The comparison of the top and bottom panels in Fig. \ref{landefign2} shows that the minimum specific angular momentum for the circular motion of the charged particle gets smaller when $\omega_q$ decreases.  

\subsection{ISCO}

Standardly, we will find the radius of ISCO by the condition $\partial_{rr}V_{\rm eff}=0$. Due to the complicated form of the equation obtained above condition, we provide here numerical and graphical analysis. 

\begin{figure}[ht!]
   \centering
     \includegraphics[width=0.98\linewidth]{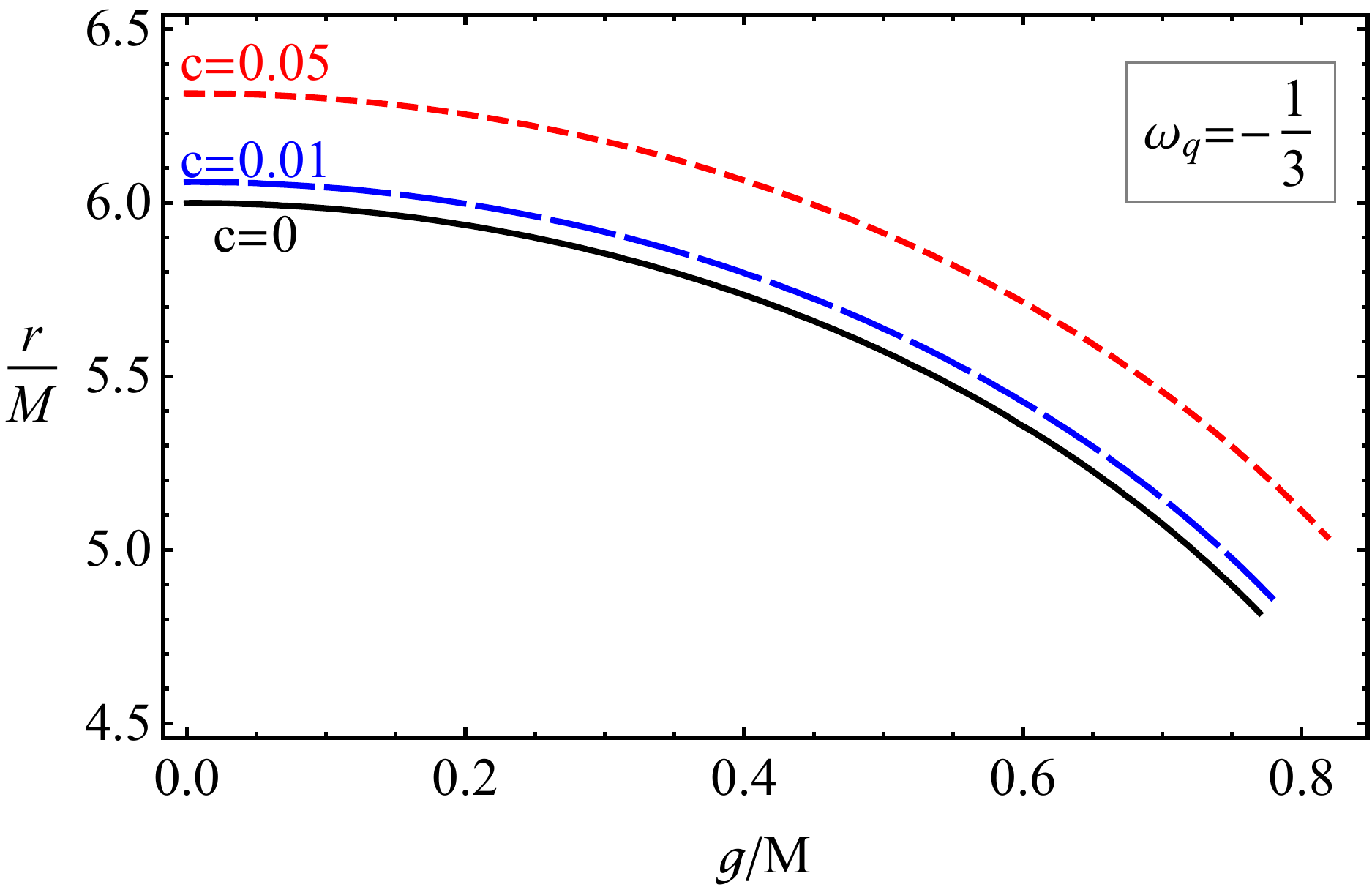}
   \includegraphics[width=0.98\linewidth]{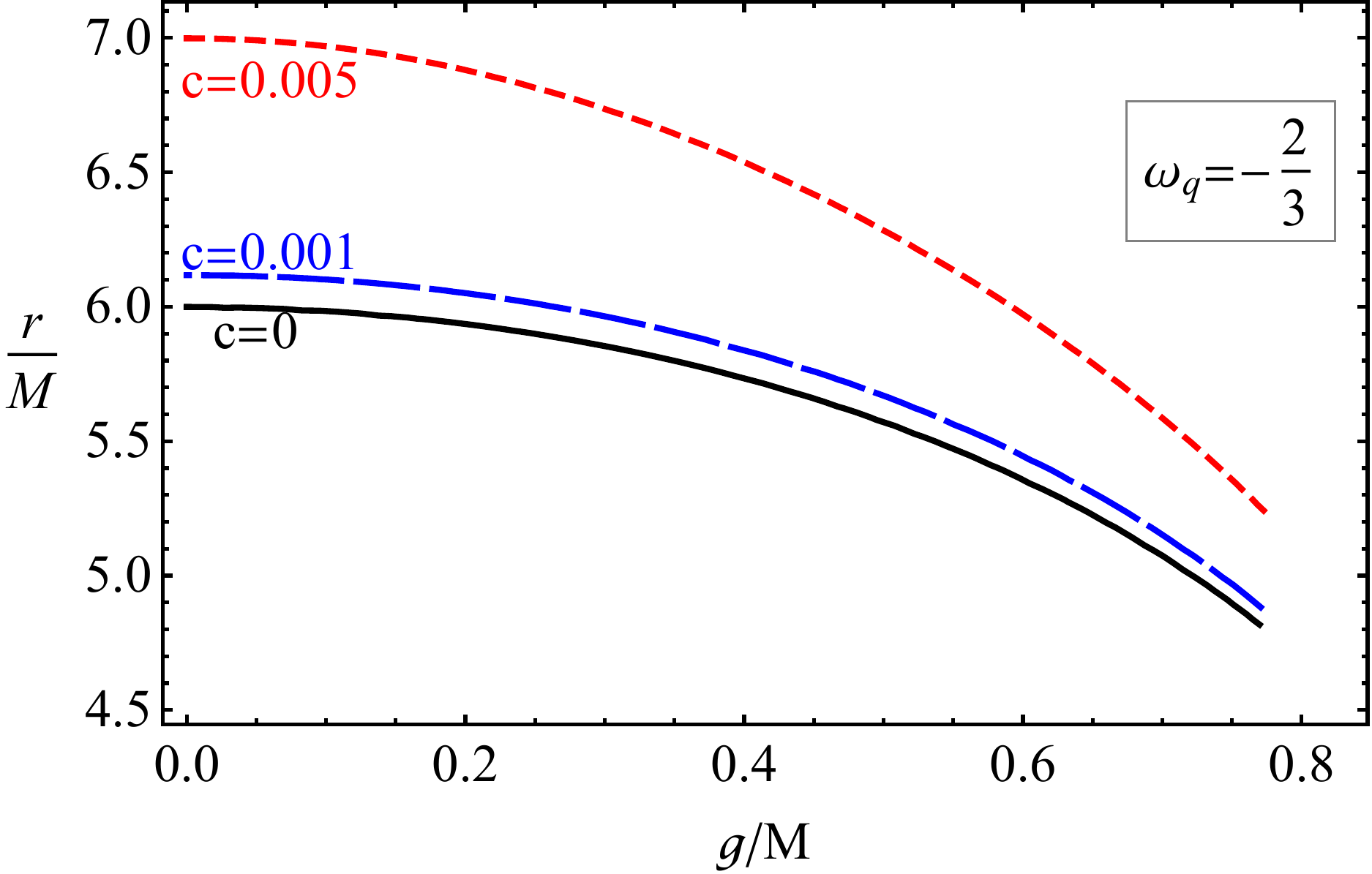}
	\caption{Dependence of the ISCO radius of test particles on the charge for the different values of the parameters $c$ and for the fixed values of the parameter $\omega_q=-1/3$ (top panel) and $\omega_q=-2/3$ (bottom panel).  \label{iscoA}}
\end{figure} 
For neutral test particles, the dependence of the ISCO on the parameters $g$, $c$ and $\omega_q$ is studied graphically in Fig. \ref{iscoA}. It is observed that as the value of $c$ increases (decreases), the radius of ISCO also increases (decreases). Also, smaller (larger) values of $\omega_q$ correspond to  the  larger (smaller) ISCOs.

\subsection{Energy Efficiency}

\begin{figure}[ht!]
   \centering
\includegraphics[width=0.98\linewidth]{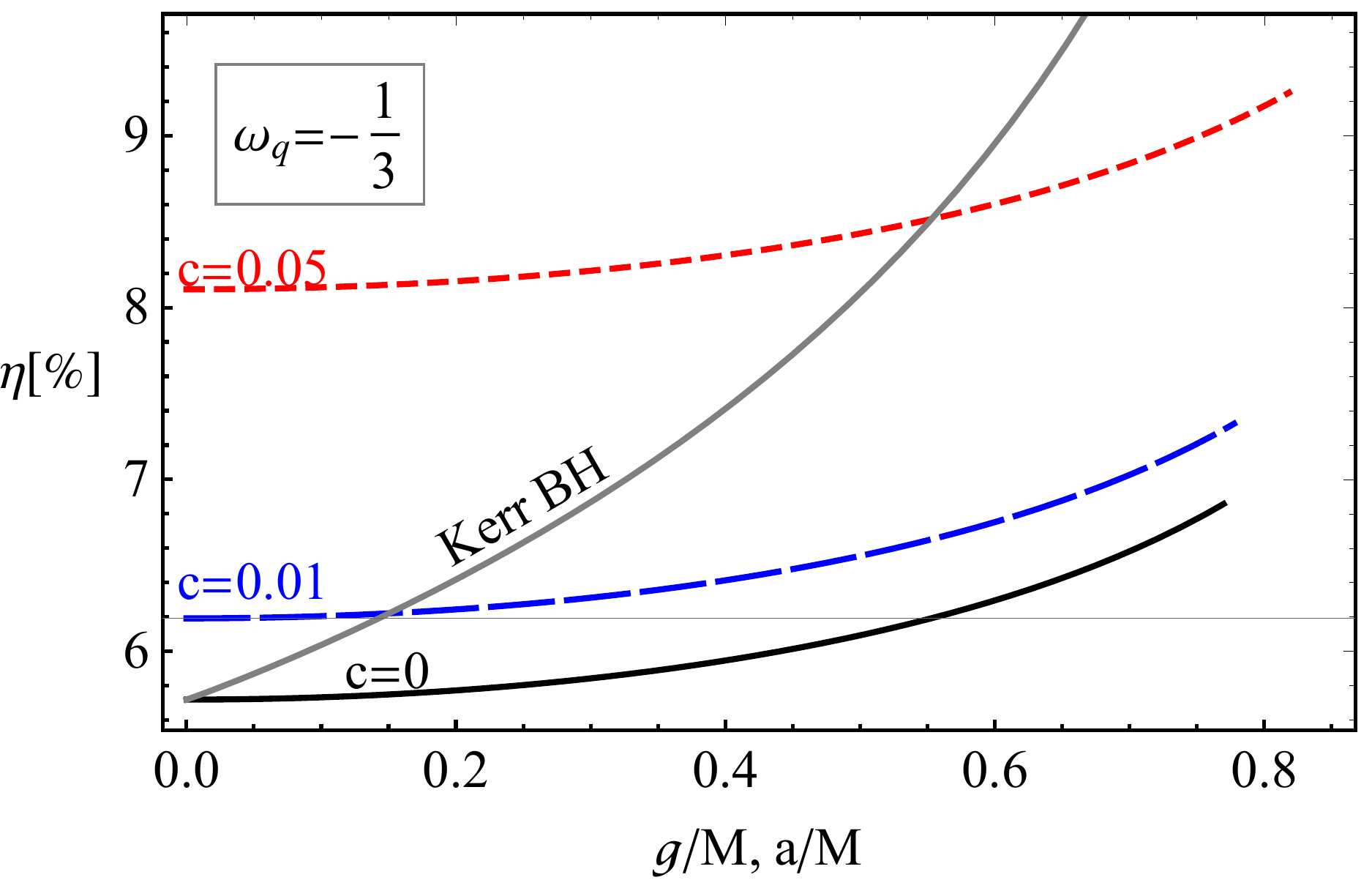}
\includegraphics[width=0.98\linewidth]{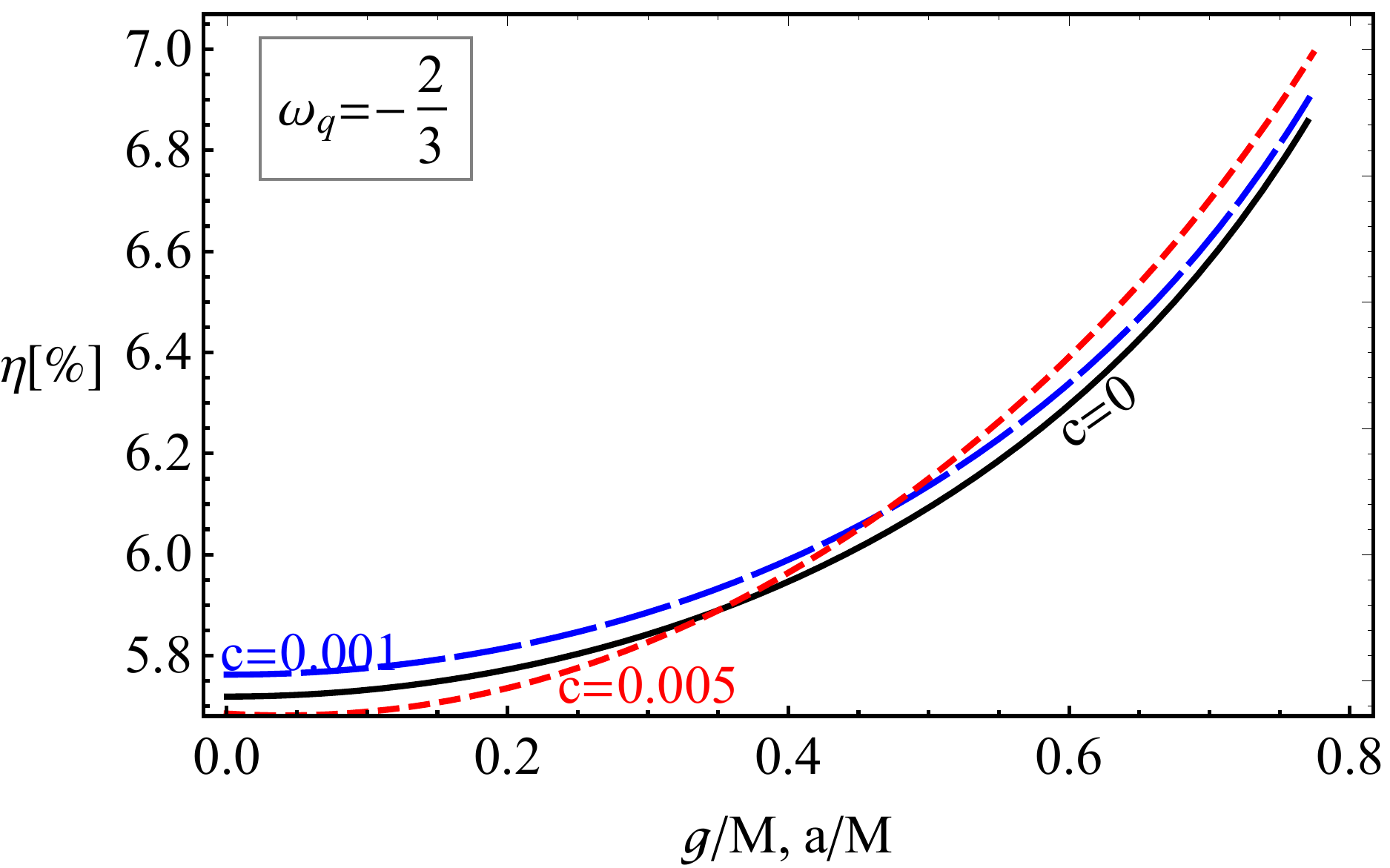}
	\caption{Dependence of the energy efficiency on the magnetic charge of the BKBH  for different values of $c$ and $\omega_q$. \label{efficiency}}
\end{figure}

The Novikov-Thorne model states that the Keplerian accretion disks formed around the astrophysical BHs are geometrically thin. The test particles moving in the circular paths around the BH can fall inside the central BH and extracts energy from it in the form of emitted radiations \cite{Novikov73}. The maximum energy which can be extracted in the form of radiation due to the infalling matter into the BH is called the energy efficiency of the accretion disk around a BH. Calculations of the energy efficiency of accretion of the test particle can be performed by the following expression
\begin{equation}
\eta=1-{\cal E}\,  \vline_{\,r=r_{\rm ISCO}},
\end{equation}
which is the difference of the rest energy of the test particle measured by local observers and ${\cal E}_{\rm ISCO}$ is the energy of the test particle at the ISCO which is given in Eq.(\ref{Energy}). The exact form of the efficiency is hard to calculate analytically due to the complicated form of the ISCO radius. However, the effects of the magnetic charge of a BKBH  on the accretion efficiency can be calculated numerically as shown in Fig.~\ref{efficiency} with a comparison to the energy efficiency for the Kerr BH.

It is clear from Fig.\ref{efficiency} that the energy efficiency around a Kerr BH and around a BKBH  is almost the same for $c= 0.01$ and $c=0.05$, which correspond to $\eta=6.2 \%$ and $\eta=8.6\%$ respectively. It also shows that the energy efficiency increases as $c$ increases. Also, for larger values of $g$, we get larger values of the efficiency, $\eta$. The energy efficiency $\eta$ becomes very large as the value of $\omega_q$ is large ($\omega_q=-1/3$) as compared to the case when it is small ($\omega_q=-2/3$). 

\section{Fundamental frequencies}

The fundamental frequencies of the test particle moving along the circular orbits around a BKBH, such as the Keplerian frequencies, frequencies of the radial and vertical oscillation, are calculated below. It is the simplest explanation for the QPO visible in the microquasars. 

\subsection{Keplerian frequency}

The angular velocity of the test particle, measured by a distant observer, also known as the Keplerian frequency, is given by 
\begin{equation}
\label{Omega}
\Omega_K=\Omega_\phi=\frac{d\phi}{dt}=\frac{\dot{\phi}}{\dot{t}} \ . 
\end{equation}
The alternate expression of the Keplerian frequency in the spacetime of a static BH, which gives the same results, is given by:
\begin{eqnarray}\label{omega}
\Omega_K&=&\sqrt{\frac{f'(r)}{2r}}\\\nonumber &=& \sqrt{\frac{c \left(3 \omega _q+1\right)}{2 r^{3 \left(\omega _q+1\right)}}+\frac{M \left(r^2-2 g^2\right)}{\left(g^2+r^2\right)^{5/2}}}. 
\end{eqnarray}  
For $\Omega_K$ to be a well-defined function, we require $f'(r)\geq0$. This would hold only if $c>0$, $1+3w_q>0$ and $r>|2g|$. Moreover, in Keplerian orbits the gravitational and centrifugal forces balance each other, and further the gravitating potential, which is  characterized by the lapse function should be an increasing function, and then its derivatives will be positive. For instance, the Keplerian frequency for circular orbits in Newtonian gravity $\Omega_K\sim \sqrt{M/r^3}>0$. In Schwarzschild geometry, two fundamental frequencies are same but differ from the third, while in Kerr geometry, all fundamental frequencies are different and positive.
For a theoretical understanding of the fundamental frequencies, one can express them in the unit of Hz: 
\begin{equation}
\nu = \frac{1}{2\pi}\frac{\mathit{c}^3}{GM} \Omega \ .
\end{equation}

Note that here we provide our calculations in converting the frequency from geometrical units ($\rm 1/cm^{1/2}$) to Hz (international unit systems, $s^{-1}$), the speed of light ${\rm c}=3\cdot10^{10} \rm cm/sec$ and the gravitational constant $G=6.67\cdot 10^{-8} \rm cm^3/(g\cdot sec^2)$. 

\begin{figure}[ht!]
   \centering
\includegraphics[width=0.98\linewidth]{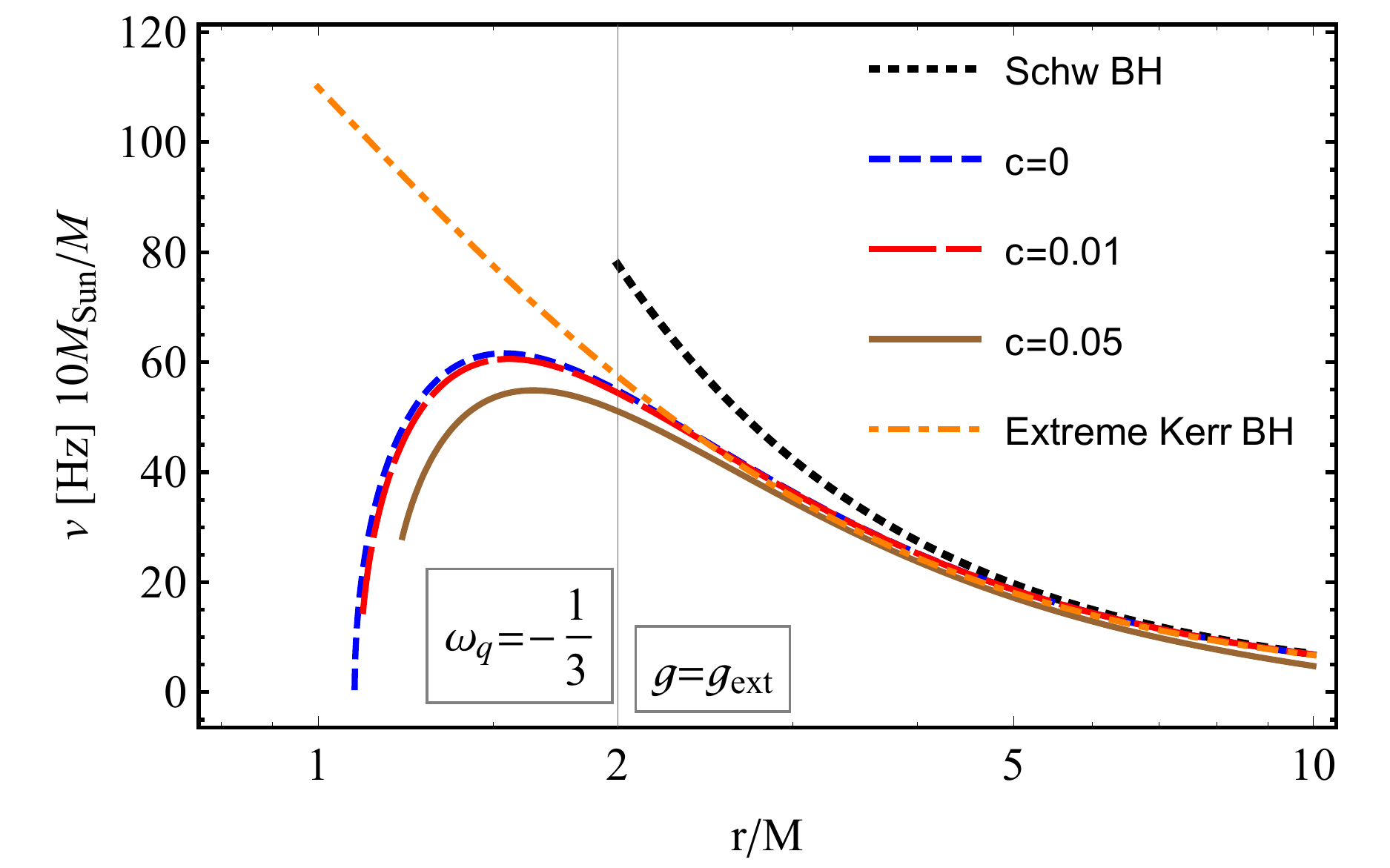}
\includegraphics[width=0.98\linewidth]{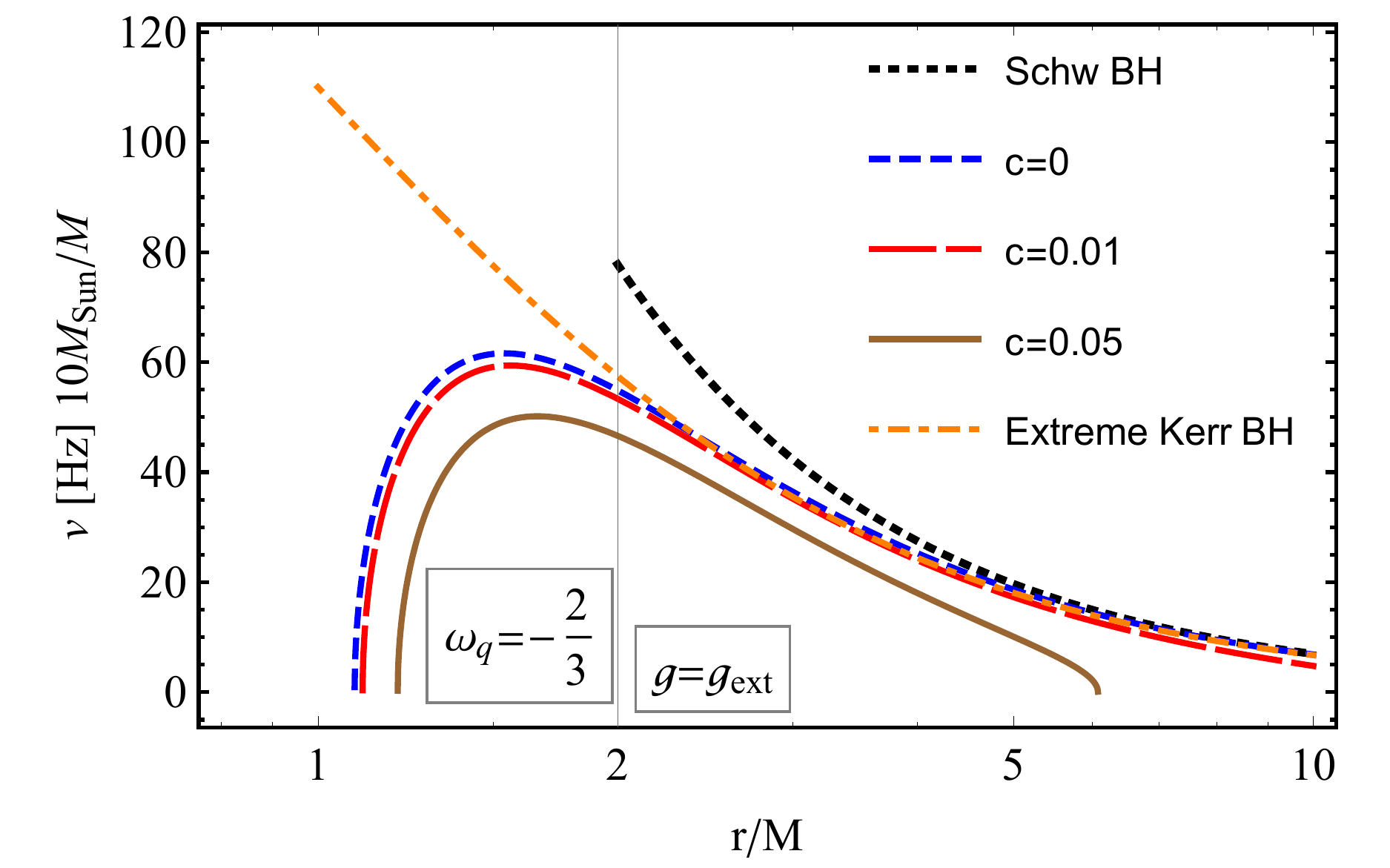}
	\caption{ The Keplerian frequencies of test particles around a BKBH  verses $r$.  \label{Keplerian}}
\end{figure}

The radial dependence of the Keplerian frequencies of test particles around a BK BH is shown in Fig.~\ref{Keplerian}. It is clear that the frequency of the radial oscillations of test particles moving around  the Schwarzschild BH gets its maximum value at $r=2M$ and in the case of the Kerr BH it gets a greater value as compared to the Schwarzschild and the maximum value appears at $r=M$. In both cases, frequencies are always  increasing. But for a BK BH  the frequency of the particle's oscillation is first increasing to its maximum value and then starts decreasing. As   $c$ increases,  the frequency of the particle decreases.  

\subsection{Harmonic Oscillations}

In the following, we will carry out a  detailed analysis of the fundamental frequencies of a neutral test particle moving around a BK BH at the equatorial plane. Since the effective potential given in Eq.(\ref{4}) reaches its extreme value at $\theta=\pi/2$ which can be found by solution of the condition $\partial_\theta V_{\rm eff}=0$. Now, we consider small perturbations of the orbital motion of the particles in radial $r\to r_0+\delta r$ and vertical (latitudinal) $\theta \to \pi/2+\delta \theta$ directions, where $r_0$ is a radial coordinate in where the
effective potential takes its extreme value \cite{stuchlik2021universe}.

\begin{eqnarray}\label{Vexpand}
\nonumber
&&V_{\rm eff}(r,\theta)=V_{\rm eff}(r_0,\theta_0)
\\\nonumber
&&+\delta r\,\partial_r V_{\rm eff}(r,\theta)\Big|_{r_0,\theta_0} +\delta\theta\, \partial_\theta V_{\rm eff}(r,\theta)\Big|_{r_0,\theta_0}
\\\nonumber
&&+\frac{1}{2}\delta r^2\,\partial_r^2 V_{\rm eff}(r,\theta)\Big|_{r_0,\theta_0}+\frac{1}{2}\delta\theta^2\,\partial_\theta^2 V_{\rm eff}(r,\theta)\Big|_{r_0,\theta_0}
\\
&&+\delta r\,\delta\theta\,\partial_r\partial_\theta V_{\rm eff}(r,\theta)\Big|_{r_0,\theta_0}+{\cal O}\left(\delta r^3,\delta\theta^3\right)\ .
\end{eqnarray}

We note that the first term of Eq.(\ref{Vexpand}) equals to zero due to limited motion of massive test particles at the radius $r_0$ and the plane $\theta_0$ (the motion determines by the energetic boundary conditions as $\dot r=0$, $\dot\theta=0$, and  $V_{\rm eff}(r_0,\theta_0)=0$). We can also drop out the second, third and sixth terms of the expansion using the stability conditions $\partial_{r,\theta}V_{\rm eff}(r,\theta)=0$.

Thus, there remains only two non-zero terms, which are the second order derivatives of the effective potential with respect to $r$ and $\theta$. Now, we replace the derivative with respect to the affine parameter in Eq.(\ref{4}) by the derivative with respect to the time (i.e $dt/d\lambda=u^t$), such that a distant observer can measure the physical quantities associated with the test particle. Substituting Eq. (\ref{Vexpand}) into Eq. (\ref{4}), we are able to derive the  equations of motion (or the Euler-Lagrange equations) for the harmonic oscillator, assuming the infinitesimal displacements $\delta r$ and $\delta\theta$, given by
\begin{equation}
\frac{d^2\delta r}{dt^2}+\Omega_r^2 \delta r=0\, \qquad \frac{d^2\delta\theta}{dt^2}+\Omega_\theta^2 \delta\theta=0,   
\end{equation}
here $\Omega_r$ and $\Omega_\theta$ are the radial and vertical angular frequencies of the test particle at the equatorial plane measured by a distant observer, defined as
\begin{eqnarray}
&&\Omega_r^2=-\frac{1}{2g_{rr}\dot{t}^2}\partial_r^2V_{\rm eff}(r,\theta)\Big |_{\theta=\pi/2},
\\
&&\Omega_\theta^2=-\frac{1}{2g_{\theta\theta}\dot{t}^2}\partial_\theta^2V_{\rm eff}(r,\theta)\Big |_{\theta=\pi/2}.
\end{eqnarray}
Finally,  the radial and the vertical frequencies in the spacetime for test particles take the form
\begin{eqnarray}\label{wr}
\nonumber
&&\Omega_r =\Omega_K \Bigg\{\left[1-\frac{c}{r^{3 \omega_q +1}}-\frac{2 M r^2}{\left(g^2+r^2\right)^{3/2}}\right] \\\nonumber && \left[3-\frac{r \left(\frac{c (9 \omega_q  (1+\omega_q)+2)}{r^{3 (1+\omega_q )}}+\frac{2 M \left(2 g^4-11 g^2 r^2+2 r^4\right)}{\left(g^2+r^2\right)^{7/2}}\right)}{\frac{c (3 \omega_q +1)}{r^{3 \omega_q +2}}+\frac{2 M r \left(r^2-2 g^2\right)}{\left(g^2+r^2\right)^{5/2}}}\right]\\&&-\frac{2 c (3 \omega_q +1)}{r^{3 \omega_q +1}}+\frac{M \left(8 g^2 r^2-4 r^4\right)}{\left(g^2+r^2\right)^{5/2}}\Bigg\}^{1/2},
\\\label{wt}
&&\Omega_\theta = \Omega_K,
\\\label{wf}
&&\Omega_\phi=\Omega_K.
\end{eqnarray}
Note that the vertical and azimuthal angular velocities are the same so  that a distant observer can not distinguish them. 
\begin{figure}[ht!]
   \centering
\includegraphics[width=0.98\linewidth]{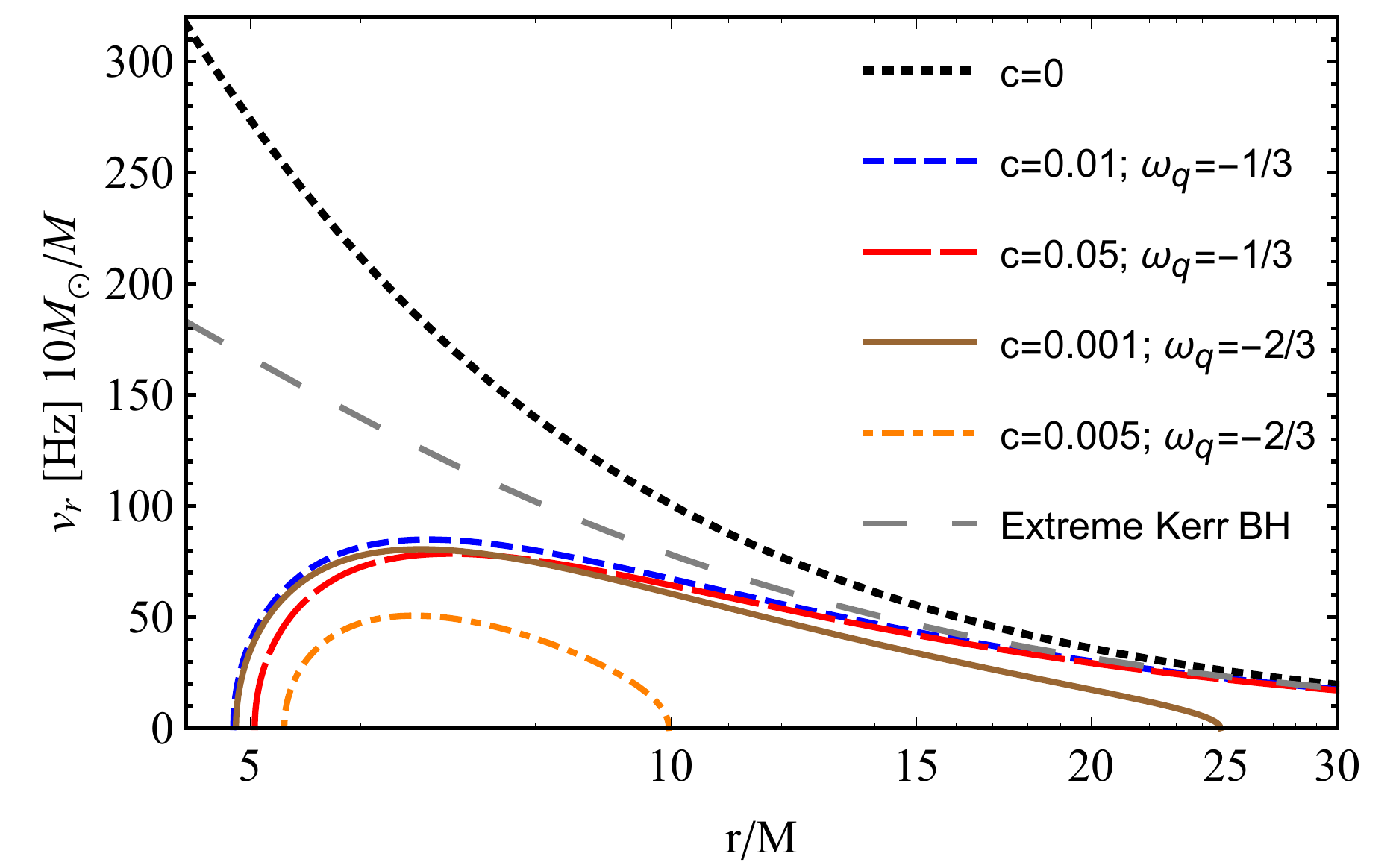}
	\caption{Radial dependence of the frequencies of radial oscillations of test particles around an extremely charged BKBH with a comparison of the frequency in the extreme rotating Kerr BH case. \label{Keplerian}}
\end{figure}

Figure~\ref{Keplerian} shows radial frequencies of test particles around a BKBH.

\section{The motion of massless particles}\label{massless}

The null geodesics around the BKBH describe the dynamics of massless particles with $m=0$. However, the photon motion can be analysed by null geodesics of the effective geometry \cite{Novello:1999pg,Stuchlik2019} due to the interaction of non-linear electromagnetic field and photons, and they do not follow the null geodesics. However, there is no such interaction for neutrino-like particles. In the next subsection, we will discuss an important feature of a BH known as the photon sphere.

\subsection{The Photon-Sphere around a BKBH }

Determinations of the characteristic radii around the BHs are one of the interesting features of BHs. The photon motion in the effective geometry of a BKBH is calculated by the geodesics of massless particles due to the interaction between the NED field and photons. Consider the following expressions:
\begin{eqnarray}\label{effmetric1}
 \tilde{g}^{\mu \nu}&=&g^{\mu \nu}-4\frac{L_{\rm FF}}{L_F}F^{\mu}_{\lambda}F^{\lambda \nu},
\\\label{effmetric2}
\tilde{g}_{\mu \nu}&=&16\frac{L_{\rm FF}F_{\mu \eta}F^{\eta}_{\nu}-(L_{\rm F}+2FL_{\rm FF}) g_{\mu \nu}}{F^2L^2_{\rm FF}-16(L_{\rm F}+FL_{\rm FF})^2},
\end{eqnarray}
where
\begin{equation}
 \quad
L_{\rm F}=\frac{\partial L(F)}{\partial F}, \quad  L_{\rm FF}=\frac{\partial^2 L(F)}{\partial F^2}.
\end{equation}
For the photons moving in the  geometry of the BKBH, the eikonal equation is given by:
\begin{equation}
\tilde{g}_{\mu \nu} k^{\mu} k^{\nu}=0,
\end{equation}
here $k^{\mu}$ is the four-wave-vector of photons given by $p^{\mu}=\hbar k^{\mu}$ (with  $\hbar = 1$, in Gaussian units). As discussed in Refs.\cite{Stuchlik2019,Dima19ApJ}, the effective potential for the photon moving in the equatorial plane in the spacetime geometry of the BKBH  is connected with the impact parameter of photon: $b=p_{\phi}/p_{t}$, which results in:
\begin{eqnarray}\label{effpotphoton} 
V_{\rm eff}&=&\frac{f}{r^2}\Big(1+2\frac{L_{\rm FF}}{L_{\rm F}} F \Big)=\left(5-\frac{7 g^2}{g^2+r^2}\right)  \\\nonumber &\times& \frac{1}{4r^2}\left(1-\frac{2 M r^2}{\left(g^2+r^2\right)^{3/2}}-\frac{c}{r^{3 \omega _q+1}}\right).
\end{eqnarray}

The photon motion around a BKBH is discussed in ample detail in \cite{Dima19ApJ}. When $L(F)=F$, the effective potential looks similar to the one describing dynamics of photons in the Reissner-Nordstr\"{o}m spacetime (see Refs.\cite{Stuchlik2000CQG1,Stuchlik2000CQG2,Stuchlik2007CQG}). The photons have circular orbits around BHs at the radius where the effective potential has its local minima. Solving $\partial_rV_{\rm eff}=0$ gives the radius, which helps in finding the impact parameter of the circular orbits,
\begin{eqnarray}
&& \frac{5 M r \left(3 r^2-4 g^2\right)}{2 \left(g^2+r^2\right)^{7/2}}+ \frac{1}{4 \left(g^2+r^2\right)^2 r^{3 \omega _q+4}} \\\nonumber
&& \times \Big\{ c \left[g^2 r^2 \left(9 \omega _q-5\right)-6 g^4 \left(\omega _q+1\right)+15 r^4 \left(\omega _q+1\right)\right]\\\nonumber && +2 \left(2 g^4+4 g^2 r^2-5 r^4\right) r^{3 \omega _q+1}\Big\}=0\ .
\end{eqnarray}

\begin{figure}[h]
    \centering
    \includegraphics[width=0.98 \linewidth]{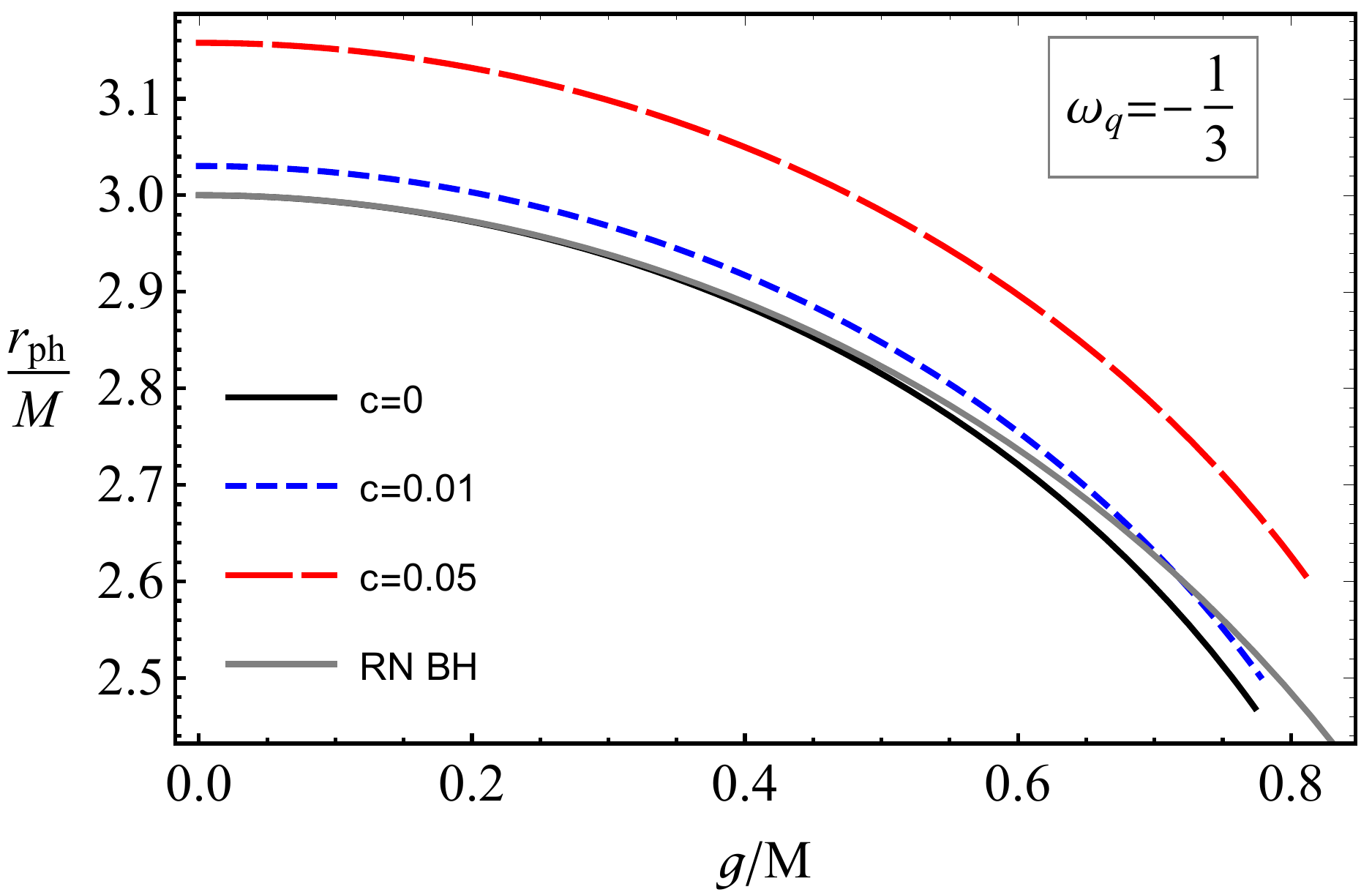}
    \includegraphics[width=0.98 \linewidth]{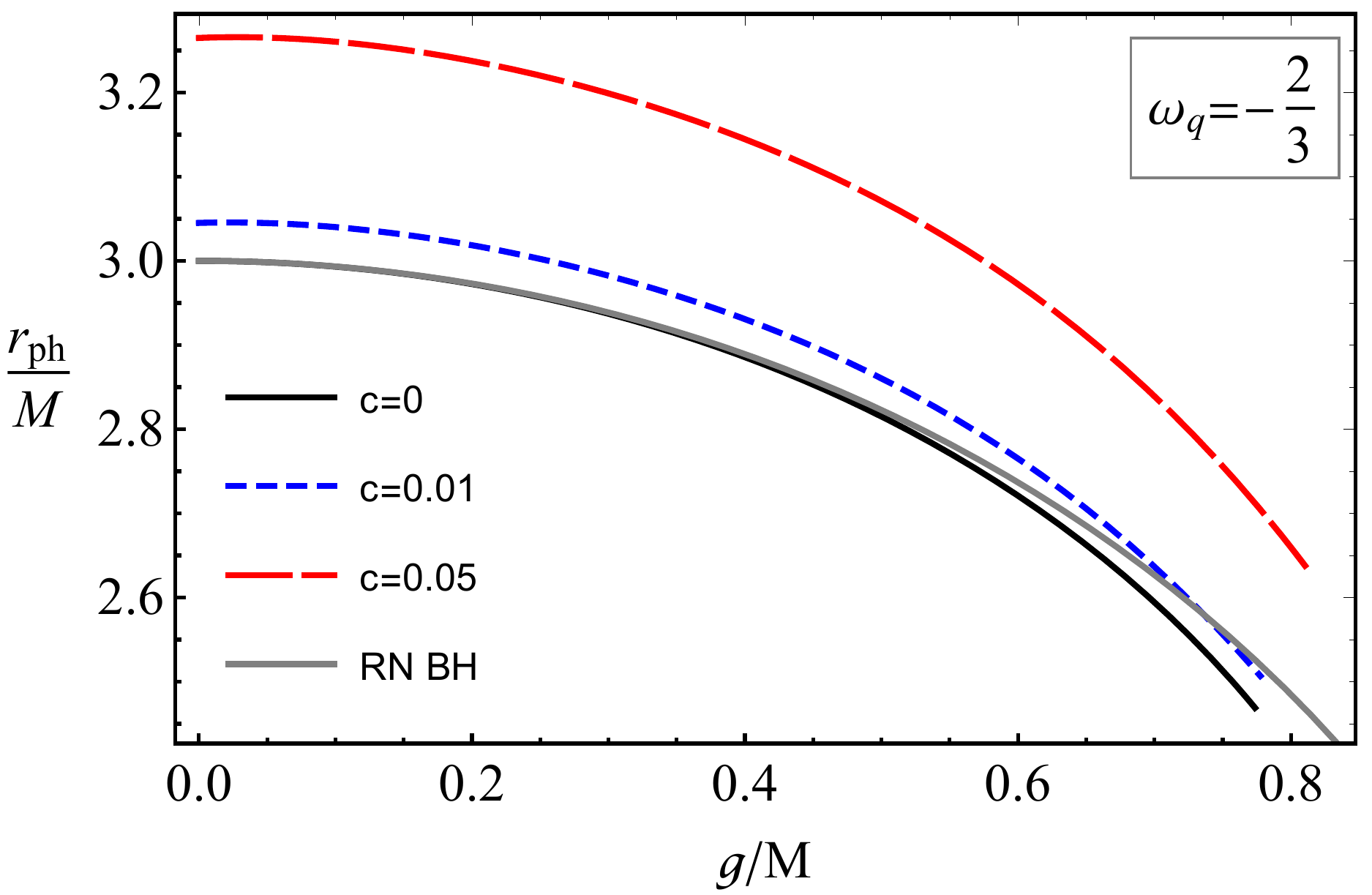}
    \caption{The radius of photonsphere as a function of the RBH charge $g$, for different values of the parameter $\omega$ and $c$.}
    \label{photonfig}
\end{figure}
Figure \ref{photonfig} illustrates the radius of photonsphere around a BKBH as a function of the magnetic charge $g$ of the BH. In the upper panel the analysis is performed for $\omega_q=-1/3$ while in the lower panel $\omega_q=-2/3$, by varying $c$, it is then observed that as $c$ increases (decreases) the radius of photon sphere increases (decreases).  
Also, for larger values of $\omega_q$ (upper panel) the radius of the photon sphere shrinks as compared to the case when $\omega_q$ is small (lower panel). 
From the astrophysical perspective, the impact parameter is important since it determines the size of the BH shadow. For a BKBH, the impact parameter is expressed as follows \cite{Stuchlik2019,Dima19ApJ,Rayimbaev2020PhRvD,Vrba2019}
\begin{eqnarray}\label{impacteq}
b^2_c &=& \frac{L_{\rm F}}{L_{\rm F}+2 F L_{\rm FF}}\frac{r^2}{f}\,\Bigg|_{r=r_{\rm ph}} \\ \nonumber &=& \frac{3 \left(g^2+r^2\right)^{\frac{5}{2}}\left(3 r^2-4 g^2\right)^{-1} r^{3 \left(1+\omega_q\right)}}{\left(g^2+r^2\right)^{\frac{3}{2}} \left(r^{1+3 \omega _q}-c\right)-2 M r^{3 \left(\omega _q+1\right)}} \Bigg|_{r=r_{\rm ph}}. 
\end{eqnarray}

\begin{figure}[h]
    \centering
    \includegraphics[width=0.98 \linewidth]{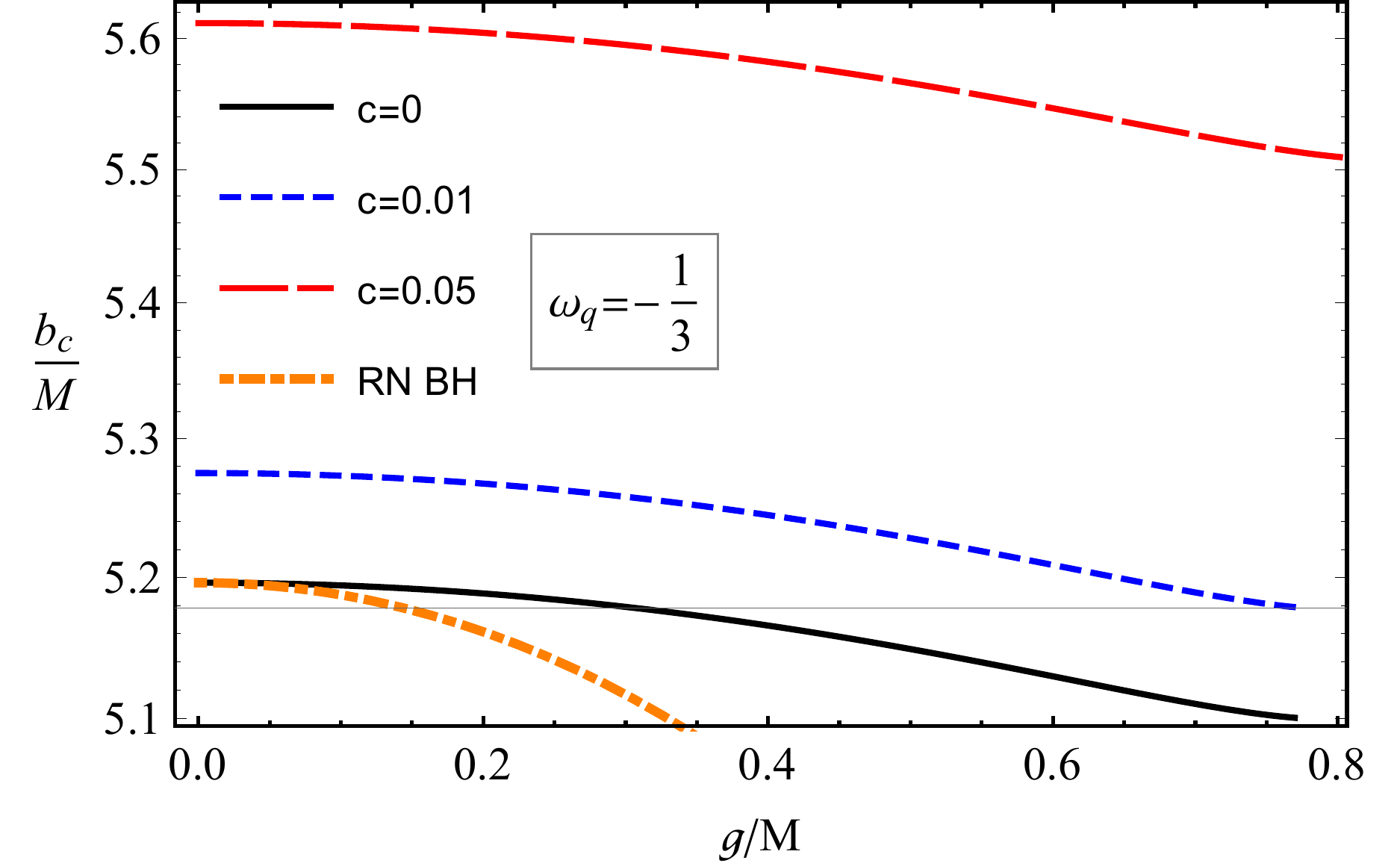}
    \includegraphics[width=0.98 \linewidth]{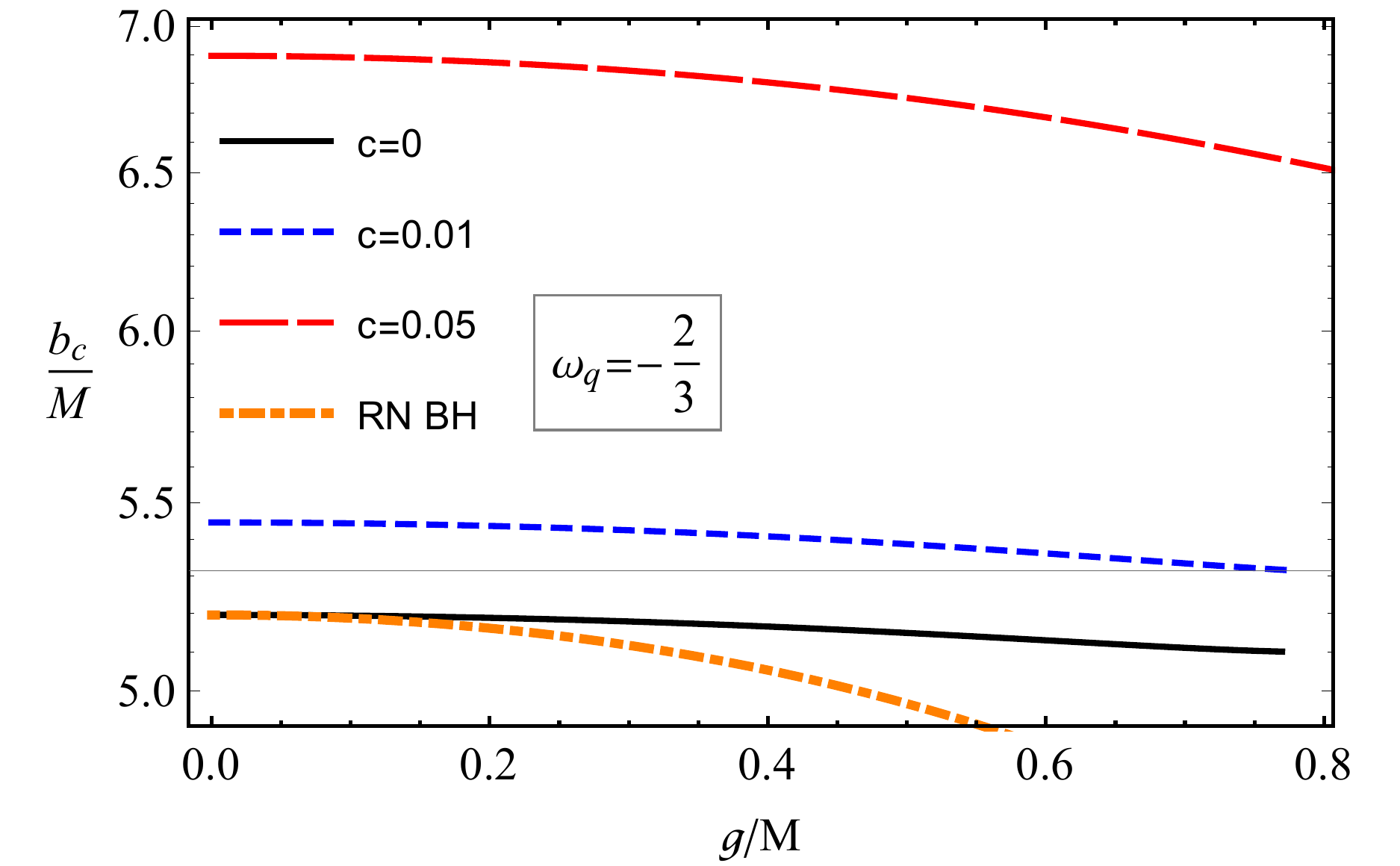}
    \caption{The impact parameter for the circular orbits of photons verses $g$.}
    \label{impactparfig}
\end{figure}

In Fig.\ref{impactparfig} an extension of the BKBH  shadow  (defined by the impact parameter $b_c$ ) as a function of the magnetic charge of the BH is illustrated. It is clear from the figure that as values of $c$ increase then the impact parameter also gets higher values as compared to the case when $c$ is smaller. For $c=0$ the values of the impact parameters for the RN  and BK BHs  become equal. Also, for larger values of $\omega_q=-1/3$ the impact parameter gets smaller, as the value of  $\omega_q$ decreases, $b_c$ gets larger. Note that as the values of the magnetic charge $g$ of the BH increase, the radius of the photon sphere decreases. 

\subsection{The Motion of Neutrino-like Particles}

In this section, we will find the equations of motion for massless  particles. One can get these equations by simply using the standard Euler-Lagrange equations for the metric of the BK BH as  given in Eq.(\ref{metric}).

Consider the Lagrangian density (in dimensionless form) for a neutral particle of mass $m$,
\begin{eqnarray}
L_{\rm p}=\frac{1}{2} g_{\mu\nu} \dot{x}^{\mu} \dot{x}^{\nu}, 
\end{eqnarray}
along with the conserved quantities of the particle
\begin{eqnarray}
\label{consts1}
 g_{tt}\dot{t}=-{\cal E}, \qquad
 g_{\phi \phi}\dot{\phi} = {\cal L}
\end{eqnarray}
where ${\cal E}$ and ${\cal L}$ are the special energy and specific angular momentum of the particle, respectively. Using the normalization condition as given below, one can derive the equation of motion of the  test particle
\begin{equation}\label{norm4vel}
g_{\mu \nu}u^{\mu}u^{\nu}=\epsilon
\end{equation}
where $\epsilon$ takes either $-1$ or $0$ for massive and massless particles, respectively.

\begin{figure}[h!]
    \centering
    \includegraphics[width=0.98 \linewidth]{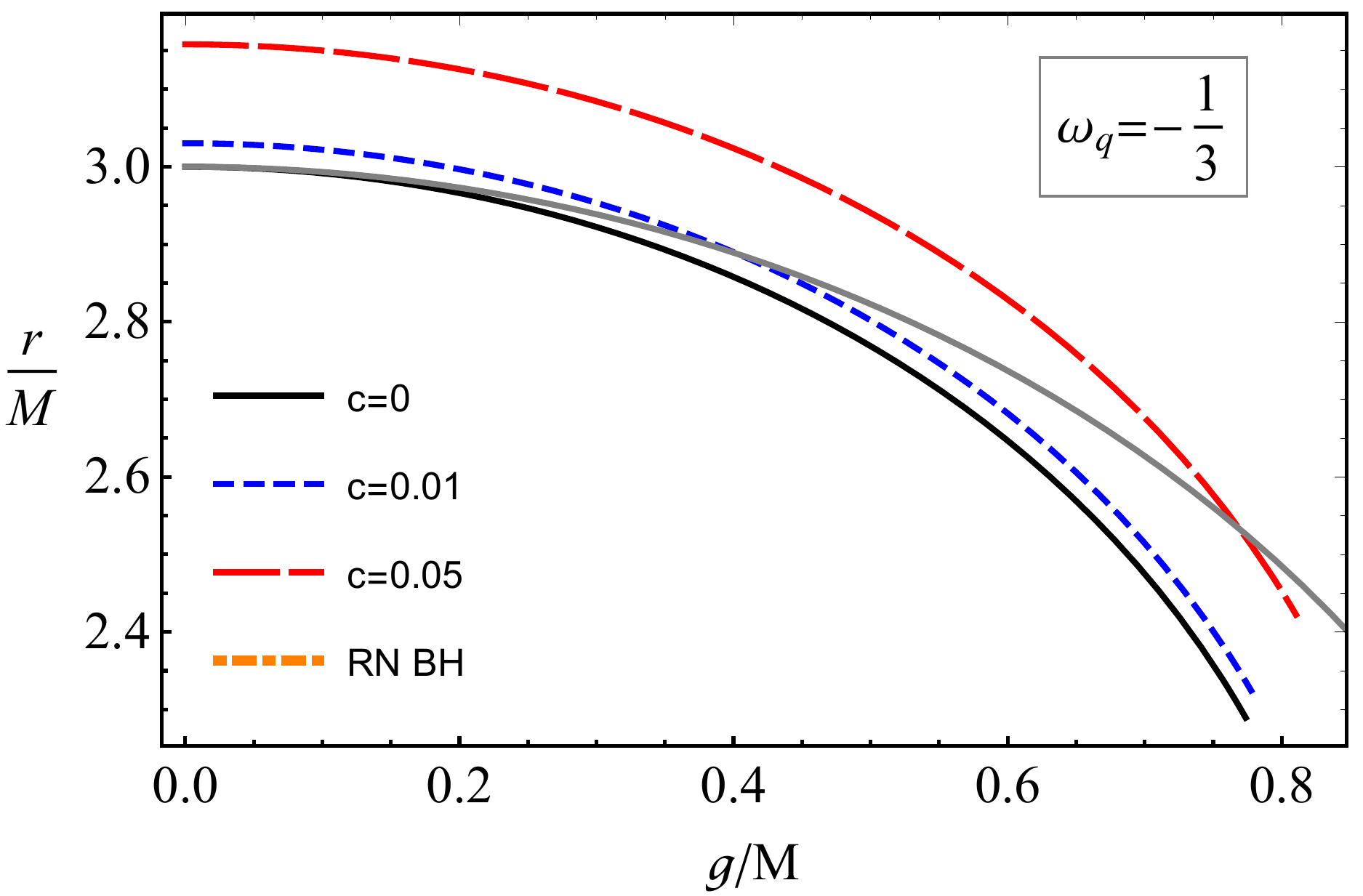}
    \includegraphics[width=0.98 \linewidth]{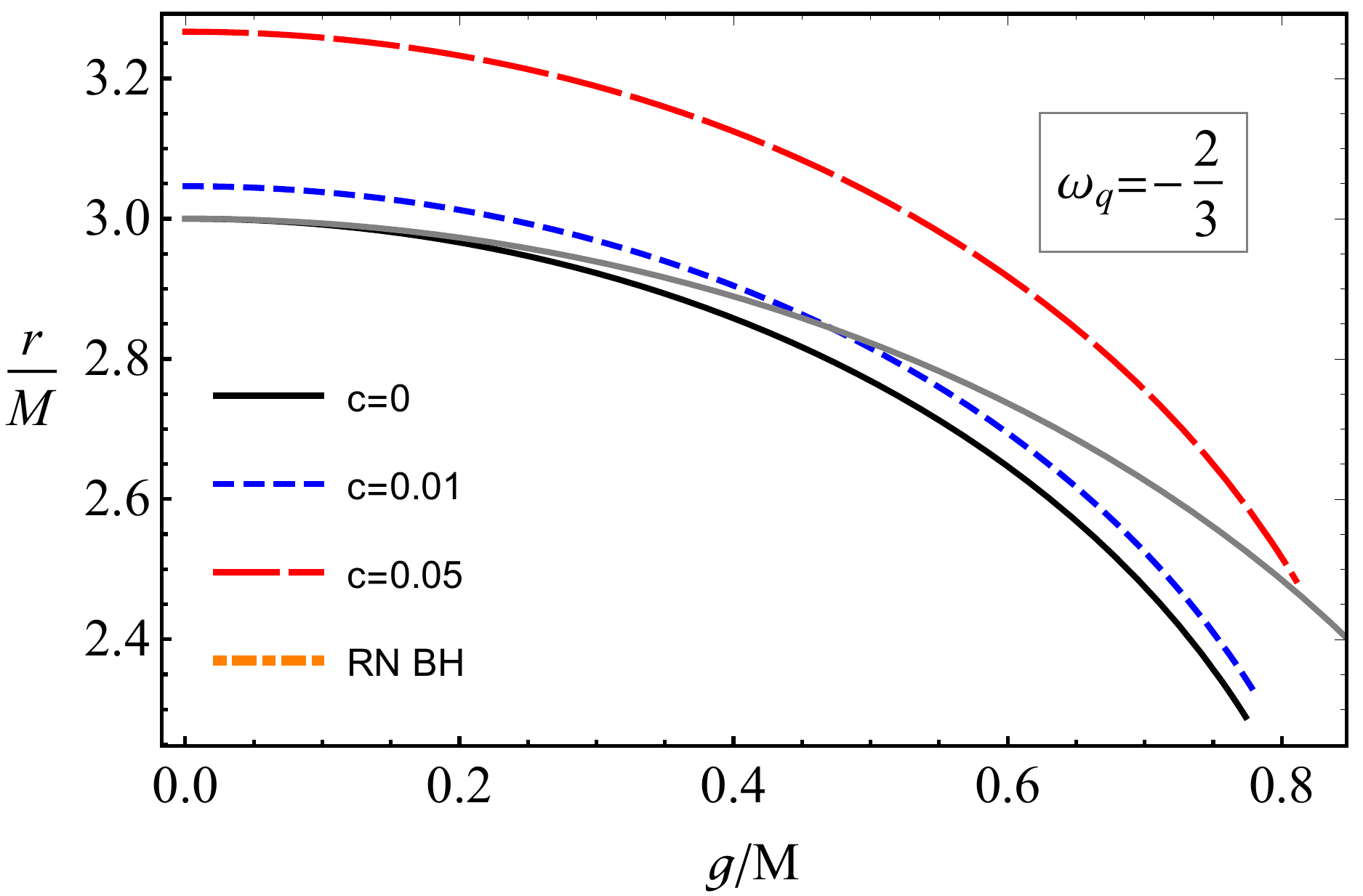}
    \caption{Circular null geodesics of massless particles around a BK BH  for different values of $c$, and a comparison with the RN BH results.}
    \label{rneutrinofig}
\end{figure}

Taking into account Eqs.(\ref{consts1})-(\ref{norm4vel}),  the effective potential for the massless particles ($\epsilon = 0$) can be obtained by  considering $$\frac{\dot{r}^2}{{\cal L}^2}+V_{\rm eff}=\frac{{\cal E}^2}{{\cal L}^2}\, ,$$ here
\begin{equation}
V_{\rm eff}= \frac{f(r)}{r^2}~.
\end{equation}
using the standard definition for circular orbits, $V'_{\rm eff}(r)=0$, one can obtain the circular orbits for the massless particle which implies 
\begin{equation}\label{rneutrino}
\frac{3 c \left(\omega _q+1\right)}{r^{3 \omega _q-1}}+\frac{6 M r^4}{\left(g^2+r^2\right)^{5/2}}-2 =0,
\end{equation}
and the impact parameter for circular null geodesics of the massless particle in the BK BH spacetime is given by the relation \cite{Dima19ApJ}
\begin{equation}
b^2_{\rm c(neu)}=\frac{ r^2}{f(r)}\,\Bigg|_{r=r_{\rm neu}} .  
\end{equation}
Eq.(\ref{rneutrino}) is hereby solved numerically, as its analytical solution is quite complicated. The numerical results for the circular orbits of massless particles are presented in Fig. \ref{rneutrinofig}. The radius of the circular orbits in the spacetime of a BK BH  depends on the magnetic charge $g$  and $c$. 

\begin{figure}[h]
    \centering
    \includegraphics[width=0.98 \linewidth]{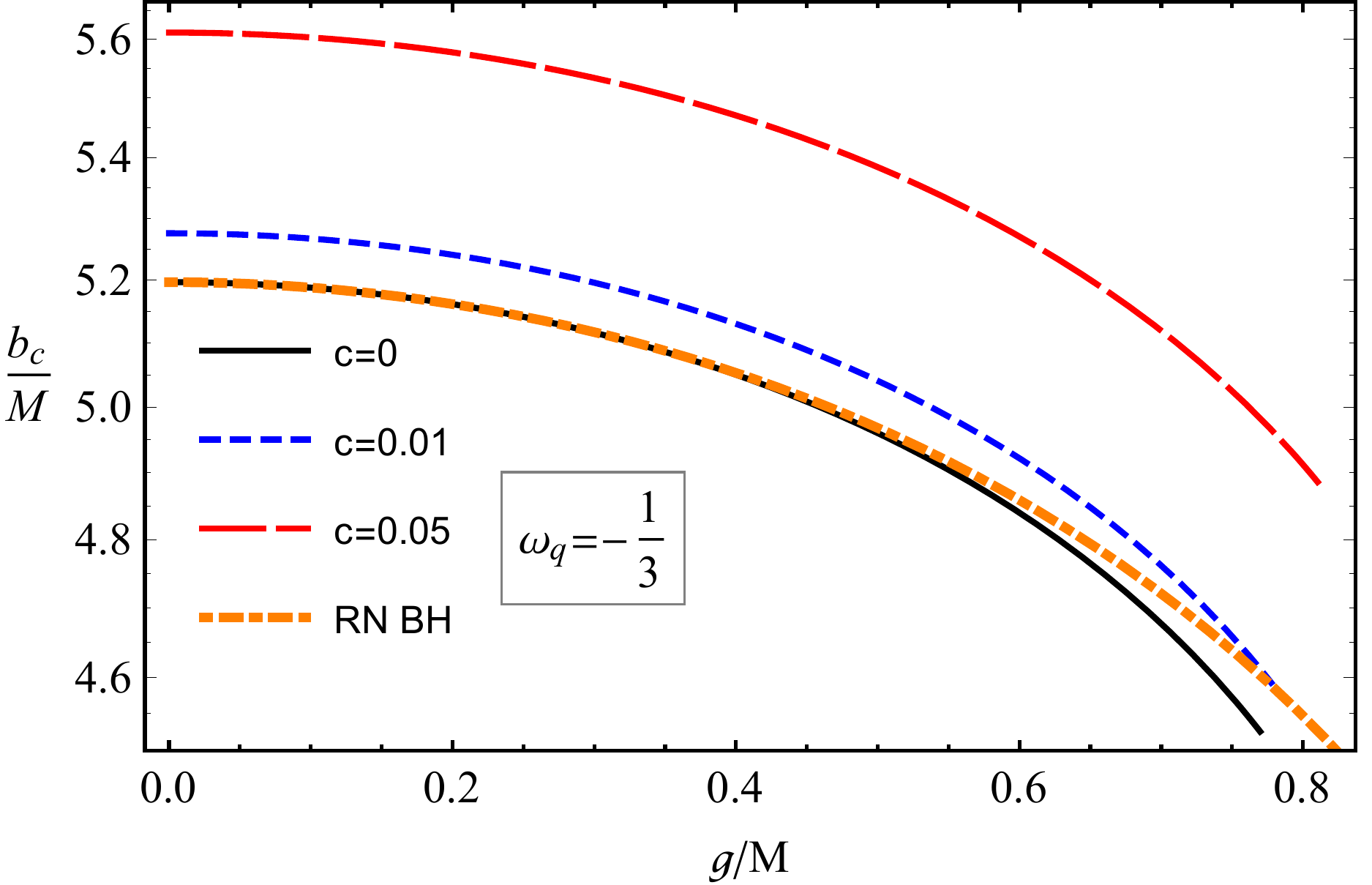}
    \includegraphics[width=0.98 \linewidth]{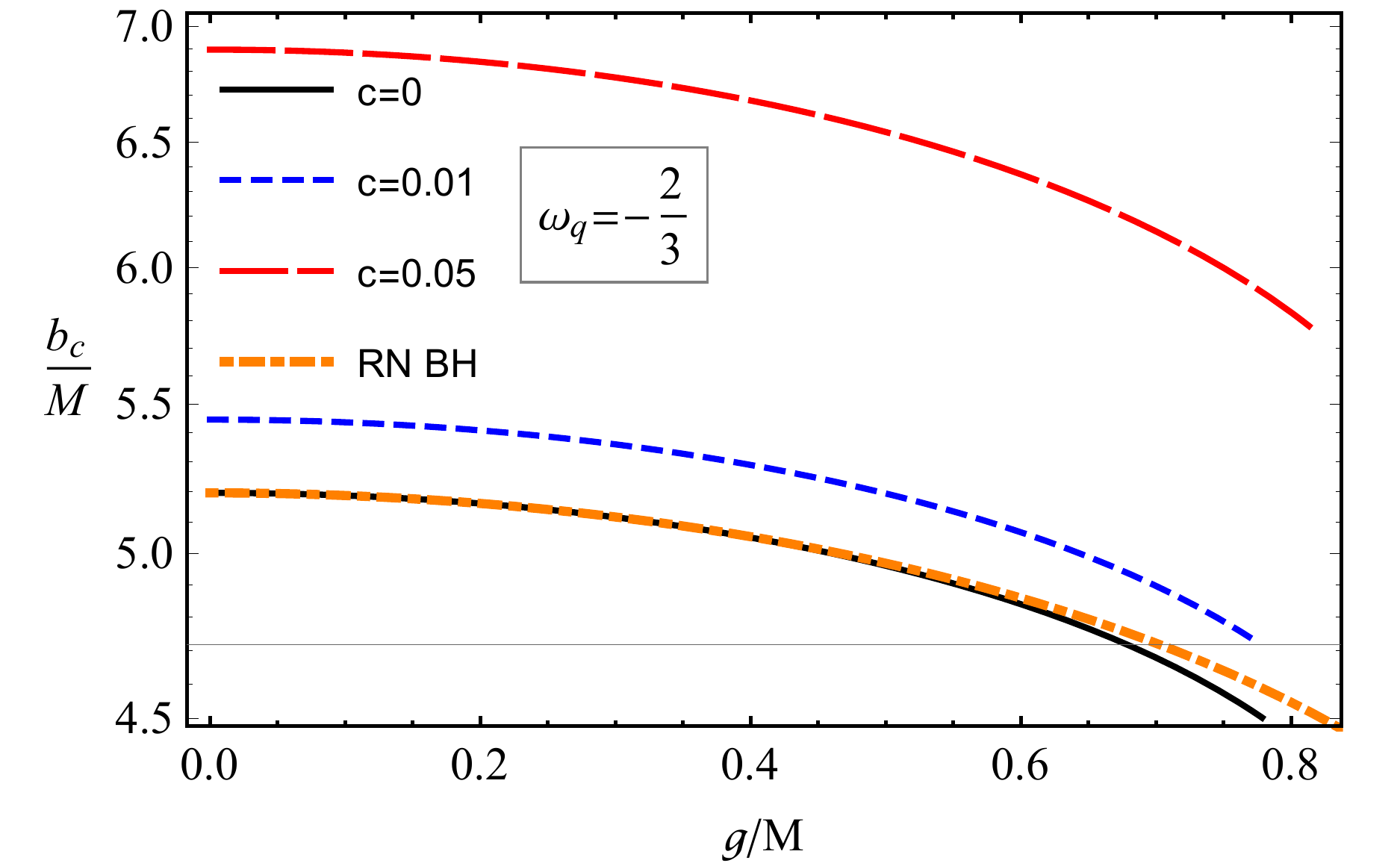}
    \caption{The dependence of the impact parameter for circular orbits of photons on the BK BH  $g$, for different values of the parameter $c$ and $\omega_q$.}
    \label{impactparfig}
\end{figure}

\section{Astrophysical Applications: BK BH  vs Kerr BH}

One of the central issues in the analysis of observational data from BHs and neutron stars is to test gravitational theories in the strong gravitational field regime.  An important concern is that different free parameters in many modified gravity theories may demonstrate similar or same effects on observational properties of particle motion. Still it is possible to differentiate a black hole from a naked singularity through various means such as gravitational lensing \cite{Virbhadra:2002ju}, geodesic precession of gyroscopes \cite{Rizwan:2018lht}, perihelion precession, shadows and accretion disk dynamics \cite{Dey:2020haf,Joshi:2013dva}. However, in order to tell an astrophysical black hole to be either spinning or not, a technique of X-ray reflection spectroscopy is quite effective \cite{Zhu:2020cfn}.

Many astrophysical black holes are modelled using Kerr black hole or Kerr-like black holes in the modified gravity theories \cite{Rayimbaev2020PhRvD,Abdujabbarov2020PDU,Juraeva2021EPJC,Abdujabbarov2020Galaxy,TurimovPhysRevD2020}, however static and spherically symmetric black holes in different modified gravity theories may still be used as model for astrophysical black holes due to their simplicity. In such cases, the astrophysical effects of black hole spin on particle motion may also mimic the effects due to modified gravity. Consequently, in this section we like to discuss and compare the gravitational effects of spin of Kerr black hole and charge of BK BH on ISCO radius, energy release efficiency, shadow size and twin pick QPO frequencies.

\subsection{In the same ISCO}

As it is observed already that the existence of the magnetic charge $g$ of a BH and the spin depreciate the size of ISCO for the test particles. If the ISCO lies much closer to the BH, than it is challenging to differentiate the measurements of the ISCO radius in the astronomical observations. The degeneracy values of the BKBH  parameters $g$, $c$  and spin $a$ of the Kerr BH provides the same radius of ISCO, will be shown in this section, and we show how to categorize the BHs through (direct or indirect) measurements of the ISCO radius.
The ISCO radius of test particles moving around rotating Kerr BHs for prograde  and retrograde orbits has the following form ~\cite{Bardeen72}
\begin{eqnarray}
r_{\rm isco}= 3 + Z_2 \pm \sqrt{(3- Z_1)(3+ Z_1 +2 Z_2 )} \ ,
\end{eqnarray}
with
\begin{eqnarray} \nonumber
Z_1 &  = & 
1+\left( \sqrt[3]{1+a}+ \sqrt[3]{1-a} \right) 
\sqrt[3]{1-a^2} \ ,
\\ \nonumber
Z_2^2 & = & 3 a^2 + Z_1^2 \ .
\end{eqnarray}
\begin{figure}[h!]
   \centering
  \includegraphics[width=0.9\linewidth]{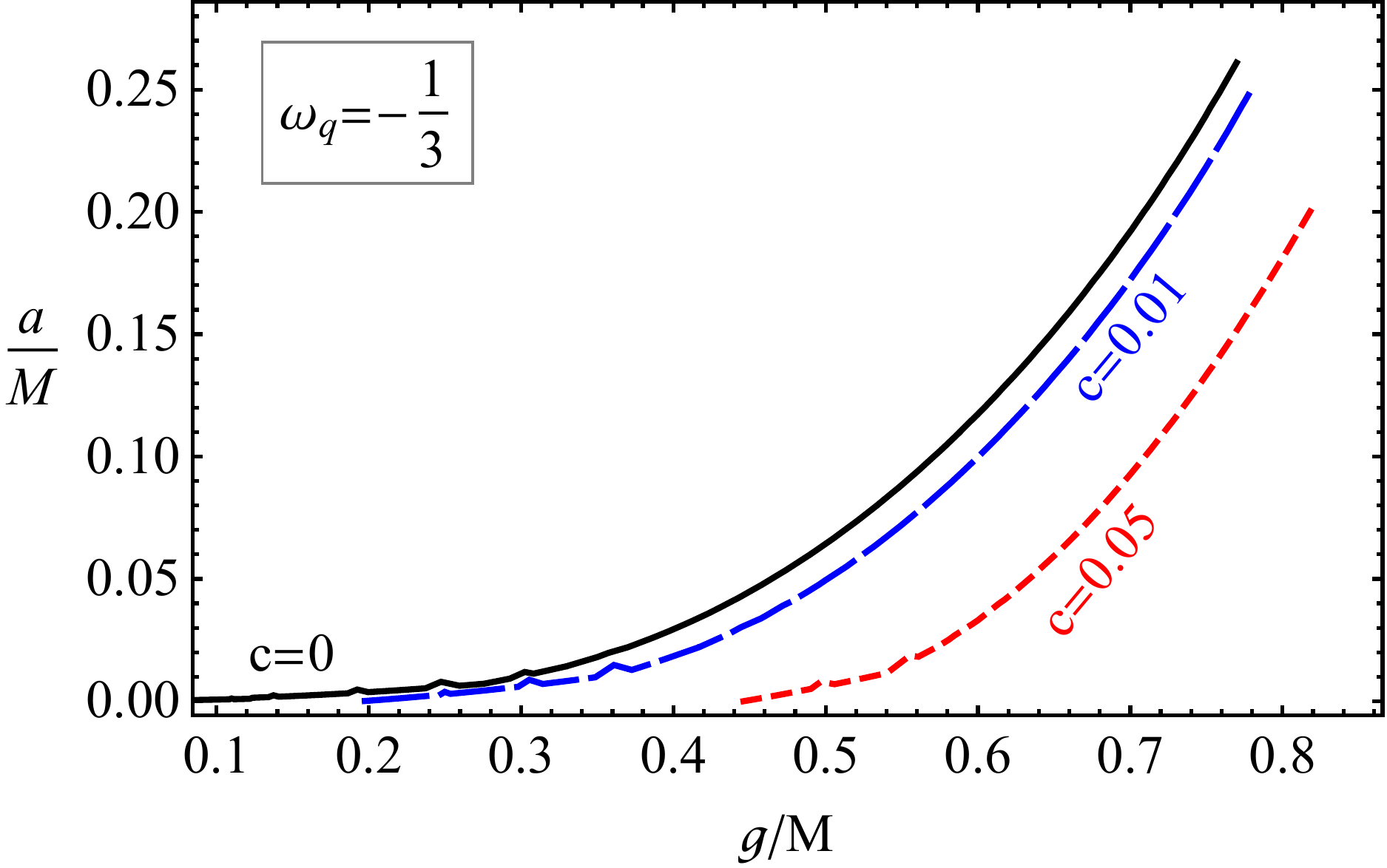}
  \includegraphics[width=0.9\linewidth]{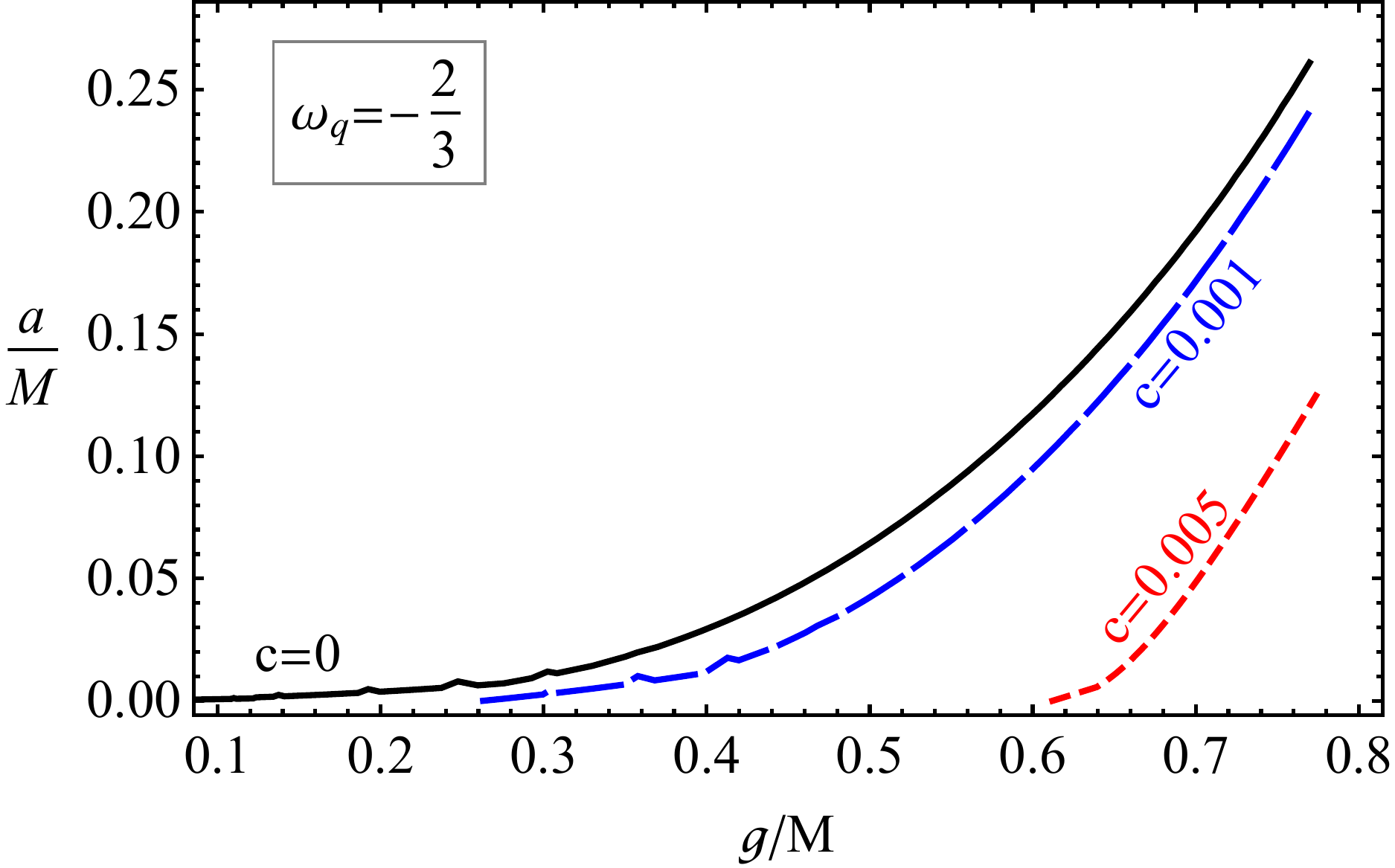}
	\caption{The degeneracy relations between the spin of a Kerr BH and a BKBH  providing the same value of the ISCO radius for different values of the BKBH parameters \label{mimic1}}
\end{figure} 

Figure~\ref{mimic1} illustrates relations between spin and the magnetic charge of the BH. Note  that the BK BH has the same size of the ISCO as that for the Kerr BH, for different values of $c$. The numerical analysis of the radius of the ISCO of a neutral particle moving around the BK BH  shows that for $c=0.01$ $g/M \in (0.2,0.78)$ and it mimics a (pin parameter of the Kerr BH), with $a\approx 0.25$ it is in the range $g/M \in (0.41, 0.82)$ and  $a \in (0, 0.2)$ (the results have also been shown in our previous works in Refs.\cite{Rayimbaev2020PhRvD,BokhariPhysRevD2020}). In our previous work Ref.\cite{Rayimbaev2020PhRvD} we have shown that the charge of a special type of regular BH in general relativity combined with nonlinear electrodynamics can mimic the spin of a rotating Kerr BH up to $a/M=0.8$, while the charge of a BH in Einstein-Maxwell-scalar gravity can mimic up to $a/M=0.936$ \cite{TurimovPhysRevD2020}, the charge of a BK BH  up to $a/M=0.9$ ~\cite{Narzilloev2020EPJC1}, and the charge of a deformed RN BH with the deformation parameter $\epsilon=6.17$ up to $a/M=0.88$~\cite{BokhariPhysRevD2020}.

\subsection{The Bolometric Luminosity}

Another astrophysics application of the studies of the energy efficiency of an accretion disk of a BH is its connection with the bolometric luminosity of the accretion disk: $L_{bol}=\eta \dot{M}c^2$, where $\dot{M}$ is the rate of the accretion matter falling  in to the central BH~ \cite{Bian2003PASJ,BokhariPhysRevD2020}. The bolometric luminosity can be calculated through astronomical observations. Our main target here is to explore whether the magnetic charge $g$ of the BK BH  can mimic the spin of the Kerr BH, resulting in the same value for the energy efficiency-- the bolometric luminosity.

\begin{figure}[ht!]
   \centering
\includegraphics[width=0.9\linewidth]{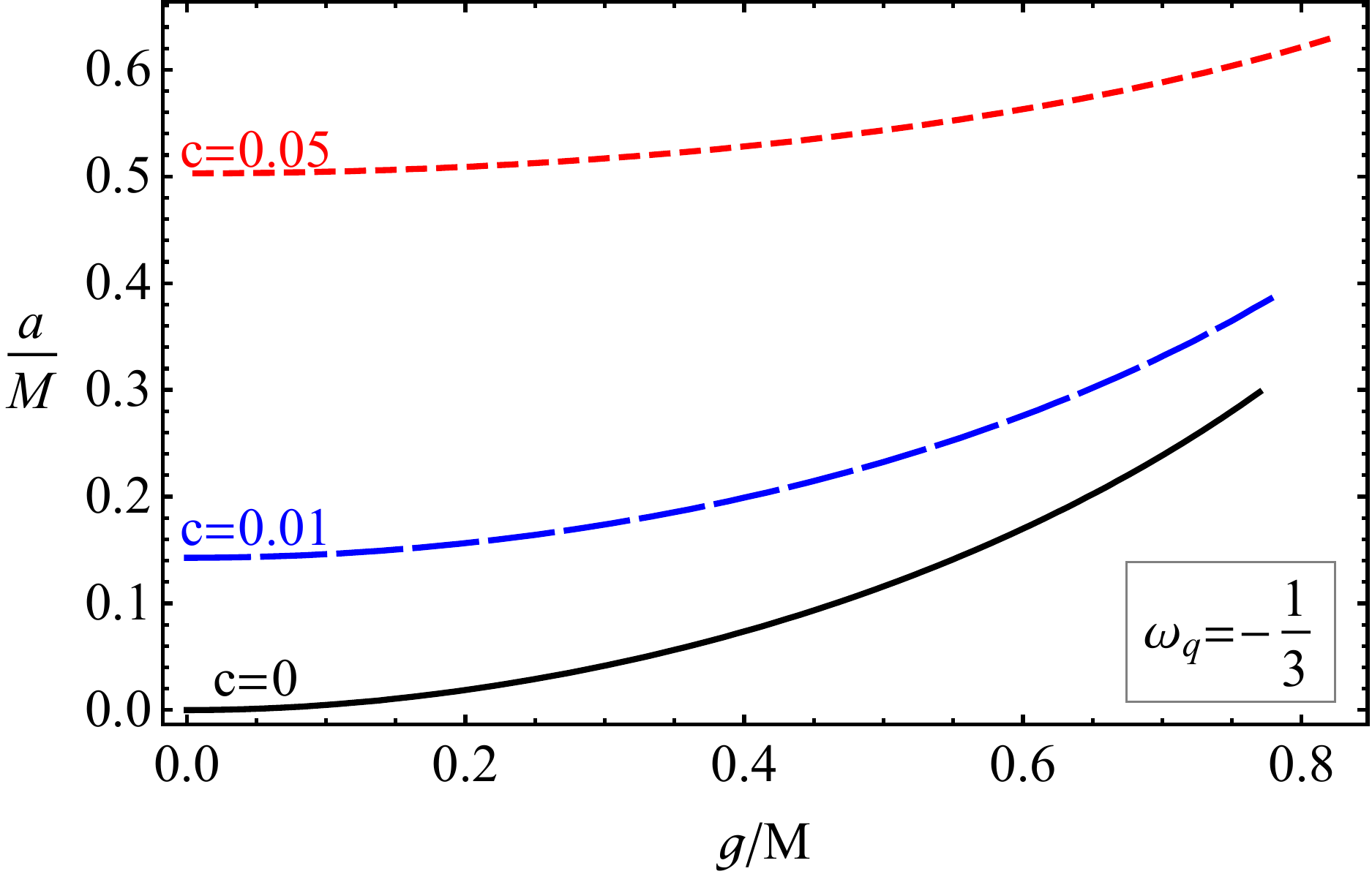}
\includegraphics[width=0.9\linewidth]{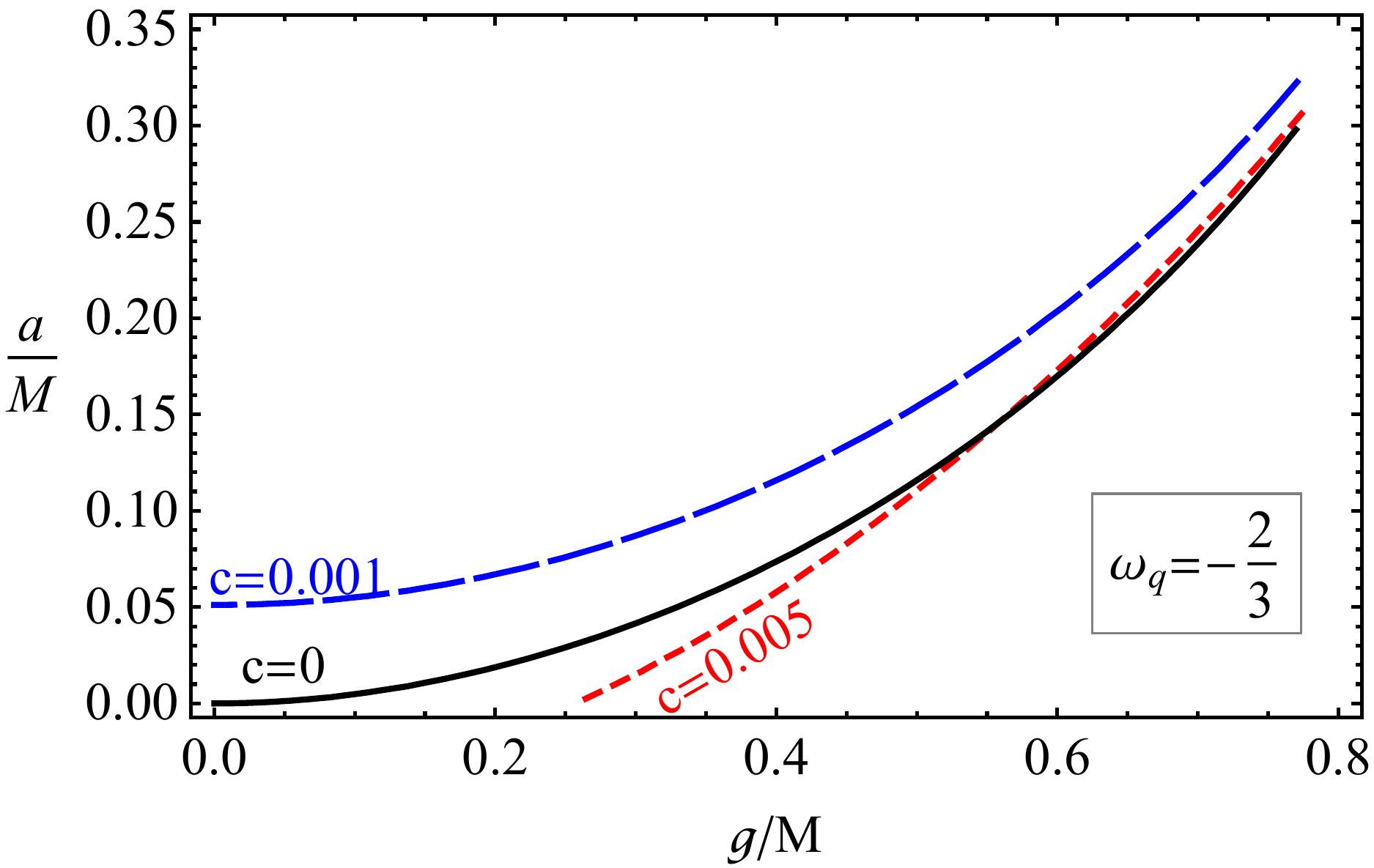}
	\caption{Degeneracy relations between the spin of the rotating Kerr BH and the magnetic charge of the BK BH  giving the same value of the energy efficiency for different values of the parameter $c$. \label{effQvsa}}
\end{figure}
One can see from Fig.\ref{effQvsa} that the magnetic charge of the BK BH  can mimic the spin of the Kerr BH up to $a/M=0.25$, an increase of the coupling parameter causes decreasing of the mimic value. 

\subsection{QPO frequencies}

In this subsection we will discuss possible frequencies of twin-pick QPOs around a BK BH with the comparison of the pure Schwarzschild, RN and Kerr BHs in various models of twin-pick QPOs \cite{Stuchlik2016AA}.
Here we discuss the relations between possible lower and upper values of twin-pick low and high frequency QPOs around a BH in its possible values of the BH parameters in various QPO models'  namely, relativistic precession (RP)\cite{Stella1999ApJ} where  the upper and lower frequencies of twin pick QPOs are identified by the radial and orbital frequencies as $\nu_U=\nu_\phi$ and $\nu_L=\nu_\phi-\nu_r$, respectively.

\begin{figure*}[ht!]
   \centering
\includegraphics[width=0.490\linewidth]{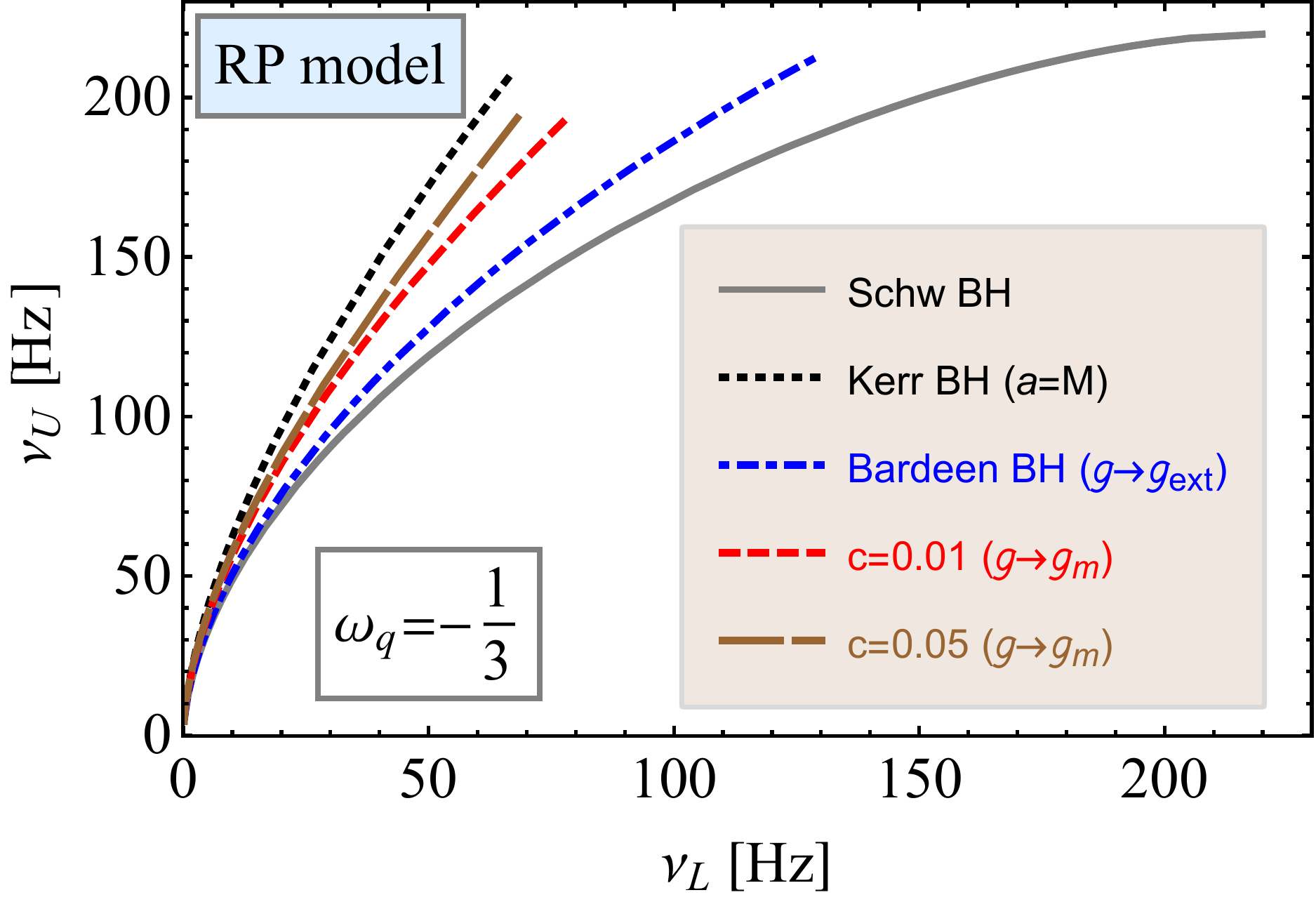}
\includegraphics[width=0.490\linewidth]{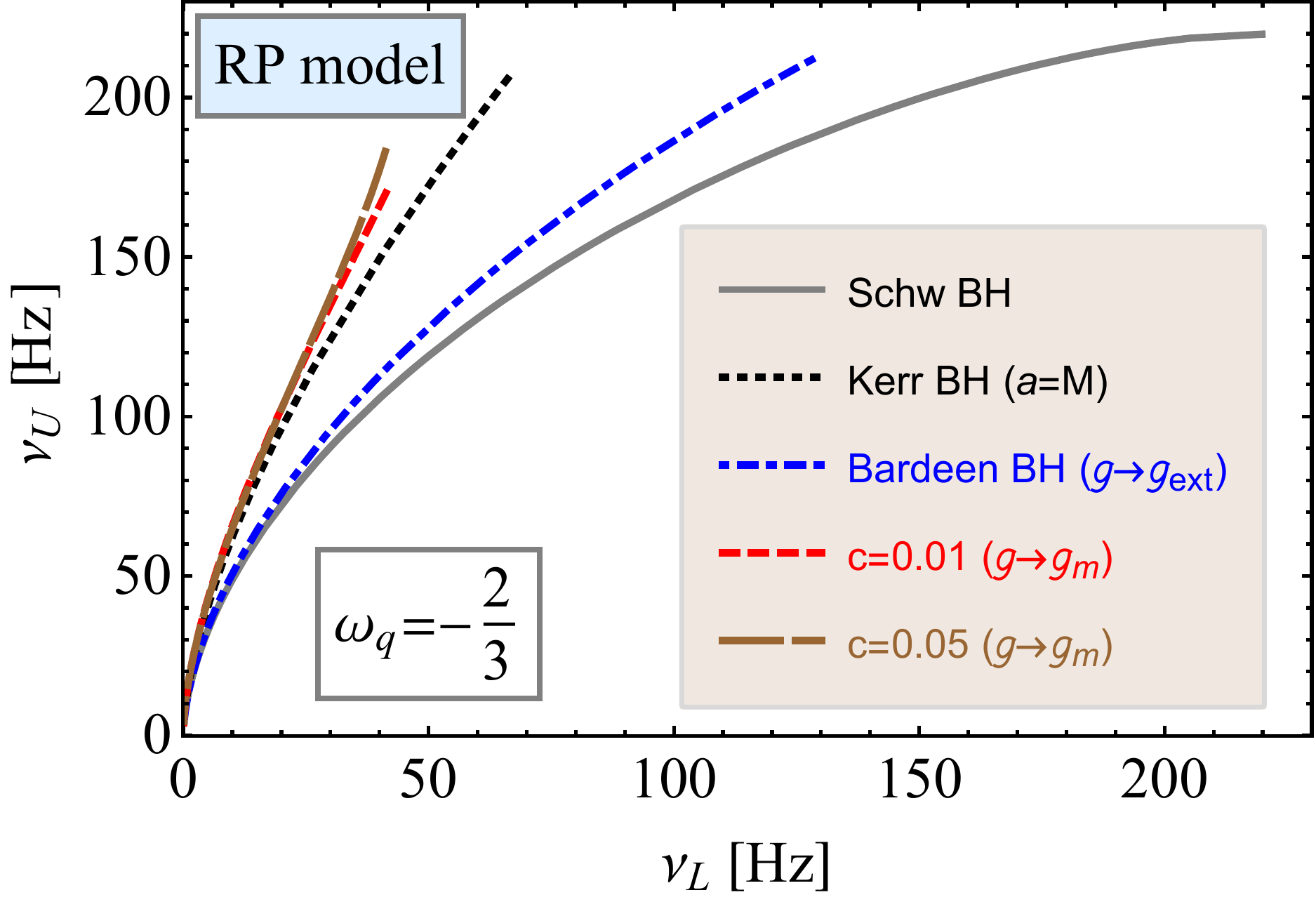}
\caption{Relations between the frequencies of upper and lower picks of twin-pick QPOs in the RP model. \label{QPO}}
\end{figure*}

Figure~\ref{QPO} illustrates relations between upper and lower frequencies of twin-pick QPOs of the RP model for QPOs analysis. Here we take $10M_\odot$ for the BH mass, when we carry out the numerical calculations. In the figure, the area between the grey and black dotted (blue dot-dashed) lines implies the possible values of the frequency of twin-pick QPOs around an extreme rotating Kerr (charged regular Bardeen) BH. 
If a QPO object having two picks at the spectrum with exact upper and lower frequencies which lies in the area between the grey and black dashed lines in $\nu_U-\nu_L$ space, the central BH could be a Kerr BH, if the QPO lies between the grey and blue dashed lines the BH could both the Kerr BH  and the regular Bardeen BH with their respectively corresponding parameters, or in other words, in this case, there is an area where the Bardeen BH can mimic the Kerr one in terms of the same values of the QPO frequencies, and it is impossible to distinguish what type of the BH is in the center. The mimic area is expanding with increasing  $c$, and it is fast with respect to the increase of $\omega_q$. Then, this important feature may be helpful to distinguish the gravity models describing the central object of twin pick QPOs.  

\subsection{The shadow size/impact parameter }

Finally, as a last application of our studies, we compare effects of the spin parameter of the Kerr BH and the magnetic charge of the regular Bardeen BH for different values of the parameters $c$ and $\omega_q$. 
The equation for photon impact parameters for co- and contour- rotating cases reduces to the following form 
\begin{eqnarray}
 &&2b_*^{\pm}=a_*^2\pm a_*+3 \left({\cal A}+\frac{9}{{\cal A}}+6\right) \\\nonumber &&+\sqrt{2 a_*^2-3 {\cal A}-\frac{27}{{\cal A}}+\frac{2 a_* \left[a^2-27\right]}{a_*^2+3 \left({\cal A}+\frac{9}{{\cal A}}+6\right)}+36}\ ,
 \label{Kerr-impact}
\end{eqnarray}
where 
\begin{eqnarray}
 {\cal A}^3&=&2a_*^2 \sqrt{a_*^4+27} -2 a_*^4-27\ ,
\end{eqnarray}
 where the notation $^*$ in the impact and spin parameters mean that they are the  normalized value to the BH mass as $b_*=b/M, \ a_*=a/M$.

\begin{figure}[h!]
   \centering
  \includegraphics[width=0.9\linewidth]{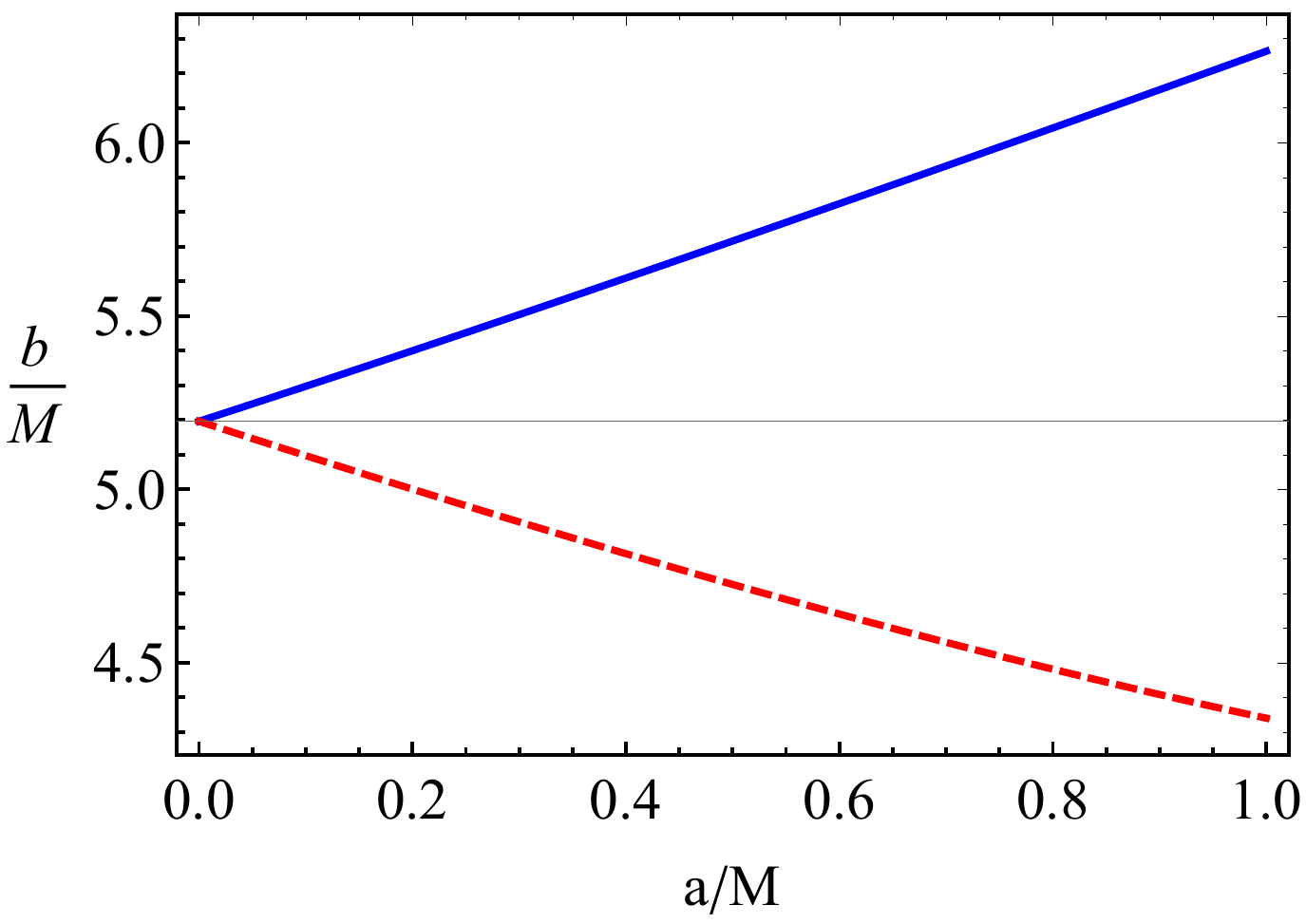}
	\caption{Impact parameter for photons as a function of the spin of the Kerr BH. The blue solid line for co-rotating and the red dashed one for contour-rotating photons. \label{impactKerr}}
\end{figure} 

Figure \ref{impactKerr} demonstrates the dependence of the impact factor for photons around a Kerr BH on the dimensionless spin parameter for co-rotating and contour-rotating photons. We can see from the figure that the impact parameter splits increasing and decreasing for prograde and retrograte orbits. A similar question arises whether the Bardeen BH charge can mimic the spin of a Kerr BH, in terms of the same values of the photon impact parameter. We have seen in Fig.\ref{impactparfig} that an increase of the quintessential parameter $c$ causes to increase the impact parameter. Now, here, we compare effects of the spin and magnetic charge parameters by equalling Eqs.(\ref{Kerr-impact}) and (\ref{impacteq}). It is impossible to get analytical relations between the two parameter, and we provide the relation numerically for different values of the parameters $\omega_q$ and $c$.     

\begin{figure}[h!]
   \centering
  \includegraphics[width=0.9\linewidth]{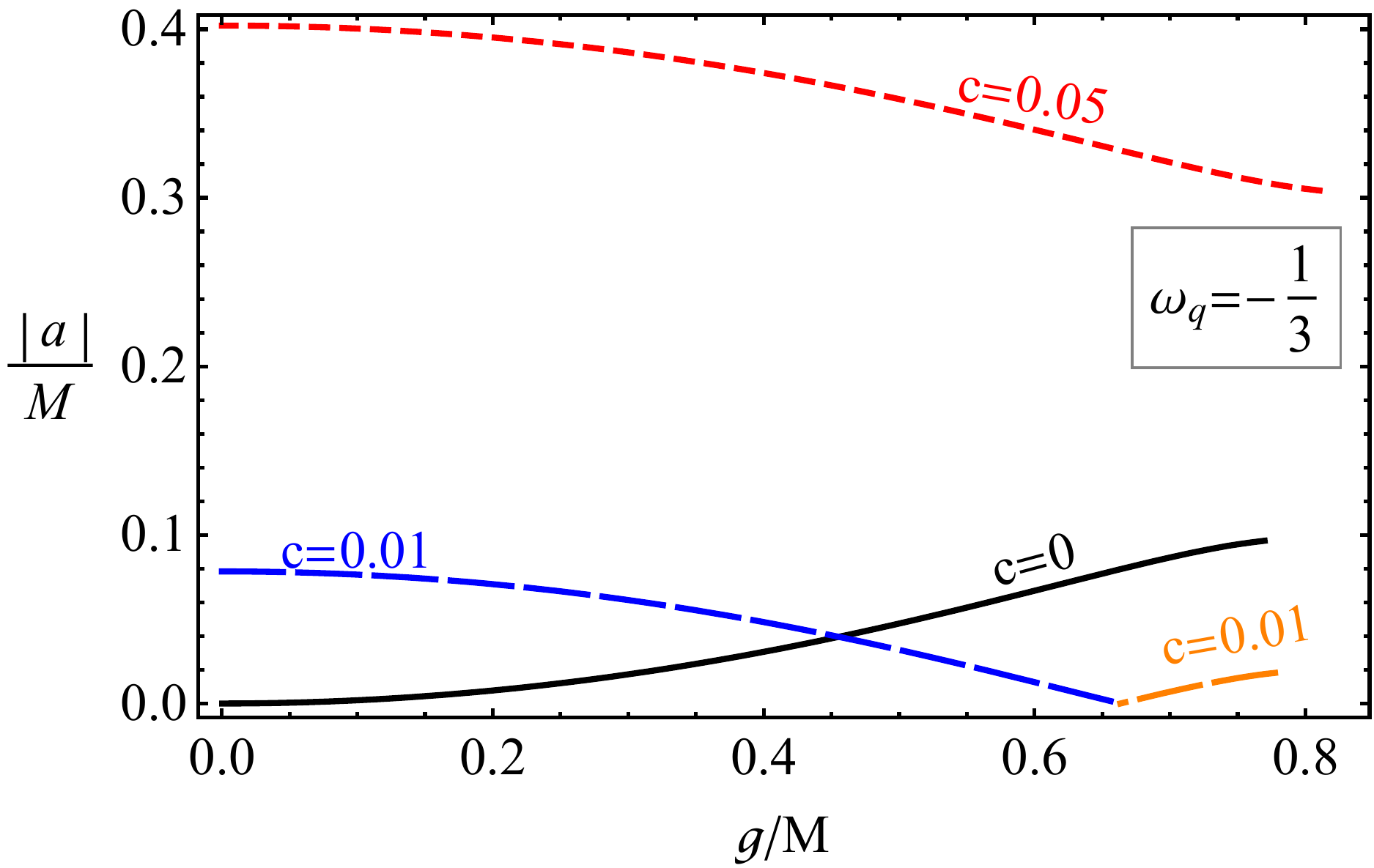}
	\caption{The degeneracy relations between the spin of a Kerr BH and a BKBH,  providing the same value of the BH shadow size for different values of the BKBH  \label{mimicimpact}}
\end{figure} 

One can see from Fig.\ref{mimicimpact} that a pure Bardeen BH charge can mimic the spin of a Kerr BH up to $a/M=0.09655$, while in the presence of the quintessential field with the parameters $c=0.01, \omega_q=-1/3$ ($c=0.05, \omega_q=-1/3$) it can mimic it in the range $a/M \in (0.3044, 0.4022)$ of contour-rotating orbits of photons.

\section{Quasinormal modes and their connection with the shadow radius}
    It is well known that QNMs encode information about the gravitational waves emitted by a BH in the  final stage, also known as the ringdown stage. It was found that if the BH radiates in the eikonal limit, then there is a link between the real part of the QNMs and the angular velocity of the last circular, null geodesic. On the other hand, the imaginary part of QNMs is linked to the Lyapunov exponent, which encodes information about the instability timescale of the orbit \cite{Cardoso:2008bp}
\begin{equation}
	\omega_{\rm QNM}=\Omega_c l -i \left(n+\frac{1}{2}\right)|\lambda|.
\end{equation}
This correspondence is expected to be valid not only for static spacetimes but also for rotating spacetimes. It was also shown in Ref.\cite{Stefanov:2010xz} that there is a correspondence between the QNMs in the eikonal limit  and the strong lensing. Following this line of thought, it was argued that the real part of QNMs can be obtained via the BH shadow radius (see, for details \cite{Jusufi:2019ltj,Cuadros-Melgar:2020kqn})
\begin{equation}\label{k1}
	\omega_{\Re} = \lim_{l \gg 1} \frac{l}{R_{sh}},
\end{equation}
which is precise in the eikonal limit having large values of $l$. It follows that
\begin{equation}
	\omega_{\rm QNM}=\lim_{l \gg 1} \frac{l}{R_{sh}} -i \left(n+\frac{1}{2}\right)|\lambda|.
\end{equation}

With the increase of the sensitivity of detectors and the possibility of detecting modes beyond the fundamental ones, this profound connection can have important observational implications in the near future. We can use the real part of QNMs to estimate the BH shadow radius or vice versa. From the physical point of view, it's not very hard to see this connection if we treat the gravitational waves as massless particles propagating along the last null unstable \cite{Stefanov:2010xz} and out to infinity. 
Let us rewrite the effective geometry given by (see also \cite{Toshmatov:2021fgm})
\begin{equation}
ds^2_{\rm eff}=-A(r)dt^2+B(r)dr^2+D(r) (d\theta^2+\sin^2\theta d \phi^2)
\end{equation}
where
\begin{eqnarray}
A(r) &=& \frac{f(r)}{L_{\rm F}} \ ,  \\
B(r) &=& \frac{1}{f(r)L_{\rm F}} \ ,   \\
D(r) &=& \frac{r^2}{L_{\rm F}+2 F L_{\rm FF}}.
\end{eqnarray}

Starting from the metric and using the condition $\dot{r}=\ddot{r}=0$, one can find that the circular null geodesics are obtained by solving the equation
\begin{eqnarray}
A(r) D(r)'-A(r)'D(r)=0.
\end{eqnarray}

If we consider a photon along null geodesics in our spherically symmetrical spacetime surrounded by matter, one can show that the Hamiltonian can be written as \cite{Perlick:2015vta}
\begin{equation}
\frac{1}{2}\left[-\frac{p_{t}^{2}}{A(r)}+\frac{p_{r}^{2}}{B(r)}+\frac{p_{\phi}^{2}}{D(r)}\right] =0.
\label{EqNHa}
\end{equation}

Due to the spacetime symmetries related to the coordinates $t$ and $\phi$, there are two constants of motion defined as follows
\begin{eqnarray}
p_{t}&\equiv\frac{\partial H}{\partial\dot{t}}=-E.\\
p_{\phi}&\equiv\frac{\partial H}{\partial\dot{\phi}}=L.
\end{eqnarray}

\begin{figure*}[ht!]
   \centering
\includegraphics[width=0.450\linewidth]{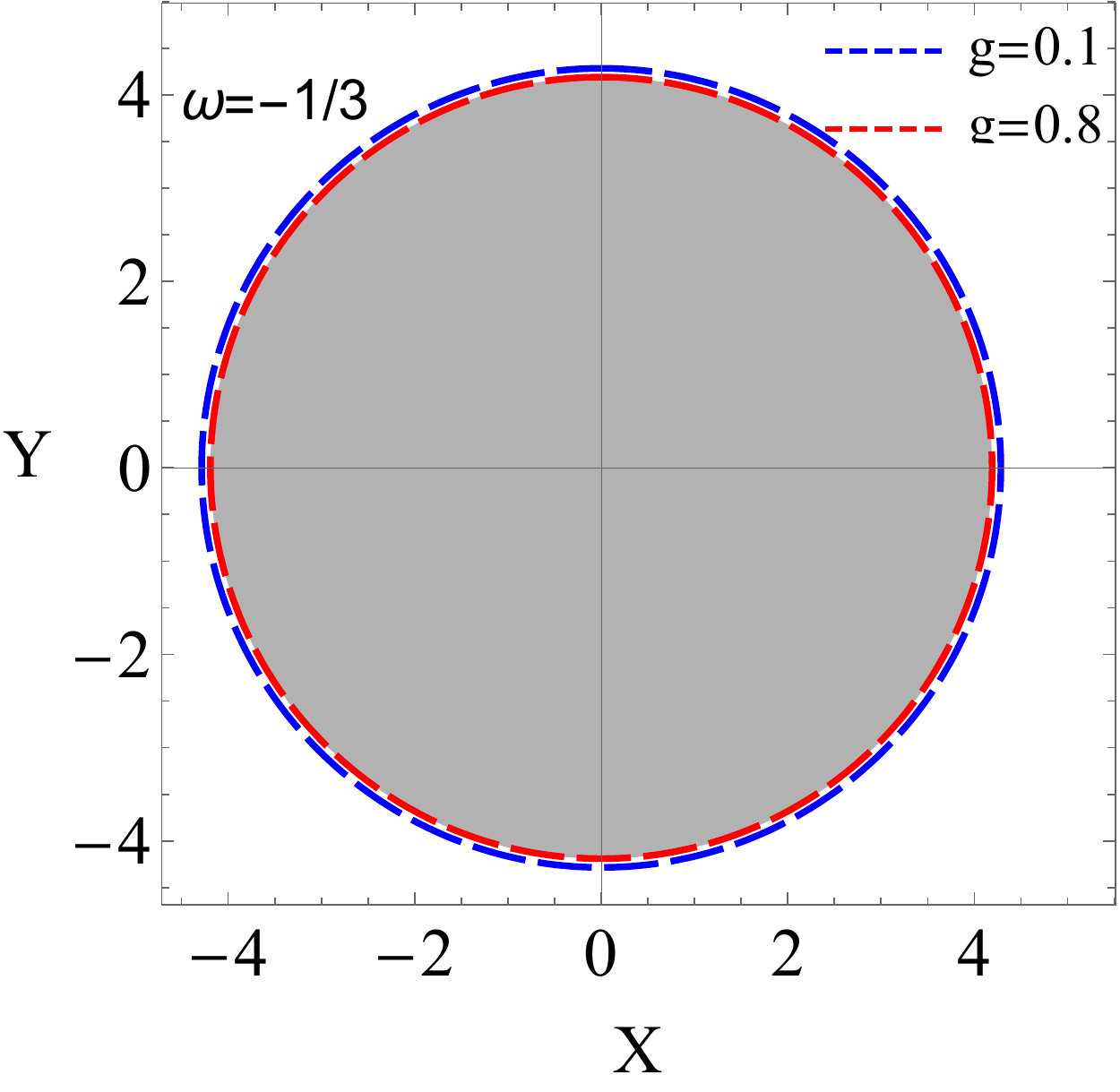}
\includegraphics[width=0.450\linewidth]{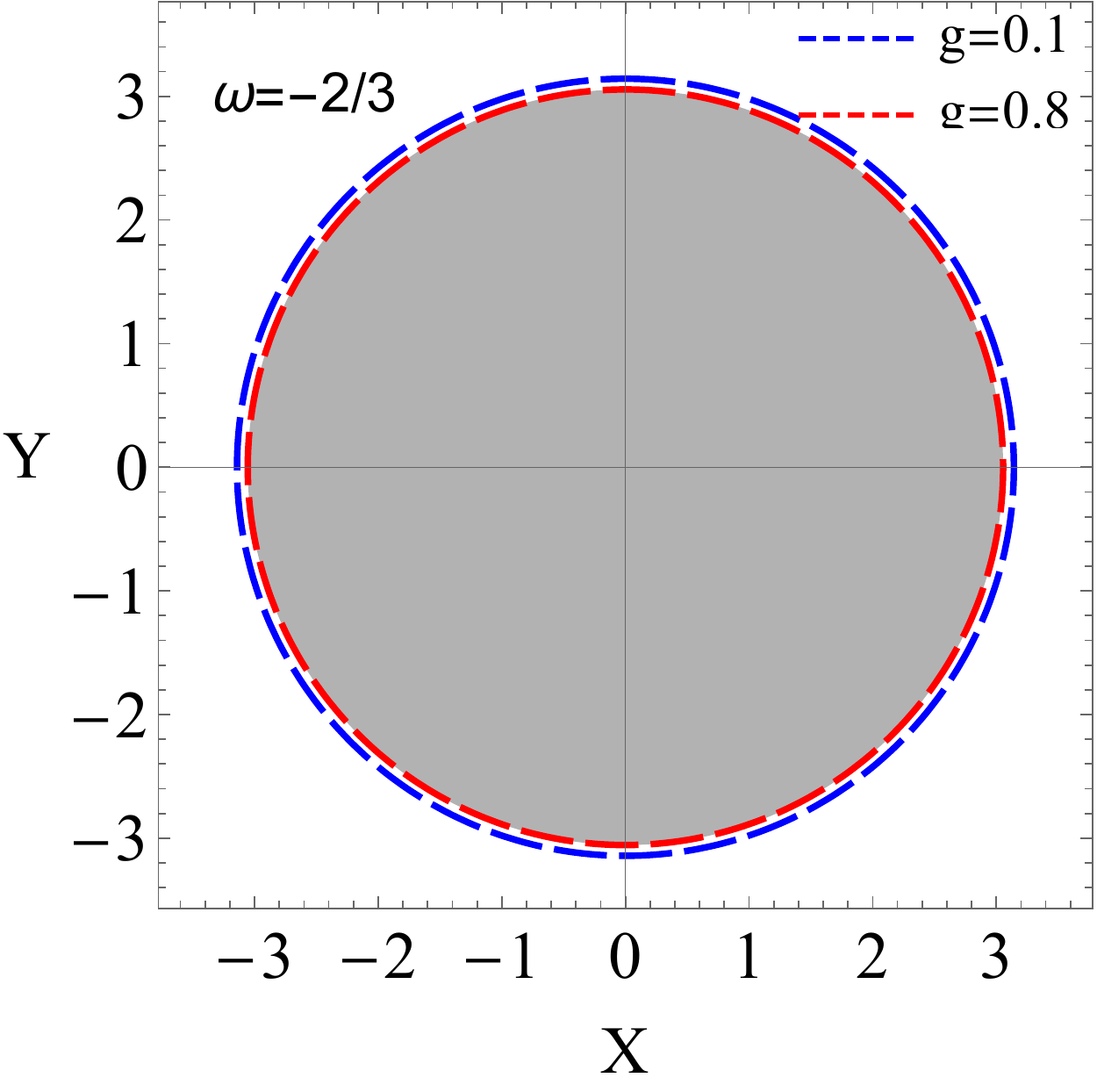}
\caption{Left panel: Shadow images of a magnetized BH where we have set $M=1$, $c=0.01$, $\omega_q=-1/3$ and $r_o=10^4M$. Right panel: Shadow images of the magnetized BH where we have set $M=1$, $c=0.01$, $\omega_q=-2/3$ and $r_o=50M$.   }
\end{figure*}

In the last two equations, $E$ and $L$ are the energy and angular momentum of the photon, respectively. Next, the  circular and unstable orbits are related to the maximum value of the effective potential in terms of the following conditions
\begin{equation}
V_{\rm eff}(r) \big \vert_{r=r_{p}}=0,  \qquad \frac{\partial V_{\rm eff}(r)}{\partial r}%
\Big\vert_{r=r_{p}}=0,  
\end{equation}
 
Without going into details here, one can now show the following equation of motion 
\begin{equation}
\frac{dr}{d\phi}=\pm \frac{\sqrt{D(r)}}{\sqrt{B(r)}}\sqrt{\left[\frac{h^{2}(r)}{h^{2}(_{\rm ph})} -1\right] }. 
\end{equation}
where $h^2(r)=D(r)/A(r)$.
Let us consider a light ray sent from a static observer located at a position $r_{0} $ and transmitted with an angle $\vartheta$ with respect to the radial direction. We, therefore, have \cite{Perlick:2015vta}
\begin{equation}
\cot \vartheta =\frac{\sqrt{g_{rr}}}{\sqrt{g_{\phi\phi}}}\frac{dr}{d\phi}\Big\vert%
_{r=r_{0}}.  \label{Eqangle}
\end{equation}

In this case the spacetime is not asymptotically flat, hence to compute the shadow radius we need to take into account the finite distance corrections yielding
\begin{eqnarray}
\nonumber
R_{sh}&=&\sqrt{\frac{f(r)(L_{\rm F}+2 F L_{\rm FF})}{L_{\rm F}}\vline_{r=r_o}}\\ &\times& \sqrt{\frac{L_{\rm F}}{L_{\rm F}+2 F L_{\rm FF}}\frac{r^2}{f(r)}\,\vline _{r=r_{\rm ph}} }
\end{eqnarray}
where $r_o$ represents the observer's position located at some large distance.
\begin{figure*}[ht!]
   \centering
\includegraphics[width=0.450\linewidth]{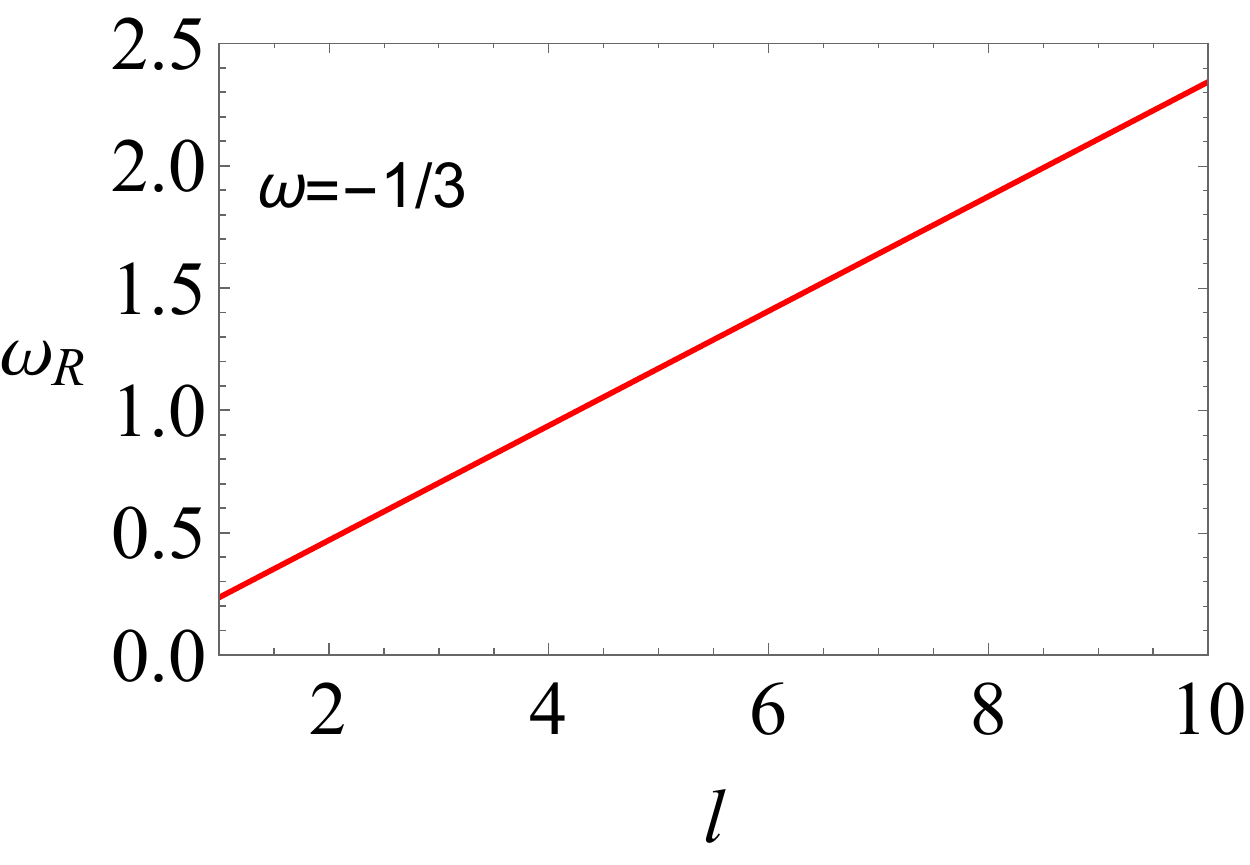}
\includegraphics[width=0.450\linewidth]{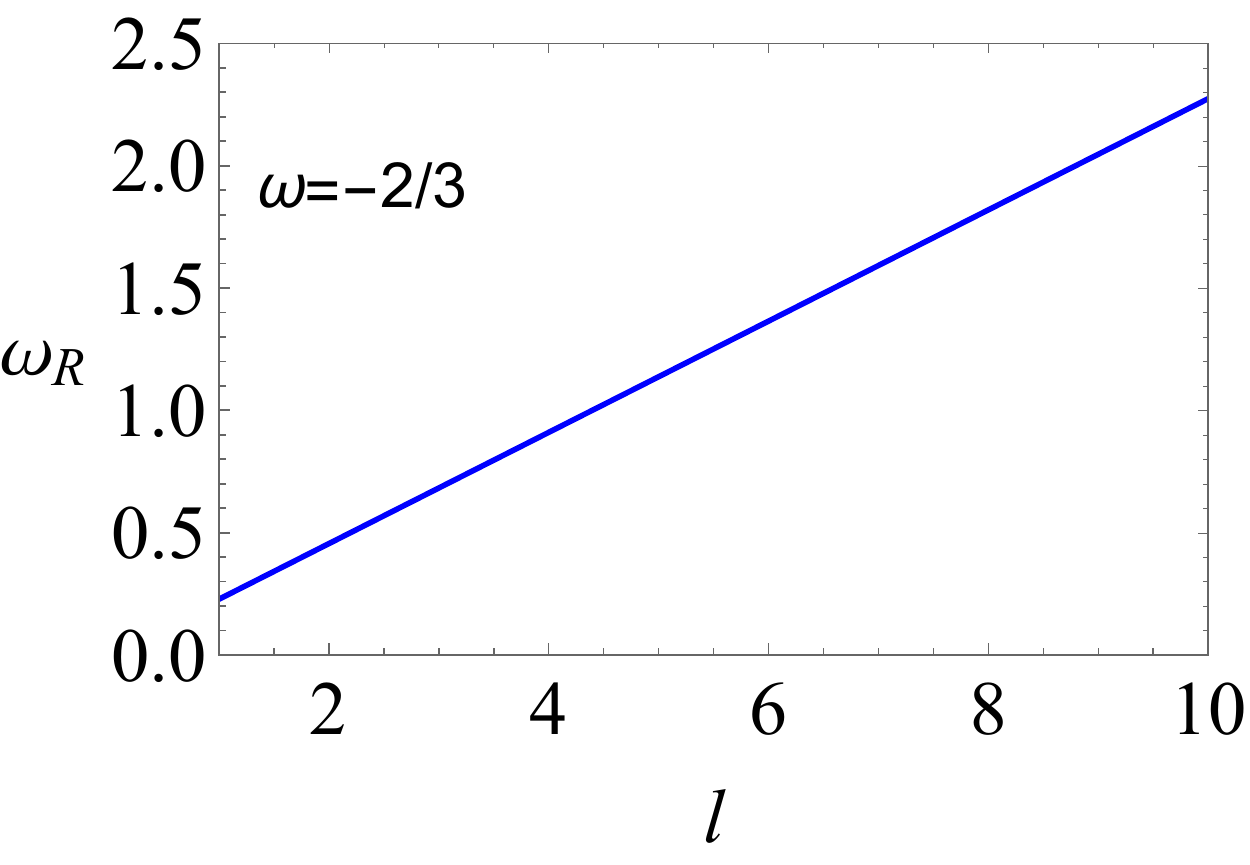}
\caption{Left panel: the real part of QNMs for the case $\omega_1=-1/3$ as a function of $l$ with $M=1$, $c=0.01$ and $g=0.5$. Right panel: the real part of QNMs for the case $\omega_1=-2/3$  as a function of $l$ with $M=1$, $c=0.01$ and $g=0.5$.  }
\end{figure*}
However, it is natural to assume that at very large distances where the observer is located, the magnetic charge vanishes $[F|_{r_o} \to 0]$, and also $2M/r|_{r_o} \to 0$, hence, the shadow radius simplifies to 
\begin{eqnarray}
\nonumber
R_{sh}&=&\sqrt{1-\frac{c}{r^{3 \omega_q+1}}\vline_{r=r_o}} \\ &\times& \sqrt{\frac{L_{\rm F}}{L_{\rm F}+2 F L_{\rm FF}}\frac{r^2}{f(r)}\,\vline _{r=r_{\rm ph}} }.
\end{eqnarray}

\begin{table}[tbp]
        \begin{tabular}{|l|l|l|l|l|l}
        \hline
   \multicolumn{1}{|c|}{ } &  \multicolumn{1}{c|}{  $g=0.1$ }
    & \multicolumn{1}{c|}{  $g=0.3$ } &   \multicolumn{1}{c|}{  $g=0.5$ }
    & \multicolumn{1}{c|}{  $g=0.7$ } \\ 
    \hline
  $l$ & \quad $\omega_{\Re}$ & \quad $\omega_{\Re}$ & \quad $\omega_{\Re}$
  & \quad $\omega_{\Re}$  \\  
        \hline
1 & 0.23226 & 0.232938 & 0.234304  &  0.236319 \\
2 &  0.46452 &  0.465876 &  0.468608 &  0.472639   \\
3 &  0.69678 &  0.698814 &  0.702912 &  0.708958 \\
4 &  0.92904 & 0.931751 &  0.937217  &  0.945278 \\
5 &  1.16130 & 1.164690 &  1.171520 &  1.181600  \\
6 & 1.39356  & 1.397630 &  1.405820 &  1.417920  \\
8 & 1.85808 & 1.863500 &  1.874430 & 1.890560  \\
10 &  2.32260 & 2.329380 &  2.343040  & 2.363190   \\
        \hline
        \end{tabular}
         \caption{ \label{table4} Numerical values of the shadow radius and the
    real part of QNMs frequencies obtained  with $c=0.01$ and $\omega_q=-1/3$. }
\end{table}

\begin{table}[tbp]
        \begin{tabular}{|l|l|l|l|l|l}
        \hline
   \multicolumn{1}{|c|}{ } &  \multicolumn{1}{c|}{  $g=0.1$ }
    & \multicolumn{1}{c|}{  $g=0.3$ } &   \multicolumn{1}{c|}{  $g=0.5$ }
    & \multicolumn{1}{c|}{  $g=0.7$ } \\ 
    \hline
  $l$ & \quad $\omega_{\Re}$ & \quad $\omega_{\Re}$ & \quad $\omega_{\Re}$
  & \quad $\omega_{\Re}$  \\  
        \hline
1 &  0.225028 & 0.225824  & 0.227449  &  0.229925  \\
2 &  0.450057  & 0.451648 & 0.454898 &  0.459850     \\
3 & 0.675085  & 0.677472 & 0.682347  & 0.689776  \\
4 & 0.900114 &  0.903297 & 0.909796  & 0.919701   \\
5 & 1.125140 & 1.129120 & 1.137240 & 1.149630    \\
6 & 1.350170 & 1.354940  & 1.364690 & 1.379550    \\
8 & 1.800230 & 1.806590 & 1.819590 &  1.839400   \\
10 & 2.250280  & 2.258240  & 2.274490 & 2.29925     \\
        \hline
        \end{tabular}
         \caption{ \label{table4} Numerical values of the shadow radius and the
    real part of QNMs frequencies obtained  with $c=0.01$ and $\omega_q=-2/3$. }
\end{table}

We have used the last equation and in Fig. (23) we show the size of the shadow images for an observer located at some distance $r_0$ from the BH. It is important to say that the finite distance corrections are very important  in the computation of the shadow radius, given the fact the effective metric of the BH is not asymptotically flat. In the case $\omega_q=-2/3$, the effect of non-asymptotically flat spacetime is stronger, hence, if we set the parameter $c=0.01$ and assume that $r_o<100 M$, in our case we have used $r_0=50M$ in Eq. (57). If we assume that the correspondence between the shadow radius and the real part of QNMs remains valid, it follows that the frequency oscillation is affected by the effective geometry of the spacetime. From Fig. (23) we see that the size of the shadow radius decreases with the increase of the magnetic charge $g$. If we assume the correspondence (43) to be true that means the real part of QNMs is also modified and affected by the non-trivial topology of the spacetime. In order to compute the QNMs we can use again Eq. (57), however, we observed that it is reasonable to rescale the expression for the shadow radius given by Eq. (57) as follows, 
\begin{eqnarray}
R'_{sh}=\sqrt{\frac{L_{\rm F}}{L_{\rm F}+2 F L_{\rm FF}}\frac{r^2}{f(r)}\,\vline _{r=r_{\rm ph}} },
\end{eqnarray}
where we have used 
\begin{equation}
R'_{sh} \to \frac{R_{sh}}{\sqrt{1-c/r_0^{3 \omega_q+1}}},
\end{equation}
as a result we obtain 
\begin{eqnarray}
R'_{sh}=\lim_{l >>1} \frac{l}{\omega_{\Re}}.
\end{eqnarray}
Since the shadow radius is measured in units of black hole mass, but if the spacetime is not asymptotically flat, the ADM mass and ADM energy of the system as measured by an observer located at $r_o$ changes. The QNMs can be computed by the following equation
\begin{eqnarray}
\omega_{\Re}=\lim_{l >>1}\frac{l}{\sqrt{\frac{L_{\rm F}}{L_{\rm F}+2 F L_{\rm FF}}\frac{r^2}{f(r)}\,\vline _{r=r_{\rm ph}} }}.
\end{eqnarray}

Using the last equation, in Tables II and III, we show the numerical values for the real part of QNMs given a specific domain of the parameters. These values corresponds to electromagnetic perturbations having the effect of nonlinear
electrodynamics. We clearly observe that the value of QNMs frequency increases with the increase of $g$. Note that the gravitational QNMs in spacetime of the Bardeen BH with quintessence were investigated in \cite{Saleh:2018hba} by means of the WKB method.  It is also important to note that, in the present work, in the correspondence (43) we have considered the effect of nonlinear electrodynamics on the spacetime geometry, i.e. effective geometry, otherwise, the correspondence between the shadow radius and QNMs can be violated. Previously, it was argued that the relation between QNMs and the unstable circular null geodesics can be violated in general relativity with black hole solutions combined with non-linear electrodynamics \cite{Stuchlik:2014qja,Toshmatov:2018tyo,Toshmatov:2019gxg}. Note that, there is another possibility or an alternative way to see the above problem by rescaling the $\omega_{\Re}$, that is, to identify the QNM frequency with the energy of the particle measured by an observer located at some distance $r_o$, hence
\begin{eqnarray}
\omega_{\Re} \rightarrow E|_{r_o},
\end{eqnarray}
in case of not asymptotically flat spacetime, the energy of the particle measured at infinity changes (we can call it ADM energy of the particle) such that the correspondence with the shadow radius is described via Eq. (60). 

\section{Conclusions}
In this article, we have studied the dynamics of neutral particles moving around the BK BH (BKBH). This work focuses on the effects of the normalization parameter $c$, the magnetic charge $g$ of the Bardeen BH and the equation of state parameter $\omega_q$ on the dynamics of the particles. We started the calculations with the scalar invariants (Ricci and Kretchmann scalars)  for the BH metric and analysed the spacetime structure in the vicinity of a BK BH. It is observed that all these scalars are well-defined at the centre of the BH, so there is no curvature singularity, and as the values of $c$ increase the scalar invariants at the origin of the BH also increase. 
Then we discussed the horizon structure of the BH, the horizon of the BH becomes larger as the values of the normalization parameter $c$ increase. Also, for larger values of the magnetic charge of the BH, its horizon becomes smaller. A detailed comparison of the horizon radius for different values of $c$ and $g$ is given in Table 1. 
The other findings of our calculations are as given below:
We have discussed the dynamics of neutral particles and the specific energy and specific angular momentum of the particle corresponding to the circular motion of the particle are obtained. The behaviour of the radial profile of the specific energy and angular momentum show that as the normalization parameter increases (decreases) the specific angular
momentum of the particle increases (decreases), also
as the values of $c$ increases the specific angular
momentum of the particle becomes large, and
it will eventually reach values quite closer to the BH as compared to the case
when $c$ is smaller. As the value of $c$ increases (decreases) the specific energy of the particle decreases (increases). 
The behaviour of the ISCO of the particle moving around the BH is also studied numerically. It is observed that for large values of $c$ the ISCO is at large radius as compared to the case when $c$ is small, e.g. for $c=0, g=0$ (the Schwarzschild BH) ISCO is at $r=6 M$ but for $c=0.05$ and for non-zero values of $g$ the ISCO becomes $7 M$.
Also, increasing values of $g$ the ISCO radius  reduces, which  means $g$ acts as a source of the gravitational force and brings the particle closer to the BH. Also, at larger value of $-1\leq \omega_q \leq -1/3$ the ISCO becomes smaller. This shows that the quintessence scalar field is also responsible for attracting particles towards the BH centre. 
Next we test whether BK BH can mimic the Kerr BH, we observe the effects of the BH charge $g$ can  mimic the effects of a rotation of Kerr BH. It is shown in Fig.(\ref{mimic1}) that the radius of ISCO of the particle moving around the BK BH and the rotating Kerr BH has same  behaviour for range $a\in (0,0.25), g/M \in (0.1, 0.78) $ for $c\leq 0.01$ and $\omega_q=-1/3$ and for $\omega=-2/3$ the range is  $a\in (0,0.25), g/M \in (0.1, 0.77) $.

Taking into account the finite distance corrections encoded by, $r_o$ we have shown how the parameters $c$ and $g$ affects the shadow radius. In particular, we have examined the cases $\omega_q=-1/3$ and $\omega_q=-2/3$ and shown that with the increase of $g$ the shadow radius decreases. This shows that, in one hand the  effect of effective geometry, and on the other hand the location of observer are quite important to find the shadow radius given the fact the spacetime geometry is non-asymptotically flat. Finally, we have also presented numerical values for the real part of QNMs obtained by means of the shadow radius and found that QNMs increase with the increase of $g$. If the correspondence between the real part of QNMs and the shadow radius still holds, one can further interpret the change of the QNMs frequency due to the change of the total mass of the system (ADM mass), which should modify the frequency of the QNMs observed at some location $r_0$ from the black hole. One way to see this change is to use the optic-geometric correspondence and identify the real part of QNMs with the energy of the particle, changing the ADM mass of ADM energy of the system, means the energy of the particle observed at some location $r_0$ also shift.  

\bibliographystyle{bibstyle}
\bibliography{references}

\end{document}